\documentclass[12pt]{article}
\usepackage{lscape}
\usepackage{geometry}
\usepackage[onehalfspacing]{setspace}
\usepackage{amssymb}
\usepackage{amsfonts}
\usepackage{graphicx}
\usepackage{amsmath}
\usepackage{bbm}
\usepackage{epstopdf}
\usepackage{natbib}
\usepackage{bibunits}
\usepackage{morefloats}
\usepackage{float}
\usepackage{comment}
\usepackage{pdflscape}
\usepackage{adjustbox}
\usepackage{subcaption}
\usepackage{longtable}
\usepackage{apalike}
\usepackage{xfrac}
\usepackage[dvipsnames,table]{xcolor}
\usepackage{etoolbox,siunitx}
\robustify\bfseries

\usepackage{hhline,multirow}
\usepackage{yfonts}

\usepackage{longtable}
\usepackage{threeparttablex}

\newcolumntype{d}[1]{D{.}{.}{#1}}

\usepackage{qtree}
\usepackage{multicol}
\usepackage{booktabs}
\usepackage{threeparttable}
\usepackage{tabularx}
\usepackage{dcolumn}
\newcolumntype{d}[1]{D..{#1}} 

\usepackage{siunitx}
\usepackage{mathpazo} 
\usepackage[scaled]{helvet} 
\usepackage{courier} 
\normalfont
\usepackage[T1]{fontenc}
\usepackage{listings}
\usepackage{epigraph}
\def\sym#1{\ifmmode^{#1}\else\(^{#1}\)\fi}

\DeclareMathOperator*{\argmin}{arg\,min}

\usepackage[pdftex,bookmarks,colorlinks]{hyperref}
\hypersetup{
  colorlinks,
  citecolor=darkpowderblue,
  linkcolor=red,
  urlcolor=blue}
\usepackage{cleveref}

\usepackage{etoolbox}

\definecolor{dukeblue}{rgb}{0.0, 0.0, 0.61}
\definecolor{darkred}{rgb}{0.8,0,0}

\makeatletter
\patchcmd{\epigraph}{\@epitext{#1}}{\itshape\@epitext{#1}}{}{}
\makeatother

\makeatletter
\def\munderbar#1{\underline{\sbox\tw@{$#1$}\dp\tw@\z@\box\tw@}}
\makeatother

\usepackage{graphicx}
\usepackage{wrapfig}

\usepackage{algorithm,algorithmic}

\usepackage{xspace}
\usepackage{microtype}

\definecolor{bred}{RGB}{122, 0, 0}
\definecolor{darkpowderblue}{rgb}{0.0, 0.05, 0.5}

\definecolor{dpd}{rgb}{0.0, 0.05, 0.5}
{}
 \definecolor{dpd2}{rgb}{0.0, 0.043, 0.43}

\makeatletter
\newcommand\halftiny{\@setfontsize\halftiny\@vipt\@viipt}
\newcommand\notsotiny{\@setfontsize\notsotiny{6.99}{9.2828}}
\newcommand\notsolarge{\@setfontsize\notsolarge{12}{14}}
\makeatother

\usepackage{soul}

\renewenvironment{abstract}
 {\small
  \begin{center}
  \bfseries \abstractname\vspace{-.5em}\vspace{0pt}
  \end{center}
  \list{}{
        \setlength{\leftmargin}{1.8cm}    \setlength{\rightmargin}{\leftmargin}  }  \item\relax}
 {\endlist}
\begin{document}
 \sloppy

    \title{\vspace*{0.9cm} \fontsize{18.95}{22}  \textbf{\color{dpd} \textls[-19]{An Adaptive Moving Average for Macroeconomic Monitoring \\ \phantom{.}}} \vspace*{0.25cm}}
\author{\hspace*{-0.3cm} Philippe Goulet Coulombe\thanks{%
Département des Sciences Économiques,  \href{mailto:p.gouletcoulombe@gmail.com}{\texttt{goulet\_coulombe.philippe@uqam.ca}}.  The views expressed in this paper do not necessarily reflect those of the Oesterreichische Nationalbank or the Eurosystem. This Draft: \today.}\\[-0.2cm] \hspace*{-0.3cm} \textbf{\texttt{\fontfamily{phv}\selectfont \notsolarge  Université du Québec à Montréal}} 
    \and 
    \hspace*{1.3cm} Karin Klieber \\[-0.2cm] \hspace*{1.5cm} \textbf{\texttt{\fontfamily{phv}\selectfont \notsolarge Oesterreichische Nationalbank}} 
}

\date{\vspace{0.7cm}
\small
\small 
\vspace{-0.25cm}
\large
  }
\maketitle

\begin{abstract}

\noindent The use of moving averages is pervasive in macroeconomic monitoring, particularly for tracking noisy series such as inflation.  The choice of the look-back window is crucial.  Too long of a moving average is not timely enough when faced with rapidly evolving economic conditions.  Too narrow averages are noisy,  limiting signal extraction capabilities.  As is well known,  this is a bias-variance trade-off.  However,  it is a time-varying one: the optimal size of the look-back window depends on current macroeconomic conditions.  In this paper,  we introduce a simple \textit{adaptive} moving average estimator based on a Random Forest using as sole predictor a time trend. Then, we compare the narratives inferred from the new estimator to those derived from common alternatives across series such as headline inflation, core inflation, and real activity indicators.  Notably, we find that this simple tool provides a different account of the post-pandemic inflation acceleration and subsequent deceleration.

\end{abstract}

\thispagestyle{empty}



\clearpage


\clearpage 
\setcounter{page}{1}

\newgeometry{left=2 cm, right= 2 cm, top=2.3 cm, bottom=2.3 cm}

\section{Introduction}\label{sec:intro}

Moving averages are widely used in macroeconomic monitoring. For example, the practice is so entrenched for inflation that many news outlets report year-over-year growth rates as the monthly inflation rate. However, there is little sacred about the twelve-months moving average, and the length of the window is very much a tuning parameter, as practitioners are well aware.


Figure \ref{fig:CPI_intro} below shows month-over-month US headline CPI inflation during two turbulent periods, overlaid with three-months (MA(3)) and twelve-months (MA(12)) moving averages. In both cases, the commonly used MA(12) not only lags behind MA(3) but also distorts visual interpretations of inflation dynamics. The three sharp drops in energy prices (November 2008, April 2020, July 2022) are given an extended release by the longer moving average, creating an impression of gradual change when, in reality, these were abrupt shifts. During the Great Recession, MA(3) inflation appears to be on target by early 2009, not 2010 as observed for MA(12). Throughout the post-Covid disinflation phase, the perceived slowdown in MA(12) is actually an elongation of a single abrupt shift in July 2022. Notably, from March 2022 to March 2024, inflation had \textit{stabilized} 1.5 percentage points above the 2\% target when viewed through MA(3). 

\vspace*{0.8em}

\begin{figure}[h!]
  \caption{\normalsize{One-sided Moving Averages for US CPI Inflation}} \label{fig:CPI_intro}
  
    \centering
    \vspace*{-0.8em}
      \includegraphics[width=0.99\textwidth, trim = 1mm 0mm 0mm 0mm, clip]{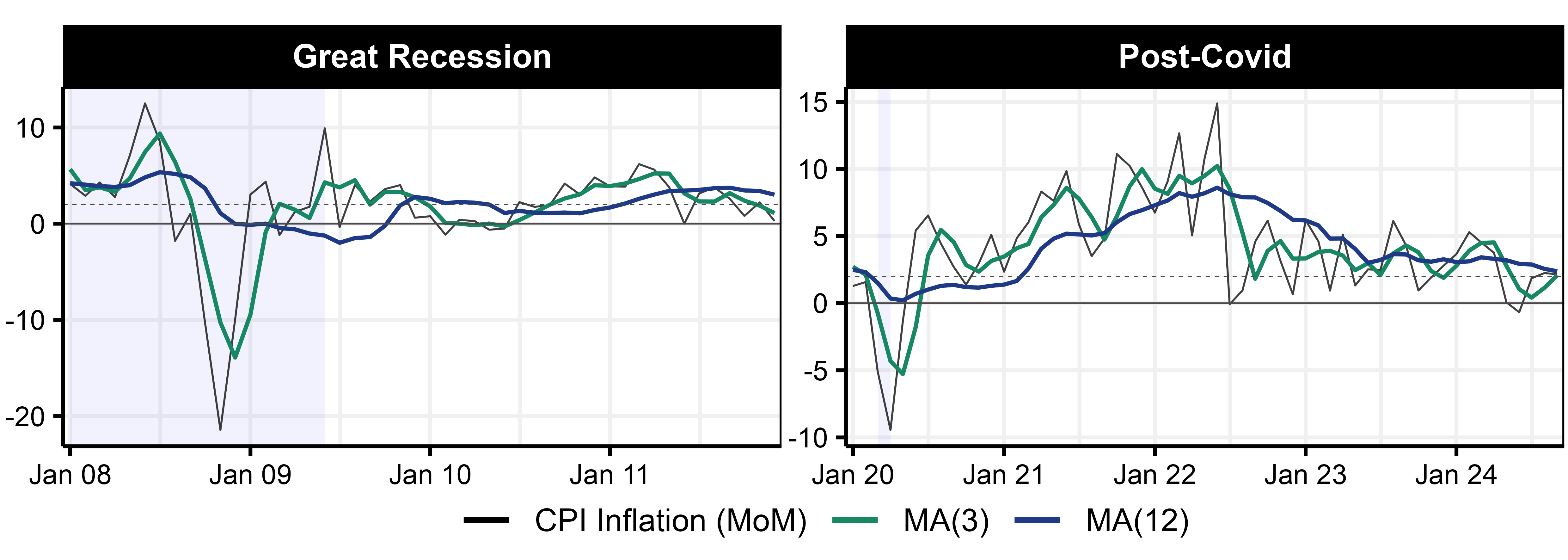}

\end{figure}
\vspace*{-1.2em}

While the benefits of shorter moving averages are self-evident in crisis periods, the longer average usually gets the upper hand in stable times, as it reduces variance without significant bias. Therefore, we are facing a time-varying bias-variance trade-off in the construction of the rolling window, and more generally, in the weighting of recent to distant realized values. 

Given this, our research question is simple: can we design a moving average estimator with a window size that adapts automatically to changing macroeconomic conditions? The answer is yes, and our proposed estimator is a Random Forest (RF) with the time series of interest as target and a time trend as sole predictor. We refer to this as \textbf{A}daptive \textbf{l}earning-\textbf{ba}sed \textbf{M}oving \textbf{A}verage, by the more compact {\selectfont \fontfamily{phv}\color{darkpowderblue} \textbf{AlbaMA}}. 

 Adaptability to smooth versus abrupt changes is procured by RF natural qualities as an adaptive nearest neighbor estimator \citep{lin2006random}. Intuitively, this occurs because if there is a clear break point, most randomized trees identify the key feature and split the sample at that point (using the trend). Therefore, their average will also feature the distinct shift.   In the opposite scenario, i.e., gradual change,  splits are randomly scattered along the underlying slope, and their aggregation elicits the slope as an average of a “staircases” distribution. \cite{MRFjae} documents this behavior in a time series context, recommending the inclusion of time trends in RF's features set to nonparametrically approximate exogenous structural change of any form in generalized time-varying coefficients. This paper leverages these insights for the simpler task of constructing adaptive moving averages, interpreted as a time-varying intercept model. As such, AlbaMA can be implemented with two or three lines of code using standard RF packages.

\vskip 0.15cm
{\noindent \sc \textbf{Results.}}  We begin by illustrating the adaptive behavior of AlbaMA using simulated data that includes both abrupt and gradual changes. As anticipated, we find that when a clear shift occurs, herd behavior dominates among the trees in the ensemble, while smoother changes lead to more dispersed splits. In the case of abrupt changes, repeated averaging of similar splits makes the break point stand out within AlbaMA. Conversely, for gradual changes, the dispersed splits average out to reflect a smooth transition, provided they are scattered evenly between the old and new state.


In our empirical analysis, we apply AlbaMA to key US and Euro Area (EA) variables commonly reported as moving averages, such as inflation, core inflation, Purchasing Managers Index (PMI), and industrial production. For each case, we visually compare AlbaMA to various benchmark averages and examine its one-sided versus two-sided behavior around economic turning points. Additionally, we analyze the weighting of recent observations implied by AlbaMA to identify when and where the weights adapt rapidly to changing conditions.

During stable periods, we find that AlbaMA’s alignment with traditional moving averages, such as MA(12) and MA(6), varies based on the target variable. For example, AlbaMA’s weighting for US headline inflation closely resembles that of an MA(6), whereas for core inflation, its “resting” average aligns almost exactly with an MA(12). When rapid changes occur, AlbaMA swiftly adjusts its weighting to emphasize recent observations. We study in depth the US headline and core inflations cases. First, we find that AlbaMA does not signal a gradual post-pandemic slowdown in US headline inflation but instead identifies a sharp decline in July 2022, followed by a stabilization around 3\%.  We report the weight allocated to the most recent observation nearly doubles at that critical  juncture. 


A similar but more pronounced pattern is observed for the core inflation surge of 2021-2022, marked by a regime shift in June 2021, followed by a lengthy, gradual decline. Analysis of the implied RF weights shows that this behavior stems from AlbaMA abandoning its MA(12) structure for a near MA(3) composition for the period between March 2021 and August 2021-- before gradually reverting back to MA(12). We observe comparable patterns in the Euro Area series and US PMI during the Great Recession.



Lastly, we document AlbaMA’s superior ability to approximate its ideal, yet unattainable, two-sided behavior. This assessment is based on comparing, across our nine series, how closely the one-sided estimate used throughout the paper matches the ex-post optimal weighted moving average in terms of mean-squared error. AlbaMA consistently performs well against all other moving averages, particularly surpassing the other adaptive option in our analysis, the Savitzky-Golay filter.


%


\vskip 0.15cm
{\noindent \sc \textbf{Literature -- Moving Averages.}} We now survey briefly the existing (and inevitably extensive) literature along three axes. The ubiquitous smoothing techniques are moving averages with a predefined, fixed window size. For macroeconomic variables, these are typically the average over a quarter, half-year, or year (3, 6, 12, respectively). A simple and equally popular extension is the exponential moving average, where past observations receive exponentially decreasing weights \citep{brown1956,holt2004forecasting}. These can be seen as one-sided exponential kernels, putting greater weight on recent observations compared to more distant ones. While the measure reduces noise and emphasizes more recent price changes, it is not adaptive as the bandwidth is fixed to a single value throughout the sample.  

The Savitzky-Golay (SG) filter is a more sophisticated, adaptive approach for smoothing noisy data \citep{savitzky1964smoothing}. It works by fitting a polynomial to a small subset of neighboring data points, allowing the filter to adapt locally to variations in the data. Smooth estimates are obtained by minimizing the sum of squared distances between the polynomial and the original points—yielding a least squares polynomial fit with closed-form linear coefficients. This filter is particularly valued in fields like chemistry and engineering, where preserving the signal’s shape and derivative information is crucial. However, it is well-known that the SG filter is not well-suited for one-sided (real-time) analysis, most notably, because of its reliance on a symmetric window around each point. As we see in our empirical results, two-sided results for AlbaMA and SG sometimes closely resemble one another, but AlbaMA has a definitive edge on one-sided results. 

In finance, although used less frequently, SG serves in technical analysis by retaining critical slopes and peaks, thereby enhancing the accuracy of, e.g., momentum assessments. Other related suggestions in the finance literature include moving averages that adjust smoothing based on the volatility of the data, i.e,. Kaufman's Adaptive Moving Average \citep{kaufman2013trading}, or leveraging fractal geometry to adapt to market price movements, as in the Fractal Adaptive Moving Average \citep{ehlers2005frama}.



\vskip 0.15cm
{\noindent \sc \textbf{Literature -- Trend-Cycle Decompositions.}} Separating time series into their trend and cyclical components has a long tradition in macroeconomics. While we focus on smoothing noisy \textit{stationary} series, which is the usual prerogative of moving averages, smoothers designed for non-stationary series can also be applied to this task. One of the most influential works is the Hodrick-Prescott filter \citep{hodrick1997postwar}, which extracts the trend by minimizing the sum of squared deviations subject to a penalty ensuring smoothness. This concept has inspired numerous variations. 


For example, $l_1$ trend filtering proposed by \cite{kim2009ell_1} adjusts the penalization by replacing the sum of squares with the sum of absolute values (i.e., an $l_1$ norm). This results in a smooth, piecewise linear trend. By being both fast and locally adaptive, the measure has beneficial features compared to wavelets and smoothing splines \citep{hastie1990generalized,donoho1998minimax}, which are fast but not adaptive, and to locally adaptive regression splines \citep{mammen1997locally}, which are adaptive but not fast \citep{tibshirani2014adaptive,tibshirani2011solution}.  Blending insights from both the machine learning and econometrics literature, \cite{phillips2021boosting} suggest iterating the HP filter in multiple steps by using the residuals from previous iterations, which is conceptually similar to $l_2$-boosting. 

Instead of relying on smoothing parameters and penalizing the smoothness of the trend, \cite{Hamilton2018} employs a linear regression-based approach. The trend component is extracted as the residuals from a direct forecasting regression of the series on its own lags for a pre-speficied horizon (a tuning parameter, often set to two years in macroeconomic context). We note that a special case of the Hamilton filter, where one includes a single lag and sets its coefficient to one (which is not always far from estimated coefficients results) implies a one-sided moving average of two years. Therefore, in this case, but also more generally, the filter's shape is not intended to be adaptive.

\vskip 0.15cm
{\noindent \sc \textbf{Literature -- Zooming on Inflation.}} While we consider various macroeconomic variables in our empirical results, the central focus will follow from our initial motivation: inflation. From the forecasting literature, an important contribution is  \cite{stock2007has} who introduce the unobserved components model with stochastic volatility (UC-SV), a time-varying trend-cycle model. This is, in fact, a time-varying integrated moving average, where the MA coefficient adapts inversely with the ratio of the variances of the permanent and transitory disturbances.  Relatedly, \cite{barunik2023dynamic} identify the heterogeneous persistence in time series via wold decomposition and localized regressions. Their proposed time-varying extended wold decomposition allows for smoothly changing persistence in economic data.  

In the spirit of exponential moving averages, \cite{eeckhout2023instantaneous} suggests to use a kernel-based measure for inflation. This ensures that greater weight is assigned to recent observations compared to more distant ones, with the degree of smoothing controlled by a bandwidth parameter. Kernel approaches to moving averages (and time-varying parameters, \citealt{giraitis2014inference}) partly deal with issues like base effects caused by hard-threshold inclusion/exclusion rules of traditional rolling windows.  While the measure reduces noise and emphasizes more recent price changes, it is not adaptive as the bandwidth is fixed to a single value throughout the sample. 


\vskip 0.15cm
{\noindent \sc \textbf{Outline.}}  The paper is organized as follows.  Section \ref{sec:method} reviews Random Forest and showcases its adaptive moving average properties on simulated data. Section \ref{sec:empirics} presents a comparison of empirical estimates derived from our method with alternative approaches, using key US and EA time series. Section \ref{sec:conclusion} concludes.

\section{Methodology}\label{sec:method}

We begin by reviewing Random Forest and discussing the specifics of the case with a single deterministic regressor, including how to retrieve AlbaMA's time-varying weights. Next, we use simulations to illustrate how the wisdom of crowds (of trees) elicits abrupt or gradual changes, depending on the underlying DGP. Finally, we demonstrate how AlbaMA's weights adapt to evolving dynamics.

\subsection{A Review of Random Forest}\label{sec:rev}

Random Forest  \citep{breiman2001} is a diversified ensemble of regression trees.  We first introduce regression trees, present their estimation algorithm,  and discuss the ensembling procedure.   
 
\vskip 0.15cm

{\noindent \sc \textbf{A Tree.}} Consider a time-series scenario where $y_t$ represents a time series for which we want to compute some sort of moving average. A simple decision tree's prediction, using a time trend as sole predictor, could look like

\vspace{1em}
\Tree[.{2022m6 to 2023m12}
[.{$t \geq \text{2023m1}$}
[.{$t \geq \text{2023m8}$}
{Leaf \( A \) : $\phantom{–} \hat{y}_t= \overline{y}_{t \in A} \phantom{-}$}
]
[.{$t < \text{2023m8}$}
{Leaf \( B \) :  $\phantom{–} \hat{y}_t =  \overline{y}_{t \in B} \phantom{-}$}
]
]
[.{$t < \text{2023m1}$}
{Leaf \( C \) : $\phantom{–} \hat{y}_t =  \overline{y}_{t \in C} \phantom{-}$}
]
]
\vspace{1em}

\noindent where $\overline{y}_{t \in A}$ denotes the average of the target's observations falling into leaf $A$. In practice, the cutting points (e.g., $t \geq \text{2023m8}$) are unknown unless we assume a window size.  \textcolor{black}{In order to build an adaptive MA, we wish to learn these window sizes and their location, which is precisely what regression trees can offer.}


\vskip 0.15cm
{\noindent \sc \textbf{Estimation.}} The usual strategy -- introduced as Classification and Regression Trees (CART) in \cite{breiman1984classification} --  is to deploy a {greedy} algorithm that {recursively} partitions the data according to
\begin{equation}\label{treesplit}
\begin{aligned}
\min\limits_{k\in \mathcal{K}, \smallskip c \in {\rm I\!R}}\Bigg[ \min\limits_{\mu_{1}} \sum \limits_{\{ t \in L| X_{i,k} \leq c  \}} \left(y_{t}-\mu_{1}\right)^{2} 
 + \min\limits_{\mu_{2}} \sum \limits_{\{ t \in L | X_{i,k} > c  \}} \left(y_{t}-\mu_{2}\right)^{2}\Bigg].
\end{aligned}
\end{equation}
Here, $\min\limits_{k\in \mathcal{K}, c \in {\rm I\!R}}$ denotes the minimization over all possible splits, 
where $k$ indexes a variable in $\mathcal{K}$, representing available features,  and $c$ is a real number representing the split point.  Note that $\mathcal{K}$ will be fairly limited in our application, as we set $\mathcal{K}=[\, t \,]$. $L$ is a leaf,  representing the sub-sample of data utilized by \eqref{treesplit} to estimate the next split.  The first $L$ in the recursion is the whole training sample,  then the algorithm proceeds recursively by using the subsamples created by the previous partition as the subsequent $L$'s in the next iteration. This process mechanically creates partitions of progressively smaller size until a stopping criterion is met, resulting in a set of terminal nodes where the prediction is the average of the $y_t$’s within each leaf. For example, the simple case illustrated above has three terminal nodes.


The overall goal of a single step is to find the optimal pair $( k^*,  \,  c^*)$ and the predicted values ($\mu_{1},  \,  \mu_{2}$)  that minimize the total within-leaf sum of squared errors.  In our application, $ k^*$ is redundant and only $c^*$ is optimized along with   $\mu_{1}$ and $ \mu_{2}$. In standard trees (such as \eqref{treesplit}),  the latter  are always the within-leaf average.  



A single tree with many splits can capture complex data structures but suffers from high variance. In the extreme, we have $\hat{y}_t = y_t$, akin to a “moving average” with a window size of 1. Pruning reduces variance by merging terminal nodes and removing insignificant splits, but this approach is limited: the pruned tree remains locally optimized, and its performance heavily depends on the extent of pruning, a sensitive tuning parameter. Even more importantly, for our goal of creating a smooth moving average, a single tree is inadequate, as it can only capture abrupt changes.


\vskip 0.15cm
{\noindent \sc \textbf{Diversifying the Portfolio.}} A highly effective strategy in machine learning is to create a diversified portfolio of trees, treating each tree ${\mathcal{T}}$ as a base learner and averaging their predictions. Turning a single tree into a forest involves three main steps:

\begin{enumerate}
\item[\texttt{\large D }:  ]  First, each tree should be allowed to grow \textit{deep}, producing a large number of terminal nodes through extensive splitting. This depth results in overfitted trees if used individually, but when averaged, their collective output smooths the series. From a moving average application perspective, this signifies that we want single trees to deliver series that are more wiggly than not, because further averaging is on the way.

\item[\texttt{\large B }:  ]  Second, we apply \textit{Bagging} (Bootstrap Aggregation,  \citealt{breiman1996bagging}), generating $B$ bootstrapped samples of the data and estimating each tree on a separate sample $b \in 1,\dots, B$. Bagging mitigates the inherent instability of individual trees and, in our MA application, improves the method’s ability to capture both smooth and abrupt changes by addressing the limitations of a single tree in representing continuous patterns.


\item[\texttt{\large P }:  ]  Third, RF typically introduces \textit{perturbation} by randomly selecting a subset of predictors at each split. However, in our application, we use a single predictor (a time-trend), so \texttt{\large P} is not applicable here. Thus, properly speaking, AlbaMA is a bagged trees estimator.
\end{enumerate}

\noindent The RF "prediction" at time $t$ is the average of the predictions from all $B$ trees: $\hat{y}_t^{\text{RF}} = \frac{1}{B} \sum_{b=1}^B {\mathcal{T}}_{b}(t)$. Therefore, it is an average of averages, and thus, a properly defined weighted average of $y_t$ and neighboring observations. It can also be interpreted as stochastic model averaging where the underlying models are MAs of different size and composition for each $t$. In the traditional bias-variance trade-off view, \texttt{\large D} lowers bias, while \texttt{\large B} (and \texttt{\large P}, where applicable) reduces variance. In our context, \texttt{\large D} and \texttt{\large B} work together to distinguish between abrupt and smooth changes, dynamically adapting the moving average window.


In this simplified case with a single regressor, the CART algorithm could theoretically be replaced by Lasso with indicator functions for each $t$, with bagging applied to the Lasso model. Nonetheless, we use Random Forest (RF), as its greedy tree optimization introduces more randomness, enabling bagging to more effectively reduce variance. Additionally, RF hyperparameters, like minimal leaf sizes, offer intuitive control for moving average applications.





\subsection{Retrieving Moving Average Weights from Random Forest}\label{sec:weights}


To gain insights into the look-back window selected by the RF, we show how to back out the weights assigned to each observation in the training set. These weights correspond to the moving average coefficients, $w_t$, which, as discussed in \cite{DUAL}, can be derived through post-processing of estimation outputs. This builds on the related literature on adpative nearest neighbors in RF \citep{lin2006random,koster2024simplifying} and  uses the insight that each individual tree's contribution to the final prediction can be expressed as:  
\[
\mathcal{T}_b(t) = \frac{1}{\sum_{\tau=1}^T I\left(\tau \in \mathcal{P}_b(t)\right)}\sum_{\tau=1}^T y_\tau I\left(\tau \in \mathcal{P}_b(t)\right) = \sum_{\tau=1}^T w_{b\tau} y_\tau.
\]
We define \(\mathcal{P}_b\) as the partition of the input space created by the tree's structure, corresponding to the specific region or leaf node where observation $t$ resides, based on the tree's splits and associated conditioning information. The final prediction, obtained by averaging over $B$ regression trees, provides the desired representation:
\[
\hat{y}_t = \frac{1}{B} \sum_{b=1}^B \mathcal{T}_b(t) = \frac{1}{B} \sum_{b=1}^B \sum_{\tau=1}^T w_{b\tau t} y_\tau = \sum_{\tau=1}^T \underbrace{\frac{1}{B} \sum_{b=1}^B w_{b\tau t}}_{w_{\tau t}} y_\tau = \boldsymbol{w}_t \boldsymbol{y}.
\]
In summary, the sequence of operations can be described as follows. First, we determine $w_\tau$ by identifying the leaf node in which observation $t$ resides for a given tree. Within that leaf, we locate the corresponding in-sample observations and assign weights to them, calculated as $\sfrac{1}{\text{leaf size}}$. These weights are attributed to the relevant in-sample observations ($w_{b\tau}$). Finally, we aggregate these contributions across all trees in the ensemble. 

By construction, the resulting $\hat{y}_t$ is a weighted average (i.e., ${w}_{\tau} \in \Delta$), as the elements of $\boldsymbol{w}_{t}$ are themselves averages of ${w}_{b\tau t}$, which are weighted average weights.



\subsection{Illustrating RF's Adaptability with Synthetic Data}\label{sec:sim}

Our simulation example comprises three distinct scenarios: (i) \textit{abrupt change}, (ii) \textit{gradual change}, and (iii) a \textit{combination} of both. In the \textit{abrupt change} scenario, the response variable undergoes a sudden shift (from -1 to 1) at the midpoint of the time series, while the \textit{gradual change} scenario features a continuous, smooth linear transition from one state to another (from -1 to 1). The \textit{combined} scenario integrates both patterns, starting with an upward slope followed by a structural break. In each scenario, our data-generating process (DGP) includes normal white noise with mean 0 and standard deviation 0.5. A formal exposition is provided in Appendix \ref{app:sim}.

In the top panel of Figure \ref{fig:Sim_exp_2sided}, we present the resulting time series observations with the AlbaMA overlay, derived from a Random Forest using 500 trees and a minimum node size of 40. In the bottom panel, we show predictions from four individual trees within the forest to illustrate whether herd behavior or dispersion prevails across our three scenarios.

\begin{figure}[t!]
  \caption{\normalsize{AlbaMA on Simulated Data}} \label{fig:Sim_exp_2sided}
    \vspace*{-0.8em}
      \centering
    \includegraphics[width=\textwidth, trim = 0mm 0mm 0mm 0mm, clip]{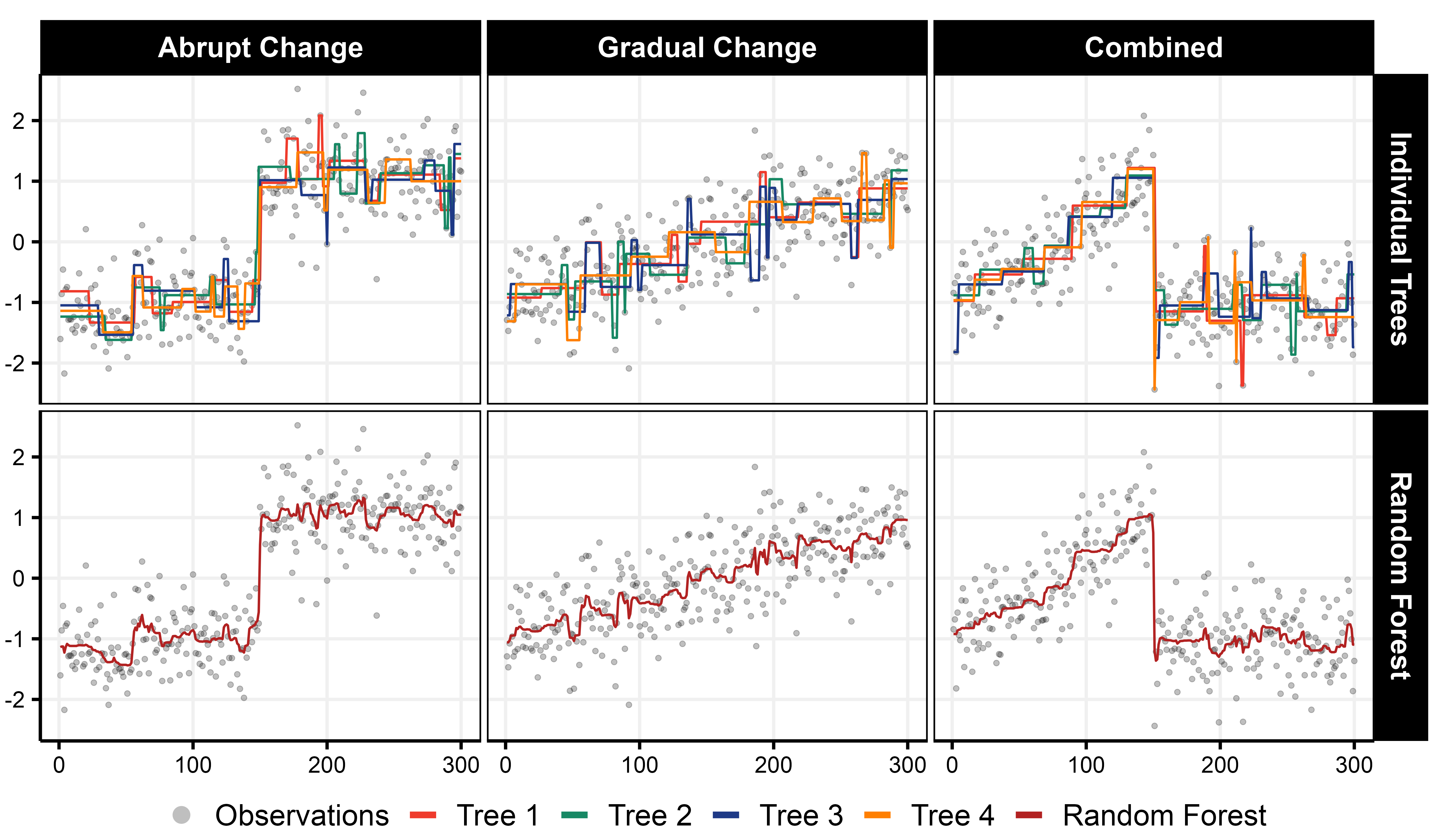}

    \begin{threeparttable}
    \centering
    \begin{minipage}{\textwidth}
      \begin{tablenotes}[para,flushleft]
    \setlength{\lineskip}{0.2ex}
    \notsotiny 
  {\textit{Notes}: The figure shows results of the two-sided RF. The upper panels illustrate four randomly selected trees for each scenario. The lower panels present the final results of the two-sided RF. Data is drawn from a normal distribution with mean 0 and standard deviation 0.5. For the \textit{abrupt change} scenario, we add a constant that suddenly shifts from -1 to 1. The \textit{gradual change} scenario features a smooth trend, which is added to the noise, and the \textit{combined} scenario uses both elements, an upward slope followed by the sudden shift back to -1. }
    \end{tablenotes}
  \end{minipage}
  \end{threeparttable}
\end{figure}

In the \textit{abrupt change} scenario, nearly all trees identify the significance of the split around observation 150. Indeed, when there is a clear regime shift in the data, all randomized trees recognize this split as a key feature. This is evident in the first column of Figure \ref{fig:Sim_exp_2sided}, where all four trees excerpts capture the break. As a result, the Random Forest fit distinctly reflects the abrupt change.



In the \textit{gradual change} scenario, there is no consensus on a specific split point because, in the true DGP, none exists. Instead, individual trees attempt to approximate a linear trend by introducing splits at various points, each creating a series of step functions to approximate a straight line.\footnote{In a noise-free setting, the ideal approximation would involve 300 steps, increasing monotonically from -1 to 1 in increments of $\frac{2}{300}$.} Bagging introduces significant diversity across the ensemble, as shown in the second column of Figure \ref{fig:Sim_exp_2sided}, where split points across the four trees are scattered between observations 1 and 300. Despite this apparent lack of coordination, some collective wisdom emerges in the shape of a nearly straight line---thanks to Bagging reducing variance by smoothing hard-thresholding rules \citep{buhlmann2002analyzing}.




\begin{figure}[t!]
  \caption{\normalsize{AlbaMA's Weights for Simulated Data}} \label{fig:Sim_exp_imp}
\vspace*{-0.9em}
      \centering
    \includegraphics[width=\textwidth, trim = 0mm 0mm 0mm 0mm, clip]{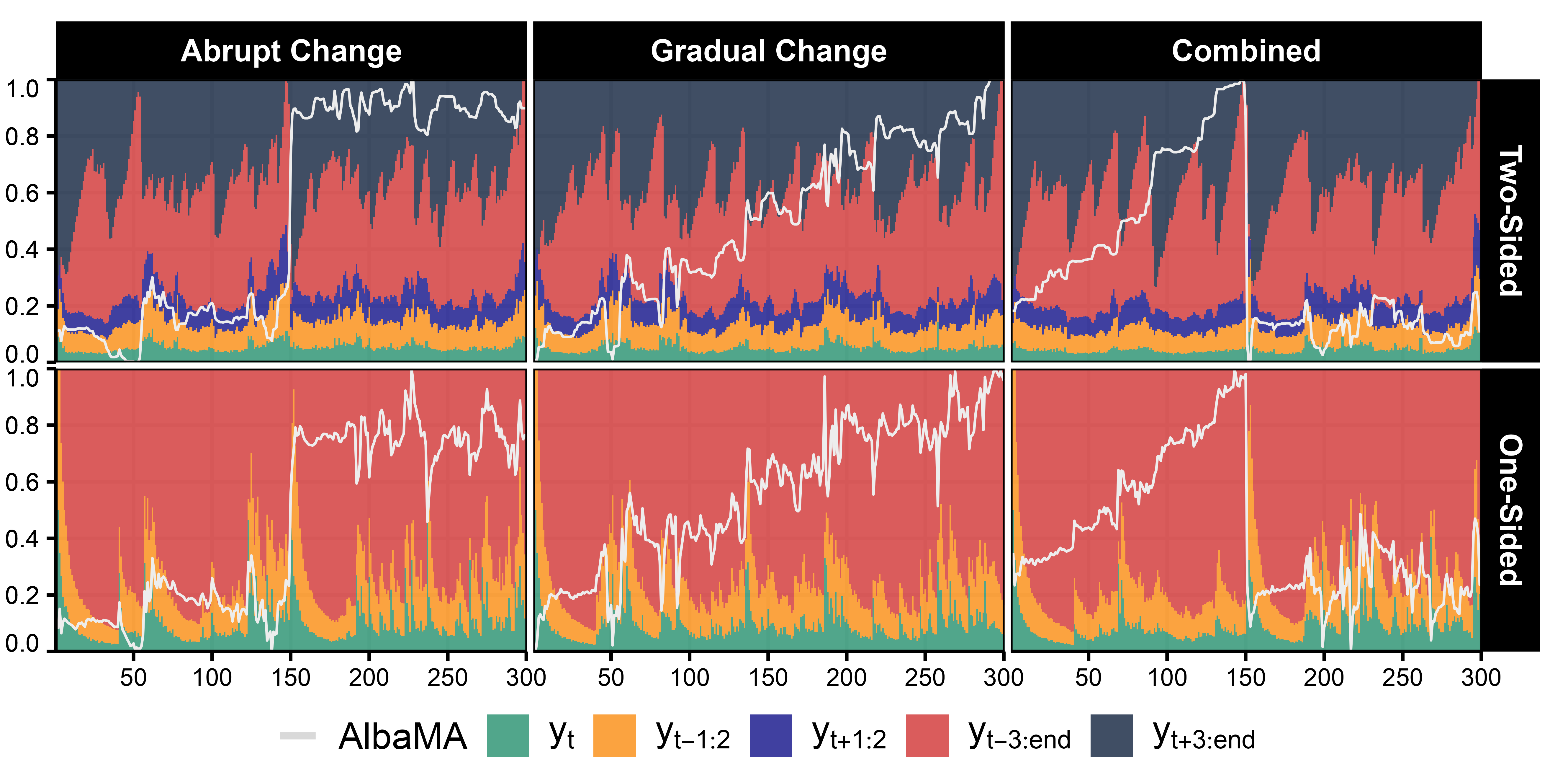}

    \begin{threeparttable}
    \centering
    \begin{minipage}{\textwidth}
      \begin{tablenotes}[para,flushleft]
    \setlength{\lineskip}{0.2ex}
    \notsotiny 
  {\textit{Notes}: The figure illustrates the weights assigned to the different observations for the two-sided RF model (upper panels) and the one-sided RF model (lower panels). Note that in the one-sided case only past data can be utilized by the RF.}
    \end{tablenotes}
  \end{minipage}
  \end{threeparttable}
\end{figure}


The final case is the \textit{combined} scenario, shown in the third column of Figure \ref{fig:Sim_exp_2sided}. First, we observe strong tree consensus around the break date. Second, there is significant dispersion among trees in the linear run-up phase, resulting in a mostly straight line. Finally, while individual trees may display extreme changes in the second half of the sample, these idiosyncratic variations average out in AlbaMA, yielding a mostly flat line that aligns with the true DGP.

Figure \ref{fig:Sim_exp_imp} illustrates the importance of lagged (and, when applicable, leading) observations in AlbaMA, shown for both one-sided and two-sided configurations. RF’s flexibility in adjusting the weighting of the look-back window over time is evident, particularly around structural breaks. In the two-sided case, we observe a sharp increase in the importance of recent lags just before an abrupt change, while leading observation weights peak right after the break. This pattern reflects RF’s ability, in the two-sided case,  to capture breaks by not mixing pre- and post-break data together when computing a moving average around the sudden shift date.

The one-sided configuration, shown in the lower panels, displays a similar tendency, with recent observations gaining weight around structural breaks before quickly reverting to a more inclusive average. In the \textit{gradual change} scenario, as well as in smooth segments of other cases, the filter’s weighting occasionally responds to what is effectively noise in the DGP. With a sample of 300 observations, AlbaMA will inevitably miss on some opportunities for smoothing.


Lastly, as a reference point, we apply traditional time series filters on the combined DGP—standard moving averages, the SG adaptive moving average, $l_1$ trend filtering, and boosted HP filters (see Appendix \ref{app:bench} for details). All of these filters, directly or indirectly, imply that $\hat{y}_t$ is a local average of $y_t$. In the \textit{combined} scenario, however, none of these benchmarks match RF’s adaptability to rapid changes (see Figure \ref{fig:Sim_comp} in the appendix), even in a two-sided application. Standard moving averages demonstrate the challenging trade-off discussed in the introduction for the case of inflation, while the SG adaptive moving average introduces unnecessary peaks in the flat portion of the DGP. For standard time series filters, avoiding the artificial smoothing of structural breaks often introduces significant noise and artificial cycles, a tendency that is evident in both the gradual change sections and the flat portions of the DGP. This arises from tuning parameters that work well for abrupt changes being unsuitable for smoother segments, and vice versa.

\section{Empirical Application}\label{sec:empirics}

We apply AlbaMA to a broad set of macroeconomic variables for the US and the Euro Area (EA). These include CPI/HICP headline and core inflation, industrial production, unemployment and the Purchasing Managers Index (PMI) on a monthly frequency.\footnote{For the US, the series are taken from FRED and span the periods 1963m1 to 2024m9 for industrial production and CPI inflation, and 2024m10 for unemployment. The PMI is extracted from Macrobond with start date 1963m1 and end date 2024m10. Data for the EA is provided by Eurostat and ranges from 1997m1 to 2024m8 for industrial production, 2024m9 for unemployment, and 2024m10 for HICP inflation.} We compare AlbaMA's solution to three-, six-, and twelve-months moving averages as well as the Savitzky–Golay filter for various historical time periods including tranquil times and periods of economic turmoil.


\subsection{AlbaMA for Inflation Monitoring}\label{sec:inflation}



\begin{figure}[t!]
  \caption{\normalsize{Results for US CPI Inflation}} \label{fig:US_CPI}
  
   \begin{minipage}[t]{\textwidth}
      \centering
      \includegraphics[width=\textwidth, trim = -9mm 0mm -19mm 0mm, clip]{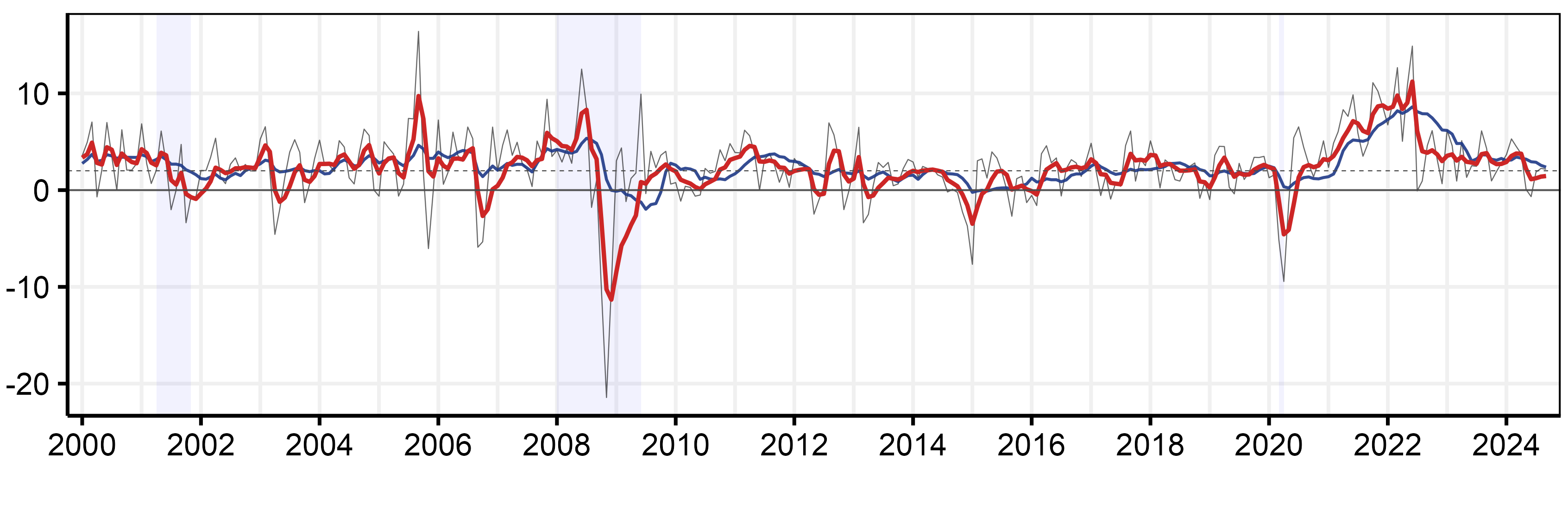}
    \end{minipage}%

    \vspace*{-1em}
     \begin{minipage}[t]{0.5\textwidth}
      \centering
      \includegraphics[width=\textwidth, trim = -13mm 0mm -7mm 0mm, clip]{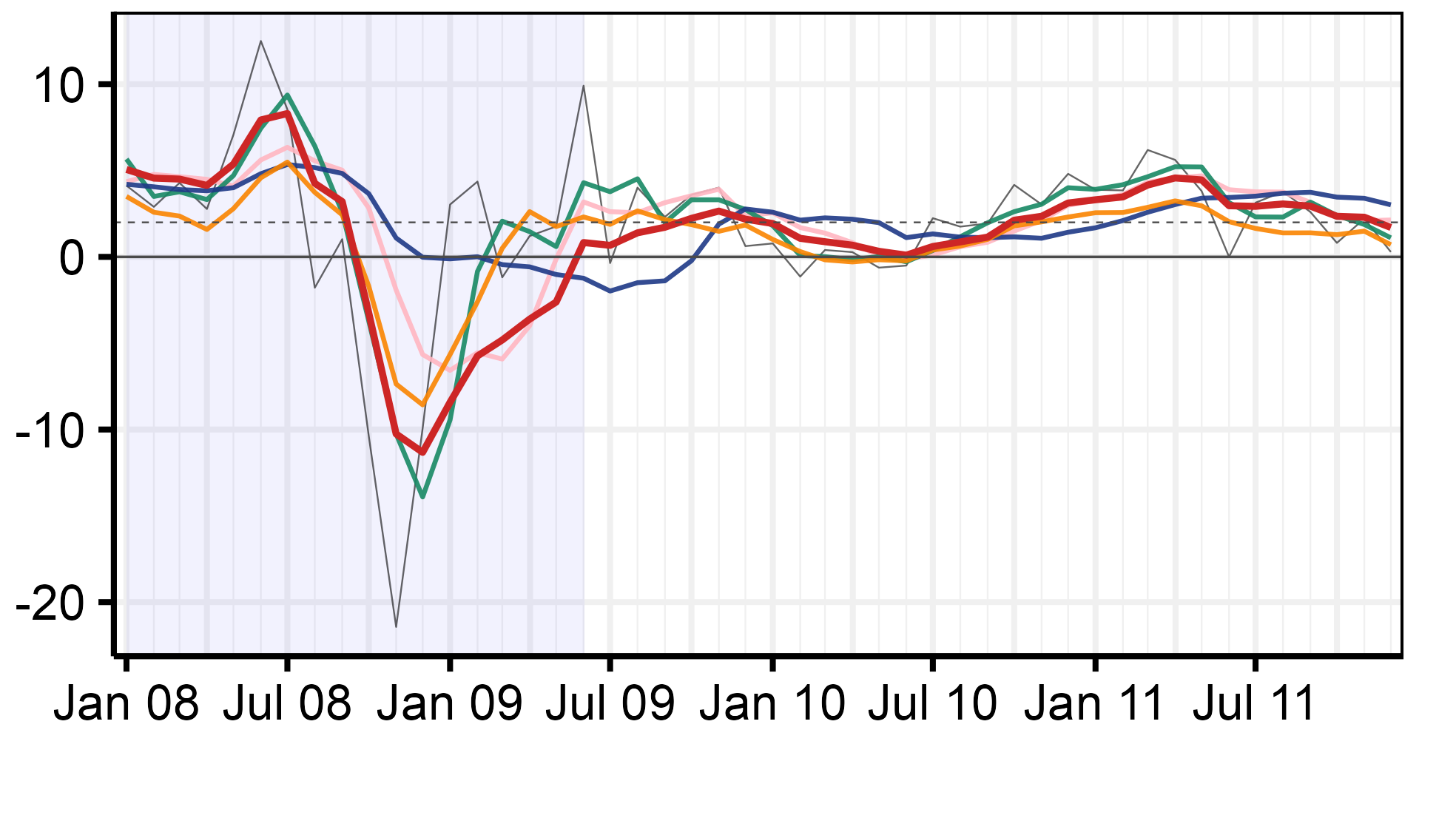}
    \end{minipage}%
    \begin{minipage}[t]{0.5\textwidth}
      \centering
      \includegraphics[width=\textwidth, trim = -3mm 0mm -15mm 0mm, clip]{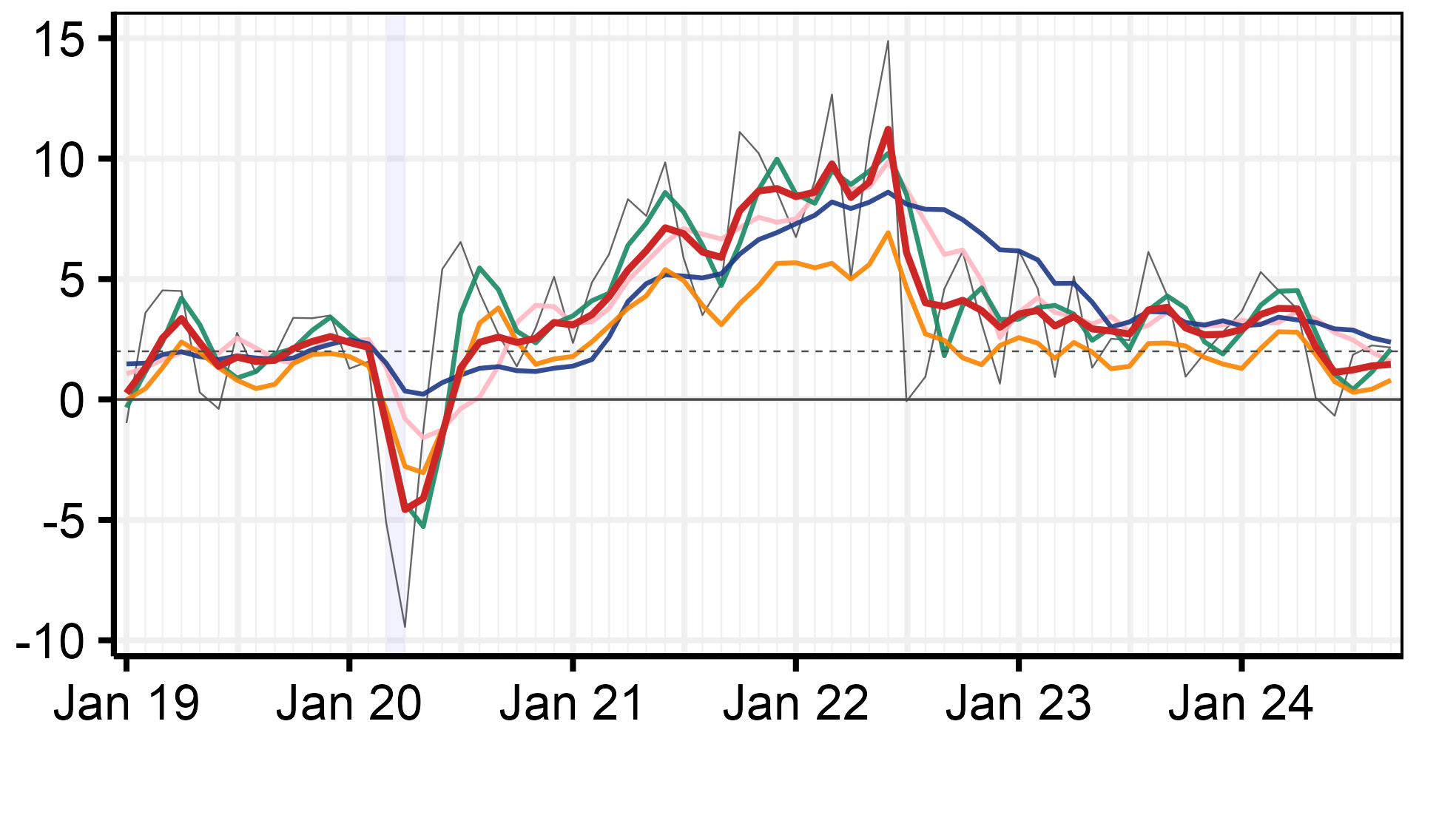}
    \end{minipage}

    \vspace*{-0.8em}
    \begin{minipage}[t]{\textwidth}
      \centering
      \includegraphics[width=0.8\textwidth, trim = 0mm 0mm 0mm 0mm, clip]{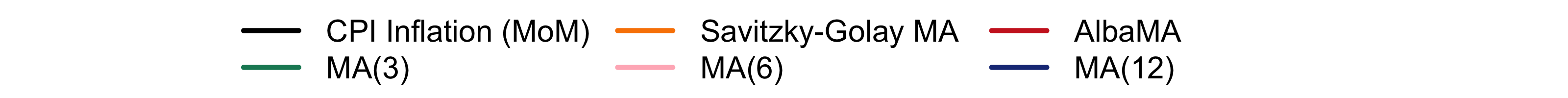}
    \end{minipage}

    \begin{minipage}[t]{0.5\textwidth}
      \centering
      \includegraphics[width=\textwidth, trim = 5mm 0mm 10mm 0mm, clip]{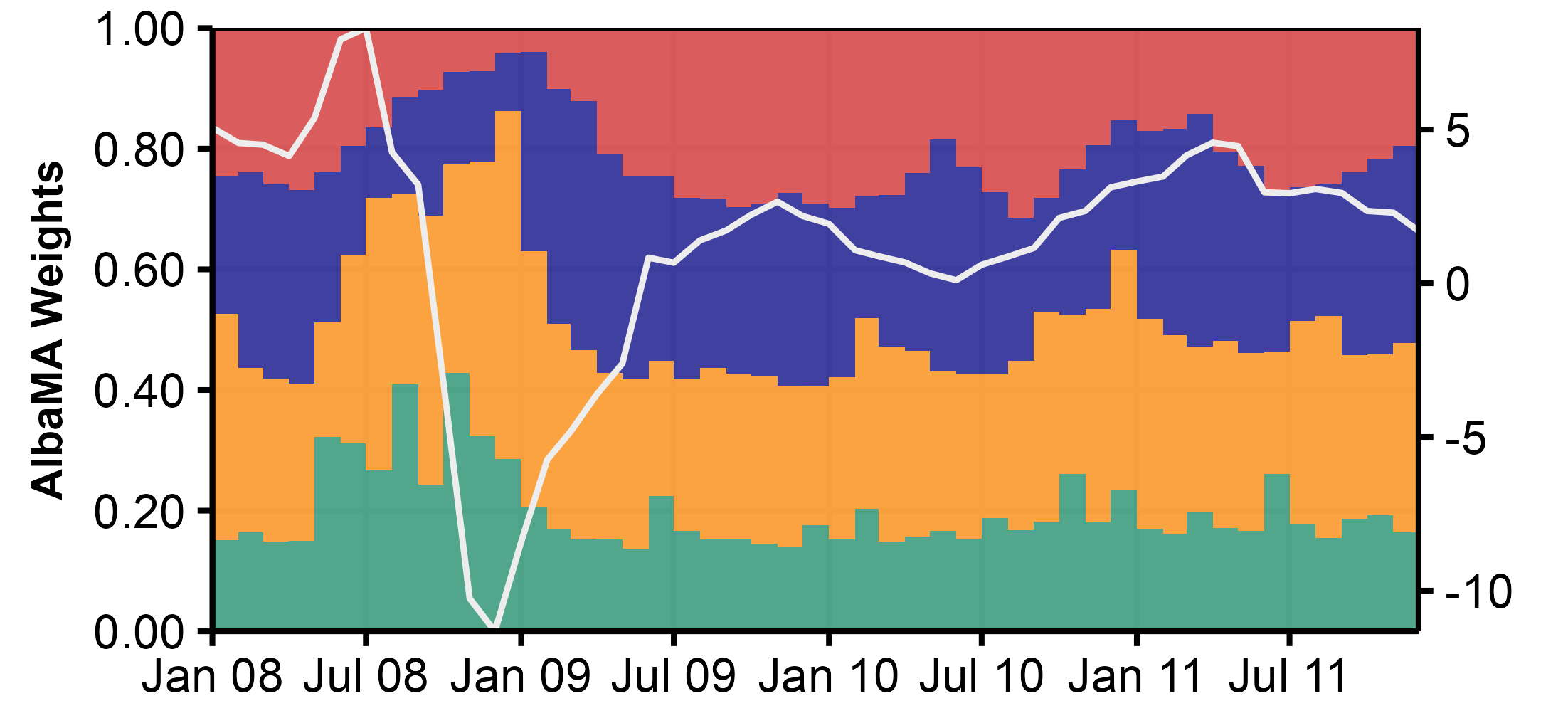}
    \end{minipage}%
    \begin{minipage}[t]{0.5\textwidth}
      \centering
      \includegraphics[width=\textwidth, trim = 10mm 0mm 8mm 0mm, clip]{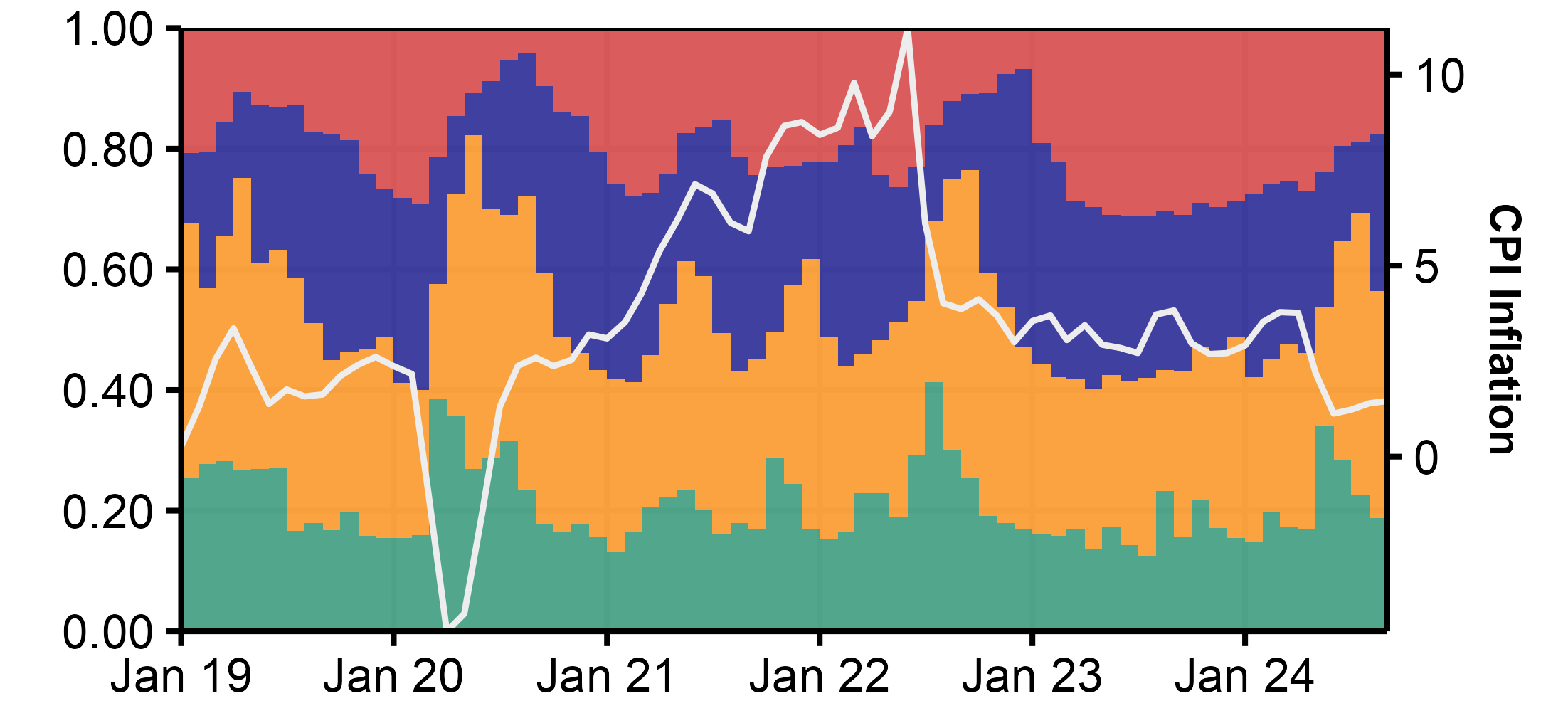}
    \end{minipage}

    \vspace*{-.5em}
    \begin{minipage}[t]{\textwidth}
      \centering
      \hspace*{1.0em} \includegraphics[width=0.95\textwidth, trim = 0mm 0mm 0mm 0mm, clip]{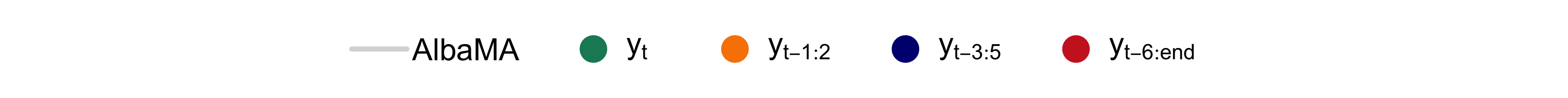}
    \end{minipage}

    \vspace*{-0.3em}
    \begin{minipage}[t]{0.5\textwidth}
      \centering
      (a) Great Recession
    \end{minipage}
    \begin{minipage}[t]{0.5\textwidth}
      \centering
      (b) Post-Covid
    \end{minipage}

    \begin{threeparttable}
    \centering
    \begin{minipage}{\textwidth}
      \begin{tablenotes}[para,flushleft]
    \setlength{\lineskip}{0.2ex}
    \notsotiny 
  {\textit{Notes}: The \textbf{upper panel} shows AlbaMA and the MA(12). The \textbf{middle panels} compare AlbaMA to standard moving averages and the Savitzky-Golay filter for (a) the Great Recession and (b) the post-Covid surge. The \textbf{lower panels} present the weights the RF assigns to past observations. All measures are one-sided.}
    \end{tablenotes}
  \end{minipage}
  \end{threeparttable}
\end{figure}

\begin{figure}[t!]
  \caption{\normalsize{Results for US CPI Core Inflation}} \label{fig:US_CPIcore}
  
   \begin{minipage}[t]{\textwidth}
      \centering
      \includegraphics[width=\textwidth, trim = -12mm 0mm -18mm 0mm, clip]{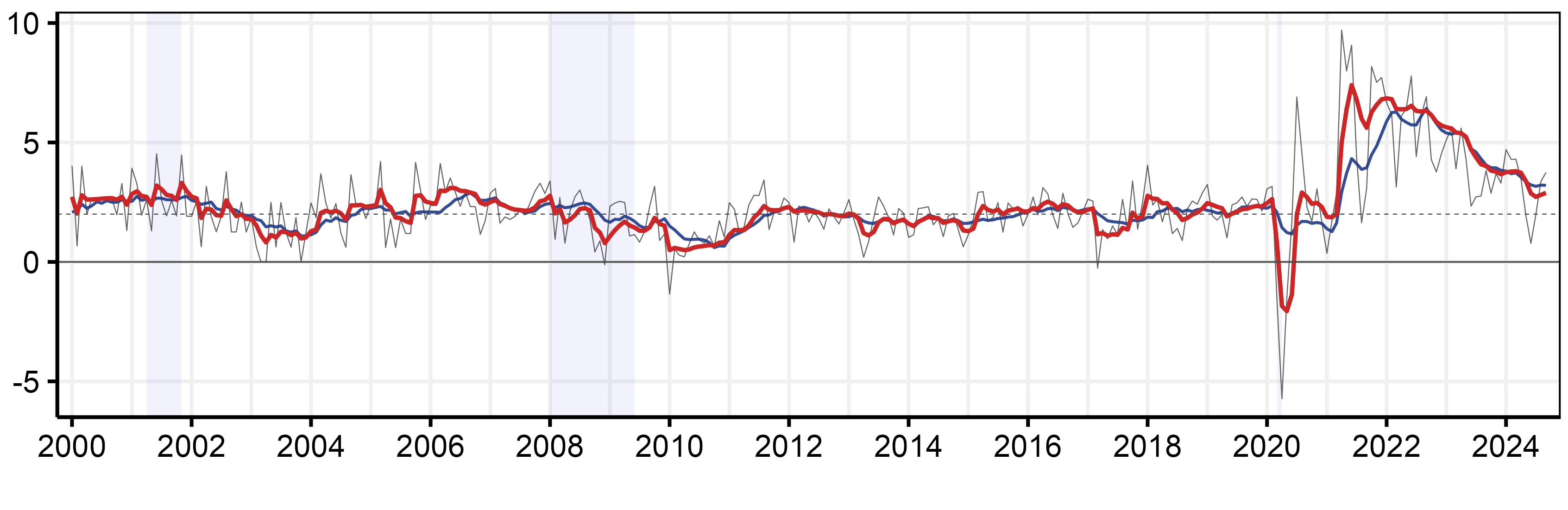}
    \end{minipage}%

\vspace*{-1em}
     \begin{minipage}[t]{0.5\textwidth}
      \centering
      \includegraphics[width=\textwidth, trim = -16mm 0mm -5mm 0mm, clip]{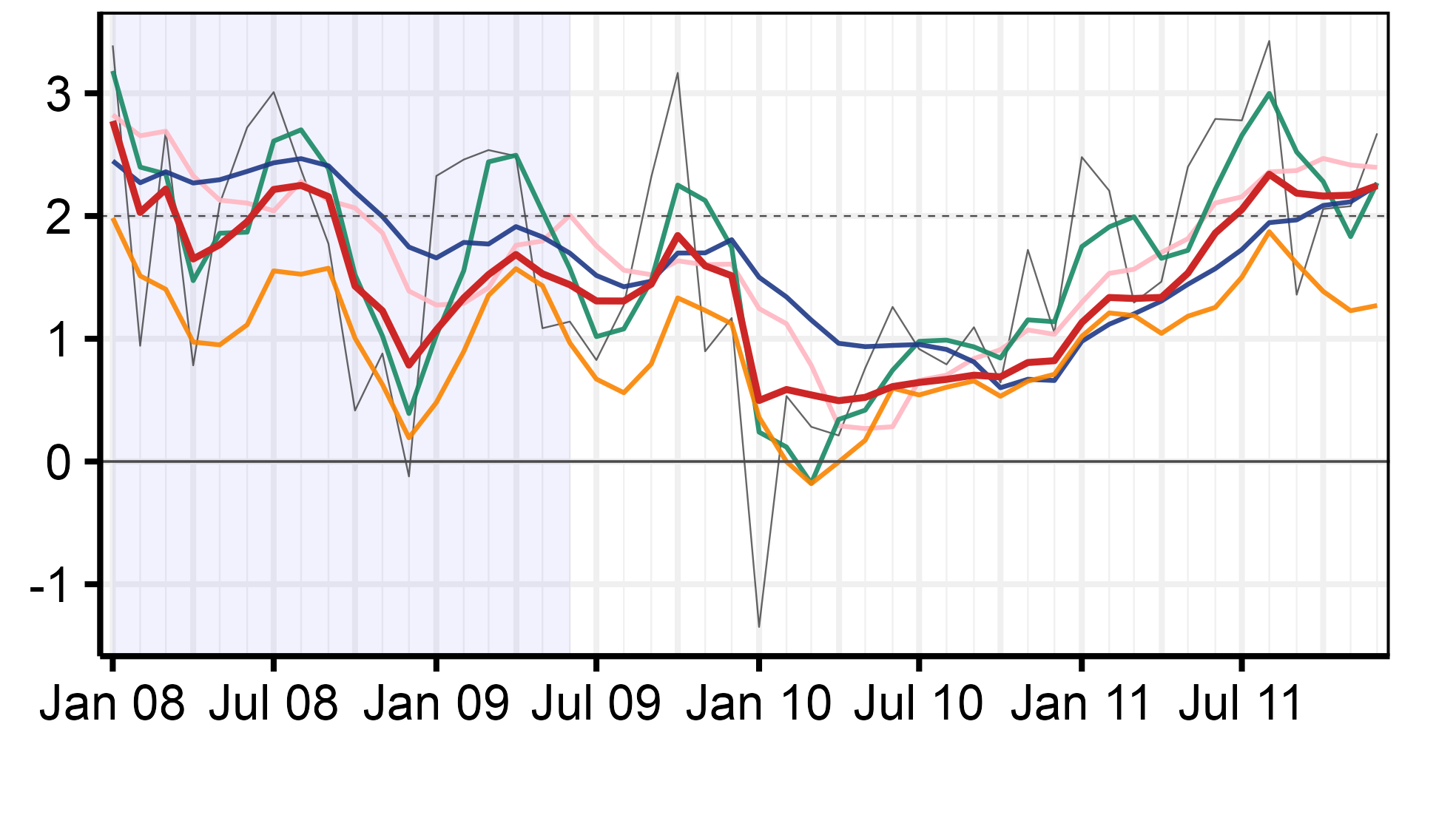}
    \end{minipage}%
    \begin{minipage}[t]{0.5\textwidth}
      \centering
      \includegraphics[width=\textwidth, trim = -5mm 0mm -13mm 0mm, clip]{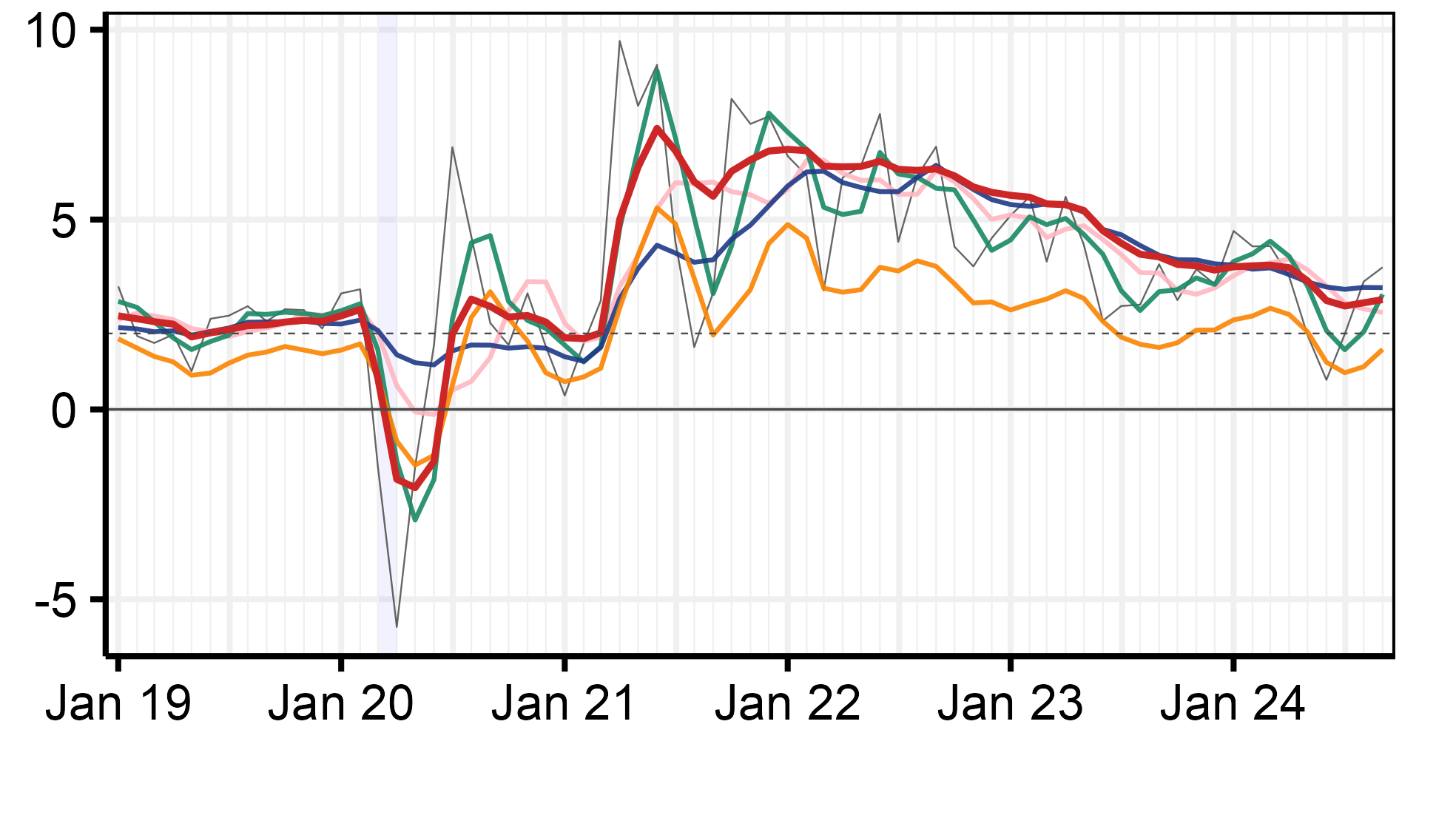}
    \end{minipage}

    \vspace*{-0.8em}
    \begin{minipage}[t]{\textwidth}
      \centering
      \includegraphics[width=0.8\textwidth, trim = 0mm 0mm 0mm 0mm, clip]{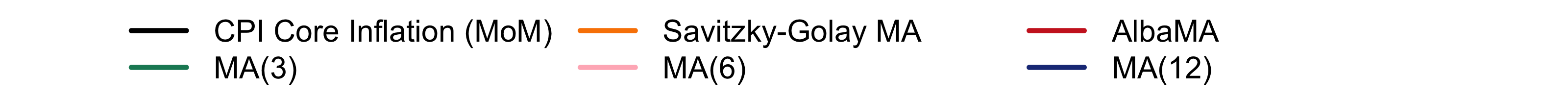}
    \end{minipage}

    \begin{minipage}[t]{0.5\textwidth}
      \centering
      \includegraphics[width=\textwidth, trim = 5mm 0mm 10mm 0mm, clip]{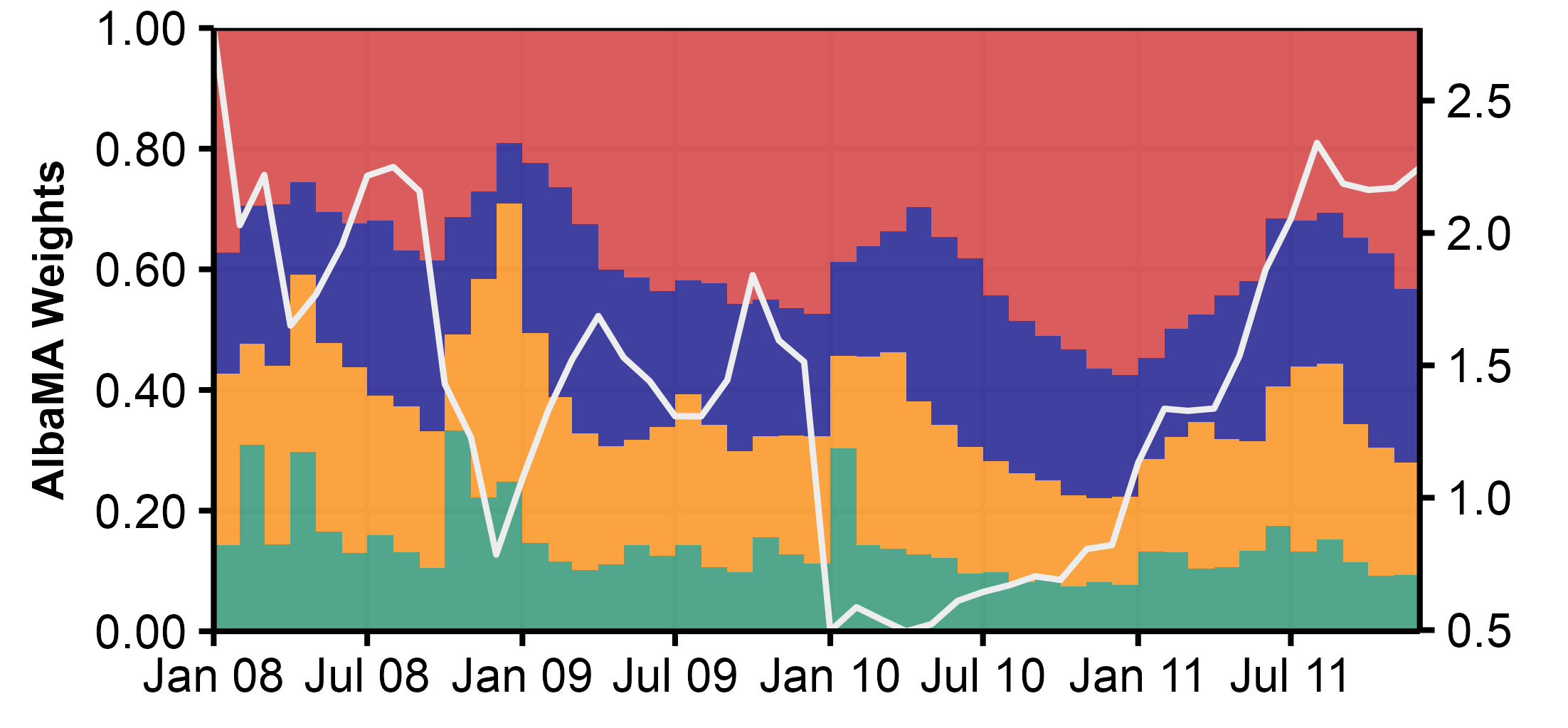}
    \end{minipage}%
    \begin{minipage}[t]{0.5\textwidth}
      \centering
      \includegraphics[width=\textwidth, trim = 10mm 0mm 8mm 0mm, clip]{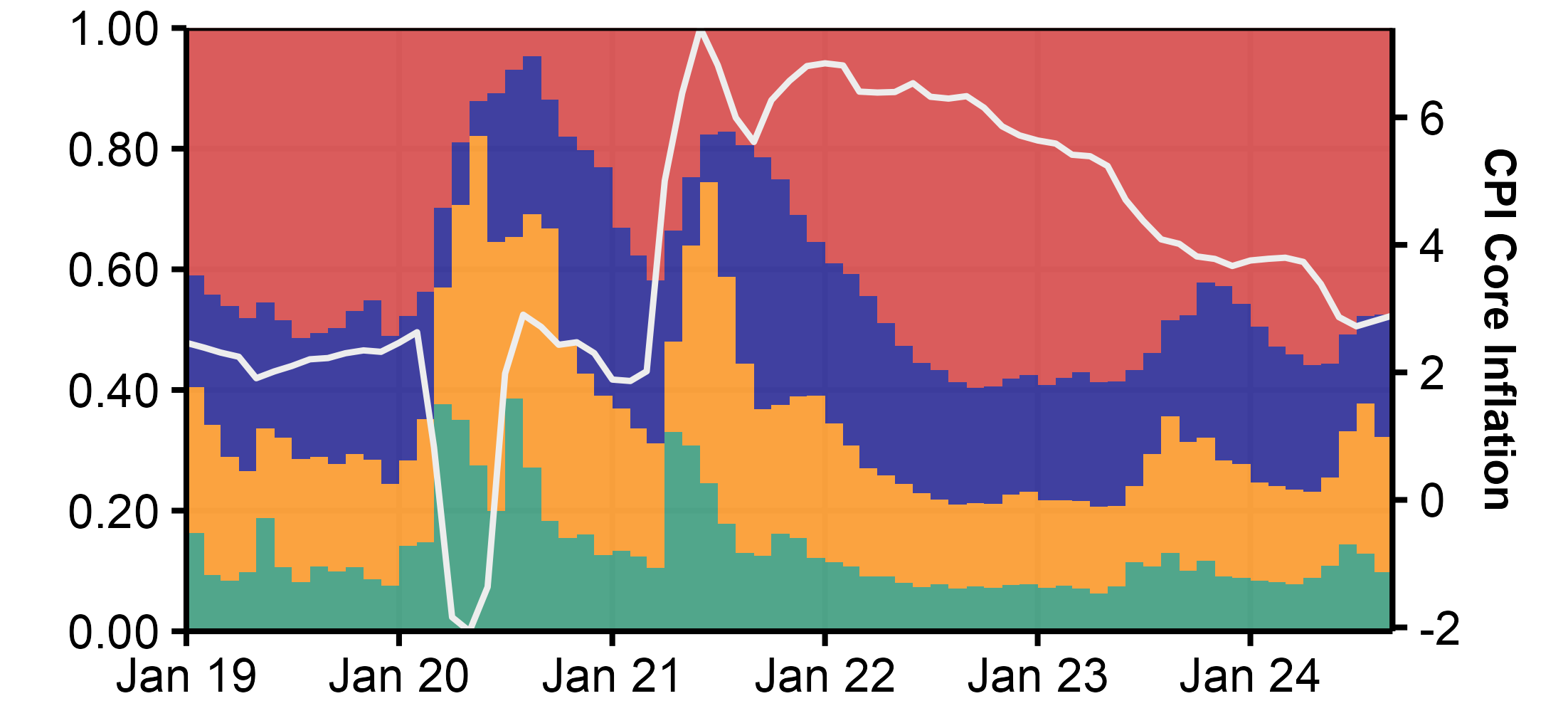}
    \end{minipage}

    \vspace*{-.5em}
    \begin{minipage}[t]{\textwidth}
      \centering
      \hspace*{1.0em} \includegraphics[width=0.95\textwidth, trim = 0mm 0mm 0mm 0mm, clip]{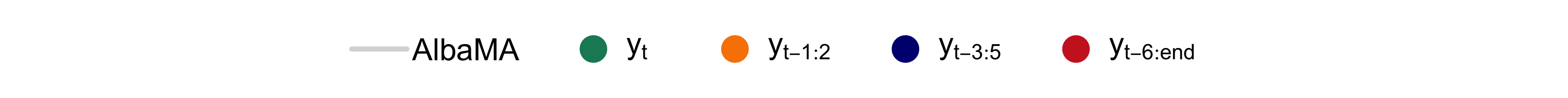}
    \end{minipage}

    \vspace*{-0.3em}
    \begin{minipage}[t]{0.5\textwidth}
      \centering
      (a) Great Recession
    \end{minipage}
    \begin{minipage}[t]{0.5\textwidth}
      \centering
      (b) Post-Covid
    \end{minipage}

    \begin{threeparttable}
    \centering
    \begin{minipage}{\textwidth}
      \begin{tablenotes}[para,flushleft]
    \setlength{\lineskip}{0.2ex}
    \notsotiny 
  {\textit{Notes}: The \textbf{upper panel} shows AlbaMA and the MA(12). The \textbf{middle panels} compare AlbaMA to standard moving averages and the Savitzky-Golay filter for (a) the Great Recession and (b) the post-Covid surge. The \textbf{lower panels} present the weights the RF assigns to past observations. All measures are one-sided.}
    \end{tablenotes}
  \end{minipage}
  \end{threeparttable}
\end{figure}

In Figures \ref{fig:US_CPI} and \ref{fig:US_CPIcore}, we present detailed results on AlbaMA for both US headline and core inflation. The time series panels illustrate a comparison among various moving average techniques, including AlbaMA, spanning from the early 2000s to the present. We focus closely on two key periods: (a) the Great Recession, and (b) the recent inflation surge beginning in 2021. Finally, we offer insights into the time-variation of the weighted average by reporting the weights AlbaMA assigns to recent observations for each $\hat{y}_t$.


\vskip 0.15cm
{\noindent \sc \textbf{Headline Inflation.}} The upper panel of Figure \ref{fig:US_CPI} contrasts AlbaMA’s readings with the widely-used twelve-months moving average of headline inflation. Notably, the RF-based solution is more responsive and exhibits greater volatility than the year-over-year (YoY) rate, particularly in the face of significant economic shocks. One illustrative episode is the sharp decline in 2008. MA(12) lags in responding to both the steep drop and subsequent mean reversion. It also shows pronounced base effects once the extreme November 2008 value falls out of its look-back window. In contrast, AlbaMA swiftly captures the downturn \textcolor{black}{from 3.2\% in September 2008 to -11.3\% in December} and quickly returns to pre-crisis levels by summer 2009, underscoring the transitory nature of the dip. Interestingly, while MA(12) indicates a significant inflation increase in late 2009, AlbaMA reveals that this uptick actually began six months earlier and, if anything, was already trending downward by that time.


During the post-pandemic inflation surge, we observe a strong and steady upward trend in AlbaMA, with its inflation readings consistently exceeding those of MA(12) and the SG filter. AlbaMA peaks at 11.2\% in June 2022, before dropping to below 4\% within the following two months. In contrast, the commonly used MA(12) shows a more gradual and persistent slowdown, taking until June 2023 to converge to the values of short-run measures. \textcolor{black}{As shown in the lower right panel of Figure \ref{fig:US_CPI}, $y_{t-1:t-5}$ account for half the weight in capturing the persistent upward trend. At the critical juncture, the role of longer lags shrinks significantly as the weight assigned to $y_t$, the most recent observation, doubles -- from 19\% in May 2022 to 41\% in July 2022. }

In comparing AlbaMA to additional benchmarks (see middle panels of Figure \ref{fig:US_CPI}), we observe similarities with the MA(3) and the SG filter around inflection points. However, the SG filter fails to capture the upward trend during the post-Covid inflation surge, whereas AlbaMA aligns more closely with the MA(6) in these trending periods.

This alignment can be examined more rigorously by analyzing the time-varying weights that the RF model assigns to past observations (see lower panels of Figure \ref{fig:US_CPI}). On average, AlbaMA assigns substantial weight—approximately 90\%—to the last five months, suggesting a general preference for the MA(3) and MA(6) benchmarks. Around key inflection points, such as the 2008 oil shock, the initial Covid shock, and the post-pandemic slowdown, the weight on recent data sharply increases. In its “steady-state,” the weight on the last three to end-of-sample observations ($y_{t-3:t-\text{end}}$) hovers around 40\%, but it drops to less than 20\% during these periods, with the difference reallocated to $y_t$ or $y_{t-1:t-3}$. As discussed in the literature, high volatility may lead to faster price adjustments, highlighting the benefits of measures that are more responsive to recent observations \citep{hall2023major,eeckhout2023instantaneous}.



The two-sided version, displayed in Figure \ref{fig:US_CPI_2sided} in the appendix, reinforces our conclusions on AlbaMA’s responsiveness. As one should expect, standard moving averages symmetrically smooth $y_t$ using both past and future data, thereby shifting turning points forward. AlbaMA behaves differently. It initially downweights future observations before a breakpoint, upweights them after it, and eventually reassigns weights to both sides after the dust has settled.  Consequently, the one-sided assessment of AlbaMA closely aligns with its two-sided counterpart—a consistency infrequently seen among time series filters \citep{orphanides2002unreliability}. This consistency is further examined in Section \ref{sec:consist}, where we quantify, for AlbaMA and various benchmarks, the average correspondence between one-sided and two-sided estimates across a broader set of series.




\vskip 0.15cm
{\noindent \sc \textbf{Core Inflation.}} It is evident that oil price volatility contributes significantly to the adaptive behavior observed in AlbaMA for headline inflation. This raises the question of whether similar gains from adaptive moving averages can be achieved for inflation measures that exclude oil prices. The answer  is yes.  In fact, we find more significant adaptive behavior when applying AlbaMA to US core inflation. We observe a close alignment between MA(12) and AlbaMA during stable periods (see upper panel in Figure \ref{fig:US_CPIcore}). This alignment is markedly disrupted, however, during the Great Recession and the post-Covid inflation surge. In these instances, the RF-based AlbaMA swiftly detects structural breaks, whereas the MA(12) responds sluggishly, smoothing out much of the impact. Notably, AlbaMA surges in April 2021, maintaining elevated levels that later converge with the MA(12).

Among other benchmark measures (see middle panel of Figure \ref{fig:US_CPIcore}), only the MA(3) matches AlbaMA’s timeliness at inflection points. While the SG filter accurately captures short-lived downturns, such as those at the end of 2009 and in 2020, it significantly underestimates inflation trends in other periods. Longer moving averages effectively track the high-inflation period in 2022 and 2023 but are slow to capture its onset. 


Examining the bottom panel of Figure \ref{fig:US_CPIcore}, we observe more pronounced swings in weights  than those seen for headline inflation. This is attributable to AlbaMA assigning during stable periods a substantial portion of its weight to distant lags, with approximately 70\% of weights allocated to  \(y_{t-3:t-\text{end}}\). This share is found to be significantly smaller in the headline inflation application. In quieter times, such as throughout 2019, AlbaMA's core inflation weight distribution closely resembles that of an equally weighted MA(12), where \(y_t\) holds about \(\sfrac{1}{12}\) of the weight, and \(y_{t-6:t-\text{end}}\) holds around \(\sfrac{6}{12}\).

AlbaMA's adaptability lies in its swift departure from the MA(12) baseline by upweighting short lags after significant shocks, before eventually reverting to a near MA(12) distribution. For instance, following the initial Covid shock, the weight on \(y_t\) triples, while the weight on \(y_{t-6:t-\text{end}}\) drops from 51\% in 2019m12 to 5\% in 2020m8. Simultaneously, \(y_{t-1:t-2}\) gains importance at the expense of \(y_{t-3:t-5}\). A similar adjustment occurs in the spring of 2021, once again transforming the effective weight distribution from an MA(12) to an MA(3). As noted in \cite{stock2007has}, the focus on recent observations intensifies during periods dominated by transitory shocks but shifts toward distant lags when the variance of permanent shocks increases. The high persistence following the post-Covid surge reflects this, with over 50\% weights assigned to lags 6 and beyond, considerably slowing the disinflation process.


These findings are closely mirrored by those from the two-sided analysis presented in \mbox{Figure \ref{fig:US_CPIcore_2sided}} (Appendix). Once again, AlbaMA generally aligns with the MA(12), except around the abrupt shifts occurring in October 2009 and April 2021. In both instances, AlbaMA continues to detect high persistence in the aftermath of those, consistent with the one-sided results.  As desired, in the two-sided case, AlbaMA assigns greater weight to recent past observations just before the break and to more distant future observations immediately after.



\begin{figure}[t!]
  \caption{\normalsize{Additional Benchmarks for US Inflation (one-sided)}} \label{fig:us_infl_ucsv}

\vspace*{-.2em}
    \begin{minipage}[t]{0.5\textwidth}
      \centering 
      (a) CPI Inflation
    \end{minipage}
    \begin{minipage}[t]{0.5\textwidth}
      \centering 
      (b) CPI Core Inflation
    \end{minipage}

    \begin{minipage}[t]{0.5\textwidth}
      \centering
      \includegraphics[width=\textwidth, trim = 0mm 0mm 0mm 0mm, clip]{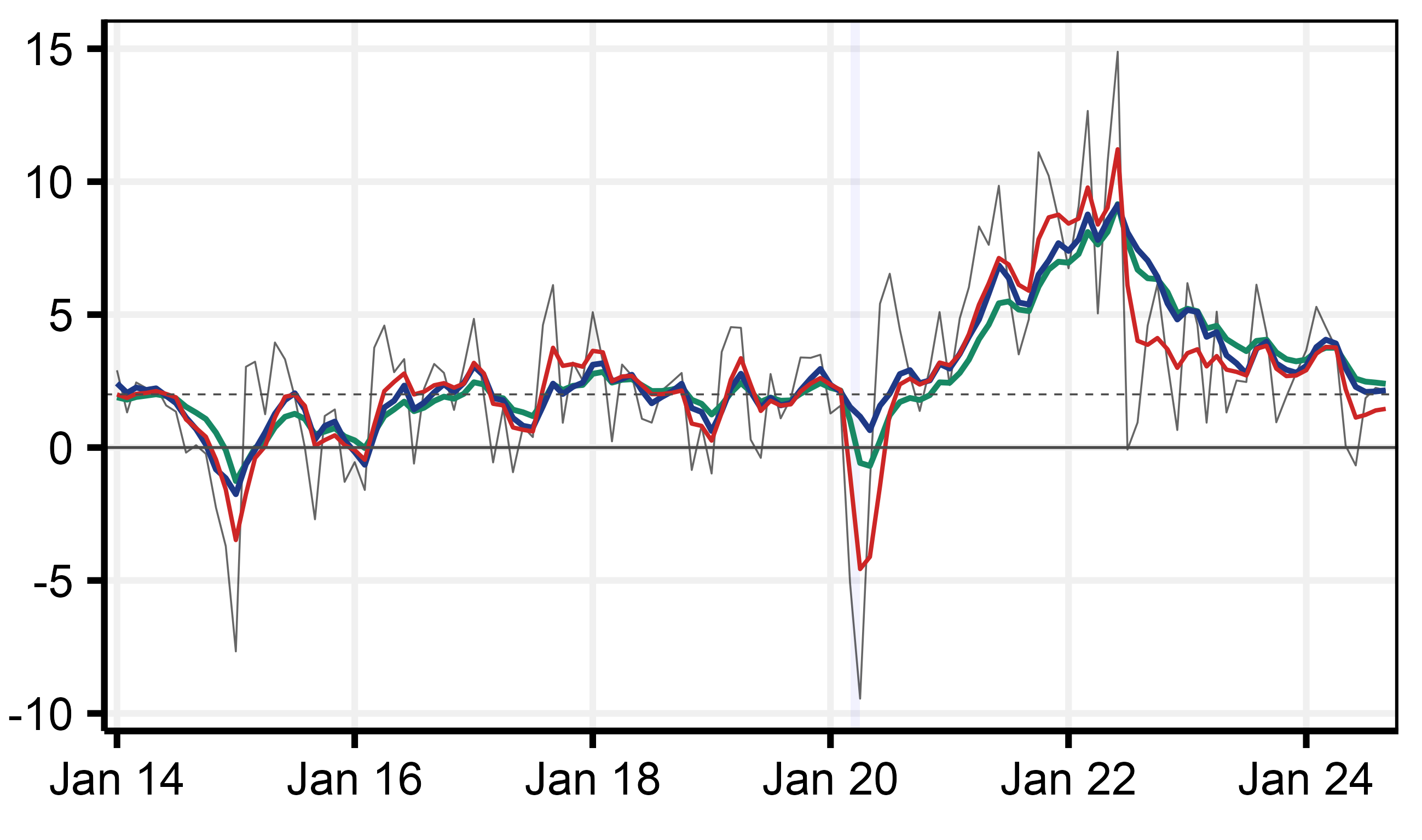}
    \end{minipage}%
    \begin{minipage}[t]{0.5\textwidth}
      \centering
      \includegraphics[width=\textwidth, trim = 0mm 0mm 0mm 0mm, clip]{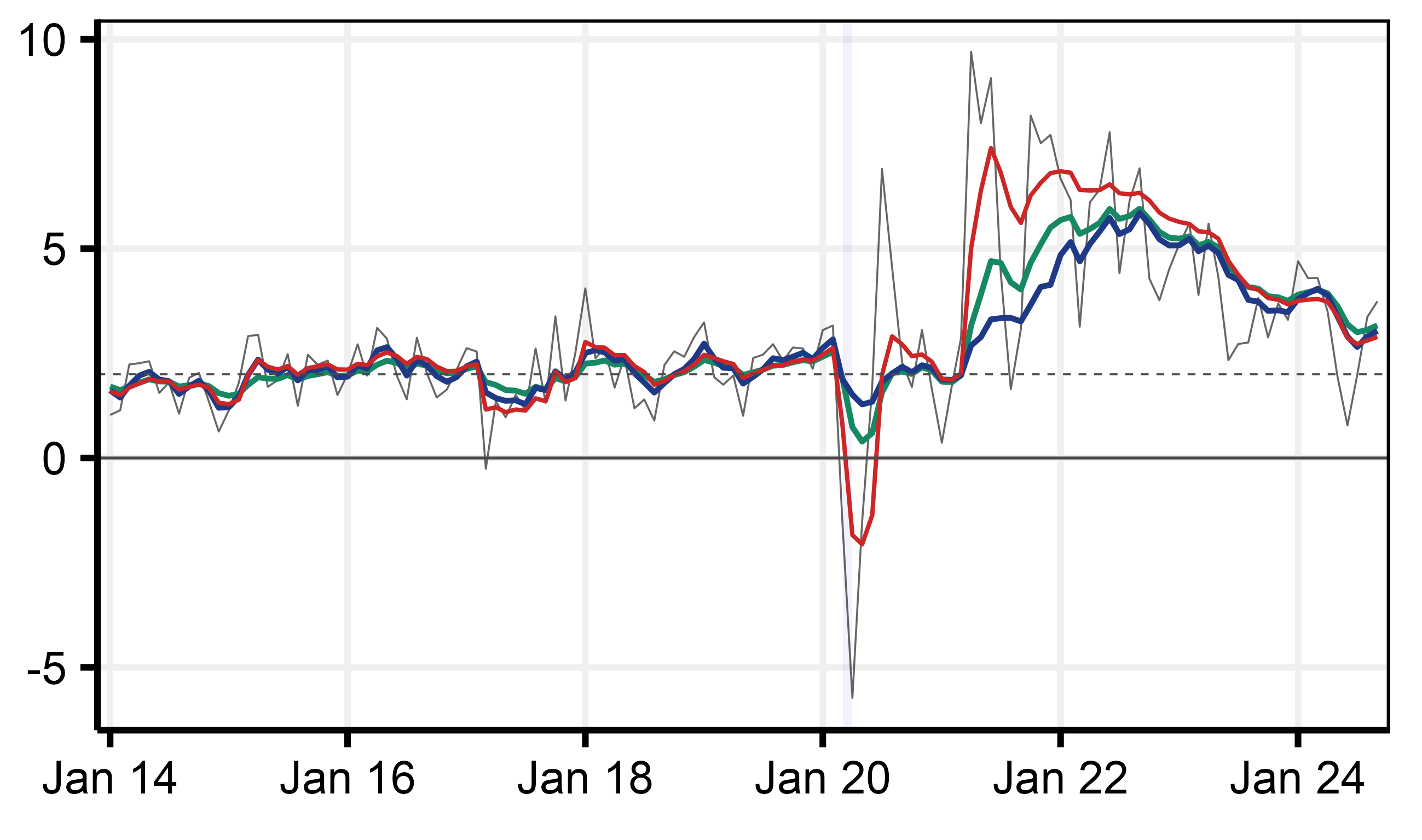}
    \end{minipage}

    \vspace*{-.5em}
    \begin{minipage}[t]{\textwidth}
      \centering
      \hspace*{1.0em} \includegraphics[width=0.95\textwidth, trim = 0mm 0mm 0mm 0mm, clip]{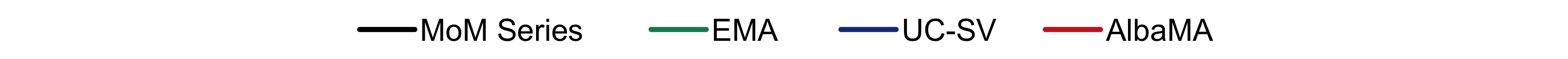}
    \end{minipage}

   \begin{threeparttable}
    \centering
    \begin{minipage}{\textwidth}
      \begin{tablenotes}[para,flushleft]
    \setlength{\lineskip}{0.2ex}
    \notsotiny 
  {\textit{Notes}: The figure compares AlbaMA to \cite{stock2007has}'s unobserved components model with stochastic volatility (UC-SV) and an exponential moving average with a smoothing factor over twelve months (EMA).}
    \end{tablenotes}
  \end{minipage}
  \end{threeparttable}
\end{figure}

\vskip 0.15cm
{\noindent \sc \textbf{Further Comparison: UC-SV and Exponential MA.}} As additional benchmarks, we compare AlbaMA to the UC-SV model of \cite{stock2007has} and an exponential moving average with a twelve-month smoothing parameter (EMA). Both models share key characteristics with AlbaMA and have been widely employed in inflation analysis. The UC-SV model decomposes inflation into trend and cyclical components, allowing the variances of their respective shocks to evolve stochastically over time. This enables the model to dynamically adjust the weighting of recent versus past observations in response to changes in volatility. In contrast, EMA imposes a fixed exponential decay structure, progressively diminishing the influence of older observations in a predetermined manner. While EMA lacks the adaptivity of UC-SV, it shares with AlbaMA the ability to upweight recent observations more sharply than a standard moving average.


As shown in Figure \ref{fig:us_infl_ucsv}, the series align closely up to the onset of the Covid-19 pandemic.\footnote{A full-sample comparison is available in Figures \ref{fig:infl_ucsv_right} and \ref{fig:infl_ucsv_center} in the appendix.} Notably, UC-SV and AlbaMA track similar paths for both headline and core inflation up to the Covid-19 shock. The primary divergence occurs around inflection points, where AlbaMA identifies abrupt shifts that the benchmarks smooth over. For headline inflation, all models capture the initial surge with comparable magnitude and speed, but UC-SV and EMA exhibit a gradual decline afterward, contrasting with AlbaMA’s abrupt downward adjustment in early 2022. For core inflation, the pattern reverses: while all models agree on a protracted disinflation process over the past three years, AlbaMA captures a sharper spike during the surge in 2021, diverging from the benchmarks’ more gradual upward trajectory.

 

\subsection{A Look at Additional Series}\label{sec:others}



\begin{figure}[t!]
  \caption{\normalsize{Results for Additional Cases}} \label{fig:results_other}
  
  \vspace*{-0.4em}
    \begin{minipage}[t]{\textwidth}
      \centering
      (a) EA HICP Inflation During the Great Recession 
    \end{minipage}

    \begin{minipage}[t]{0.5\textwidth}
      \centering
       \includegraphics[width=\textwidth, trim = 2mm -5mm 0mm 0mm, clip]{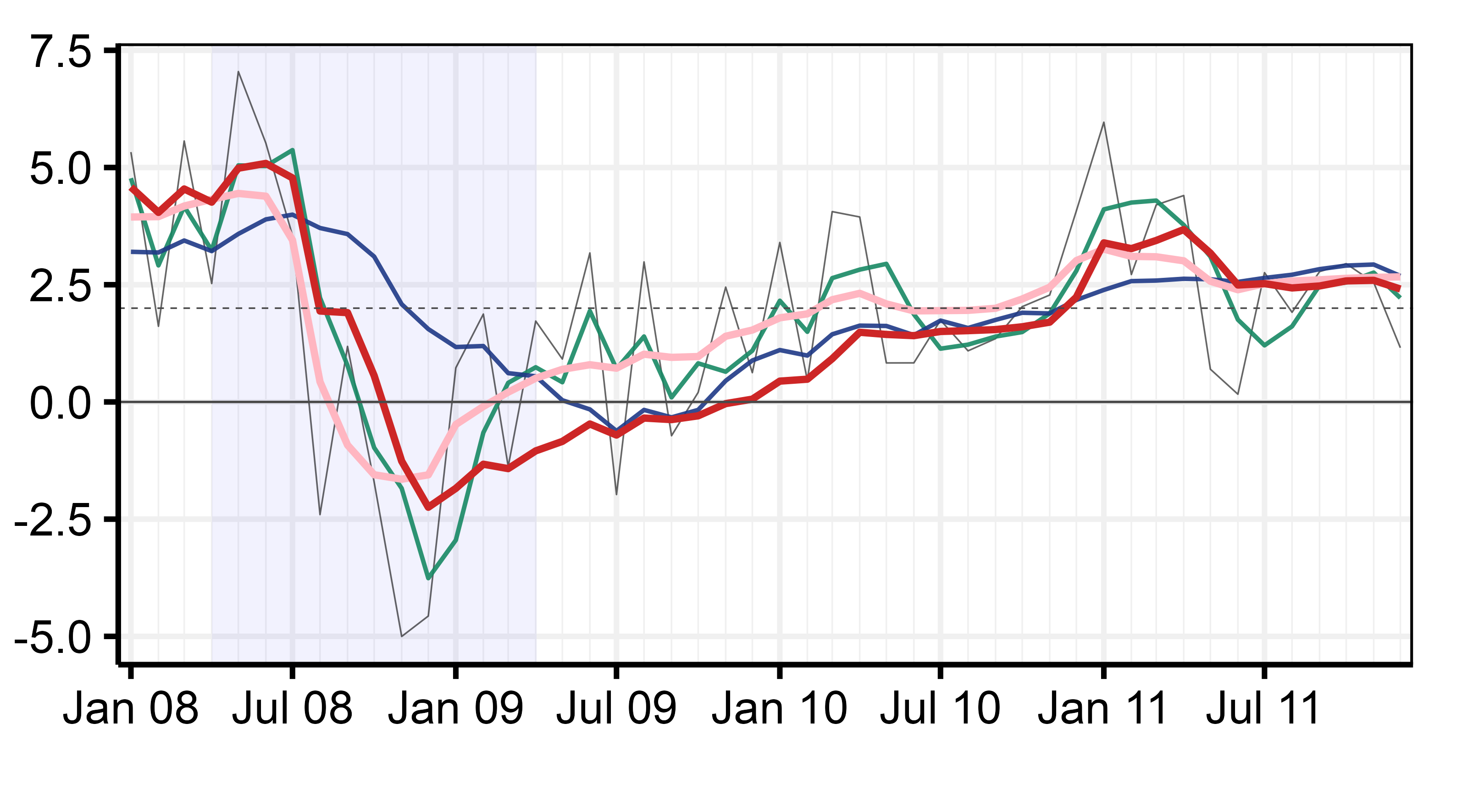}
    \end{minipage}%
    \begin{minipage}[t]{0.5\textwidth}
      \centering
      \includegraphics[width=1.06\textwidth, trim = 0mm 3mm 5mm 0mm, clip]{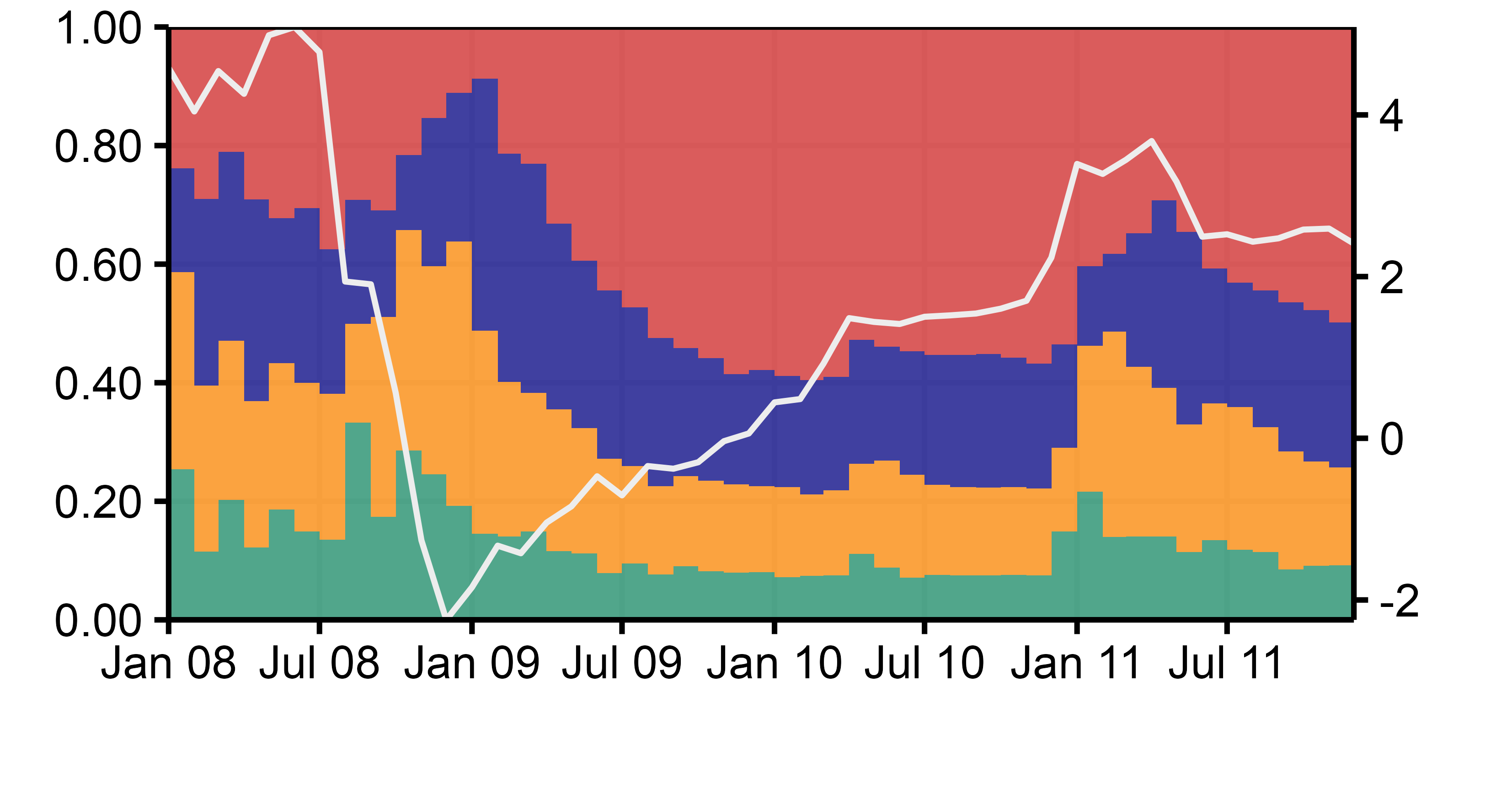}
    \end{minipage}

    \vspace*{-1.1em}
    \begin{minipage}[t]{\textwidth}
      \centering
      (b) EA HICP Core Inflation After Covid-19
    \end{minipage}

    \begin{minipage}[t]{0.5\textwidth}
      \centering
       \includegraphics[width=\textwidth, trim = -1mm -6mm 3mm 0mm, clip]{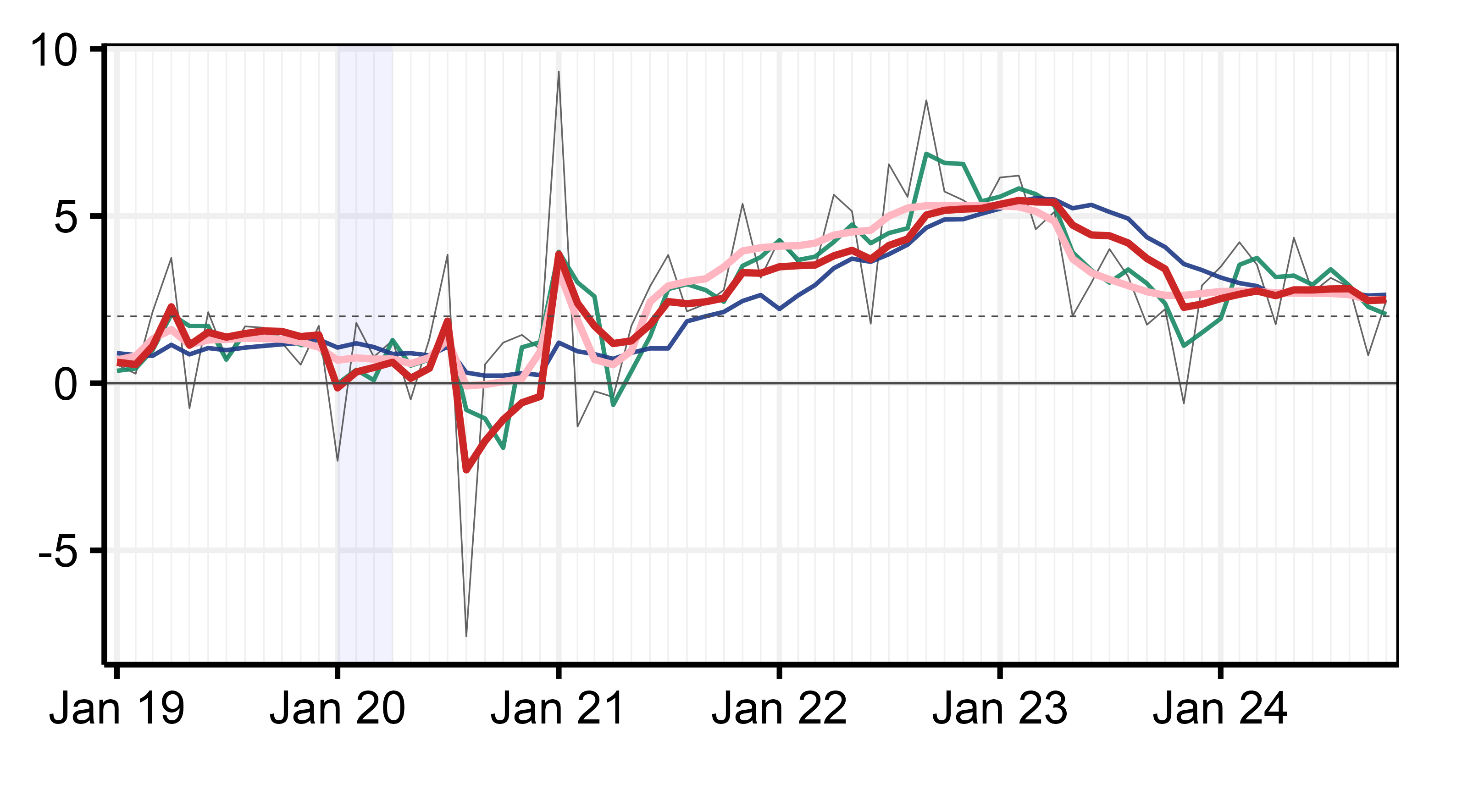}
    \end{minipage}%
    \begin{minipage}[t]{0.5\textwidth}
      \centering
      \includegraphics[width=1.08\textwidth, trim = 0mm 3mm 0mm 0mm, clip]{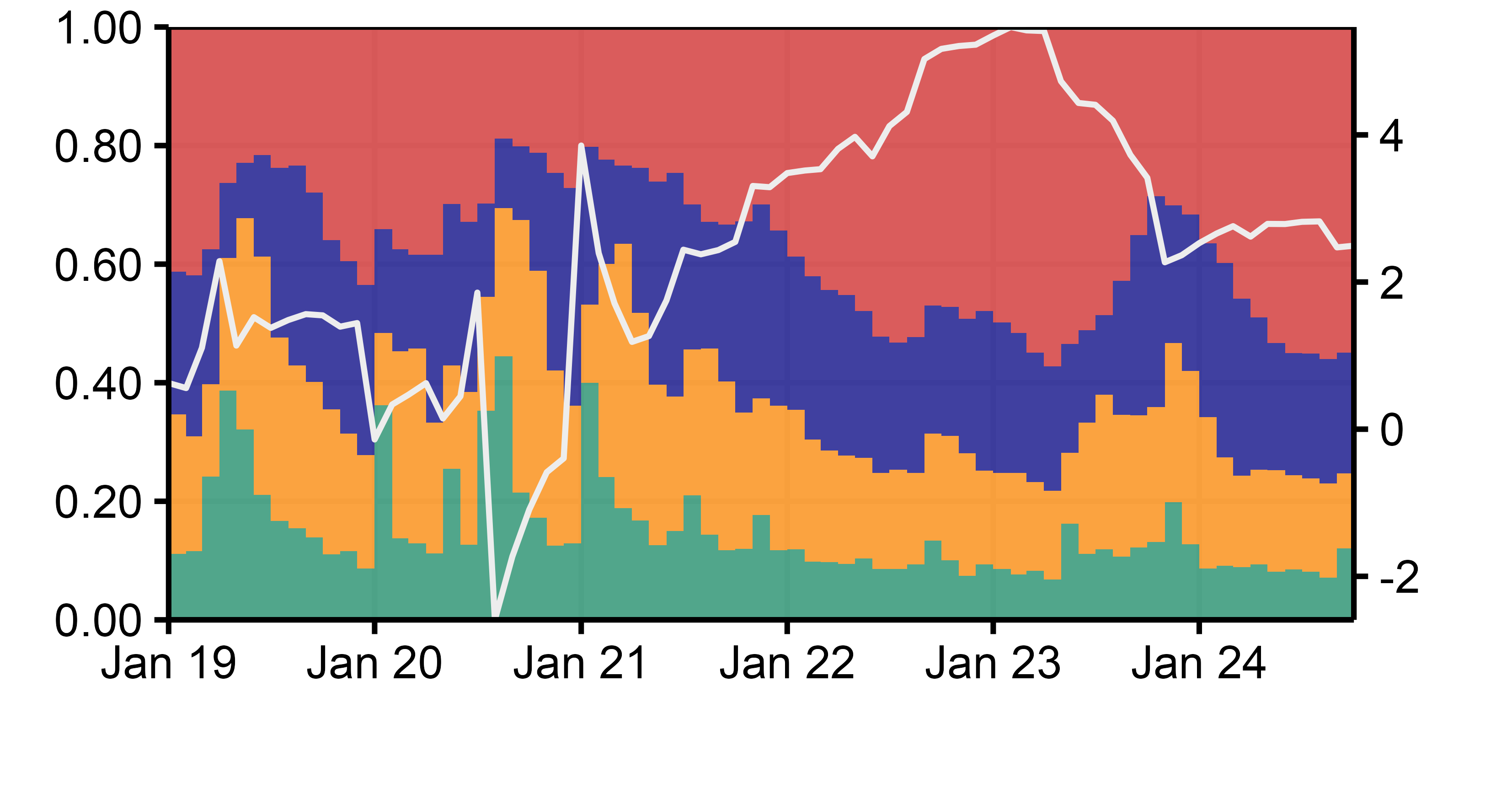}
    \end{minipage}

    \vspace*{-1.1em}
    \begin{minipage}[t]{\textwidth}
      \centering
      (c) US PMI During the Great Recession
    \end{minipage}

     \begin{minipage}[t]{0.5\textwidth}
      \centering
       \includegraphics[width=\textwidth, trim = 0mm -6mm 0mm 0mm, clip]{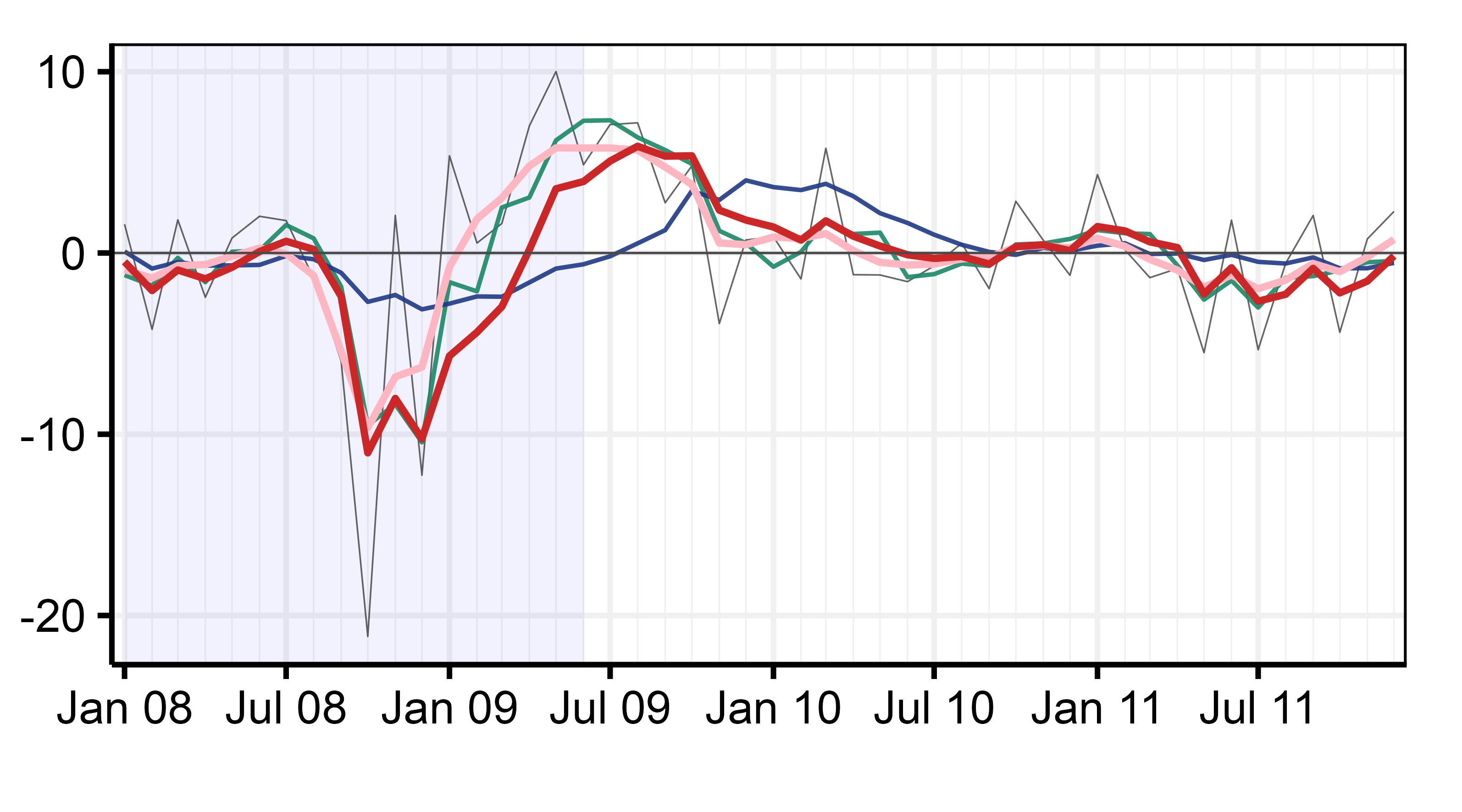}
    \end{minipage}%
    \begin{minipage}[t]{0.5\textwidth}
      \centering
      \includegraphics[width=1.1\textwidth, trim = 0mm 4mm 0mm 0mm, clip]{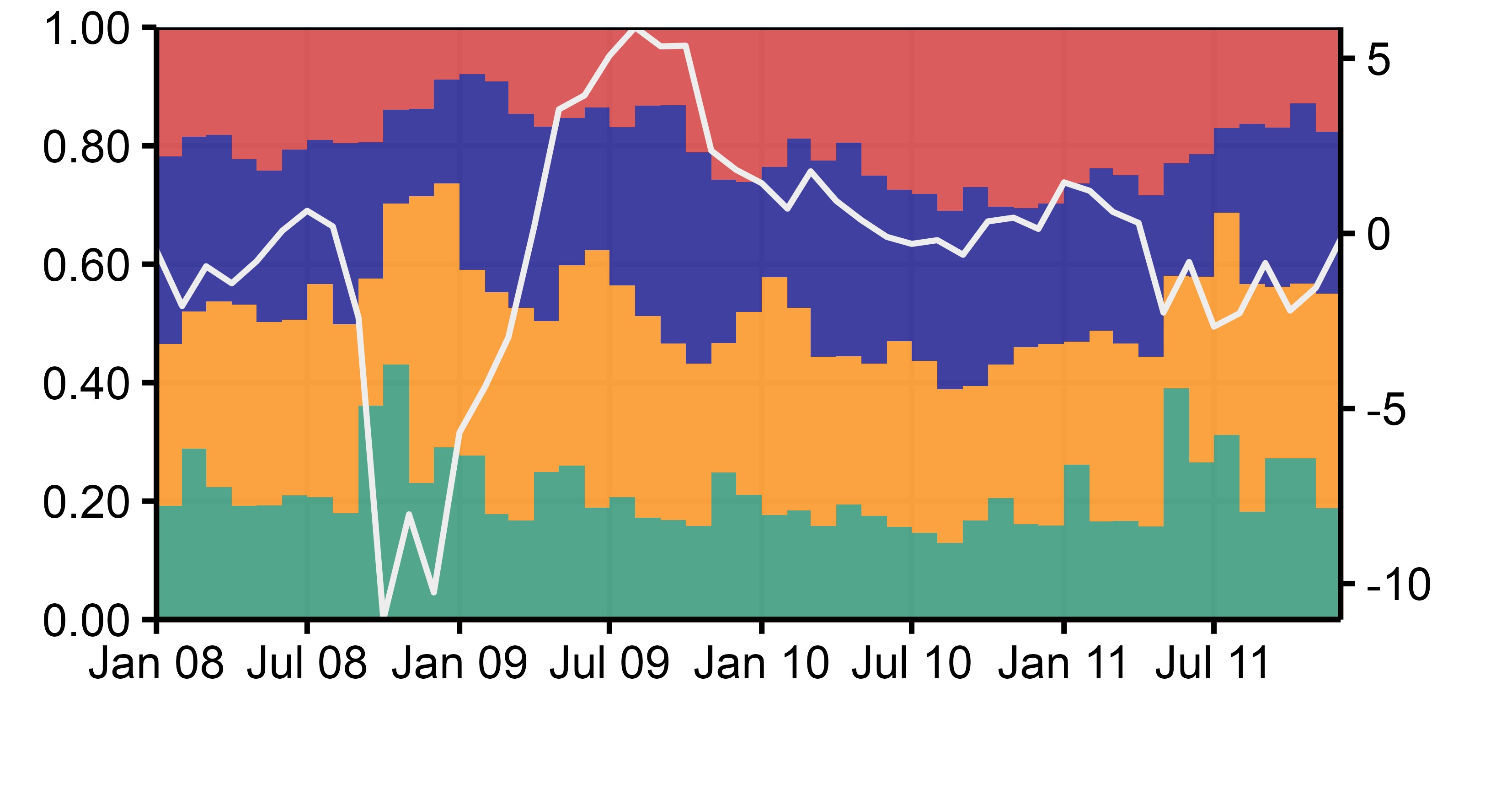}
    \end{minipage}

    \vspace*{-1.1em}
     \begin{minipage}[t]{\textwidth}
      \centering
      (d) EA Industrial Production During the Great Recession
    \end{minipage}    

    \begin{minipage}[t]{0.5\textwidth}
      \centering
       \includegraphics[width=\textwidth, trim = 1mm -5mm 2mm 0mm, clip]{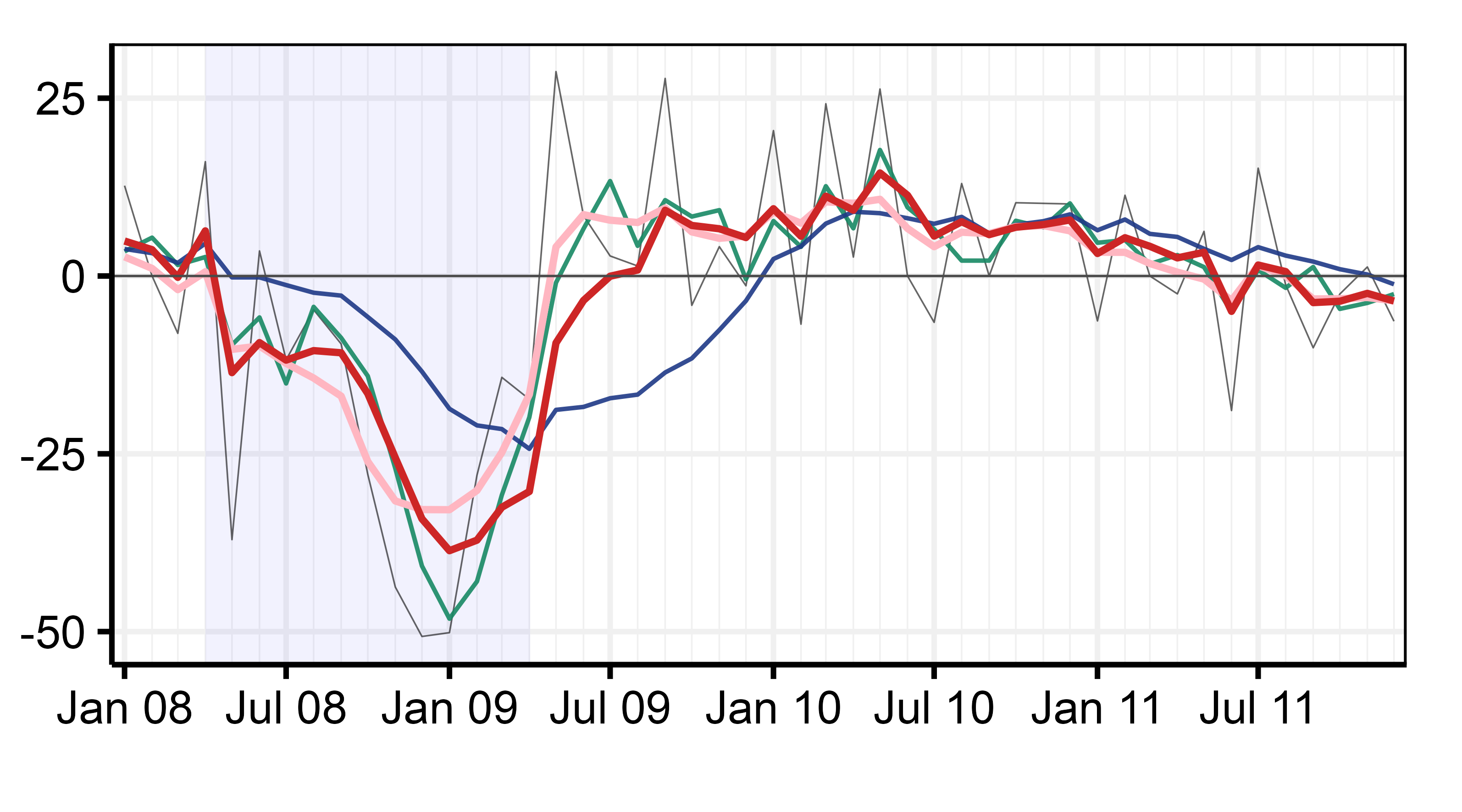}
    \end{minipage}%
    \begin{minipage}[t]{0.5\textwidth}
      \centering
      \includegraphics[width=1.08\textwidth, trim = 0mm 3mm 5mm 0mm, clip]{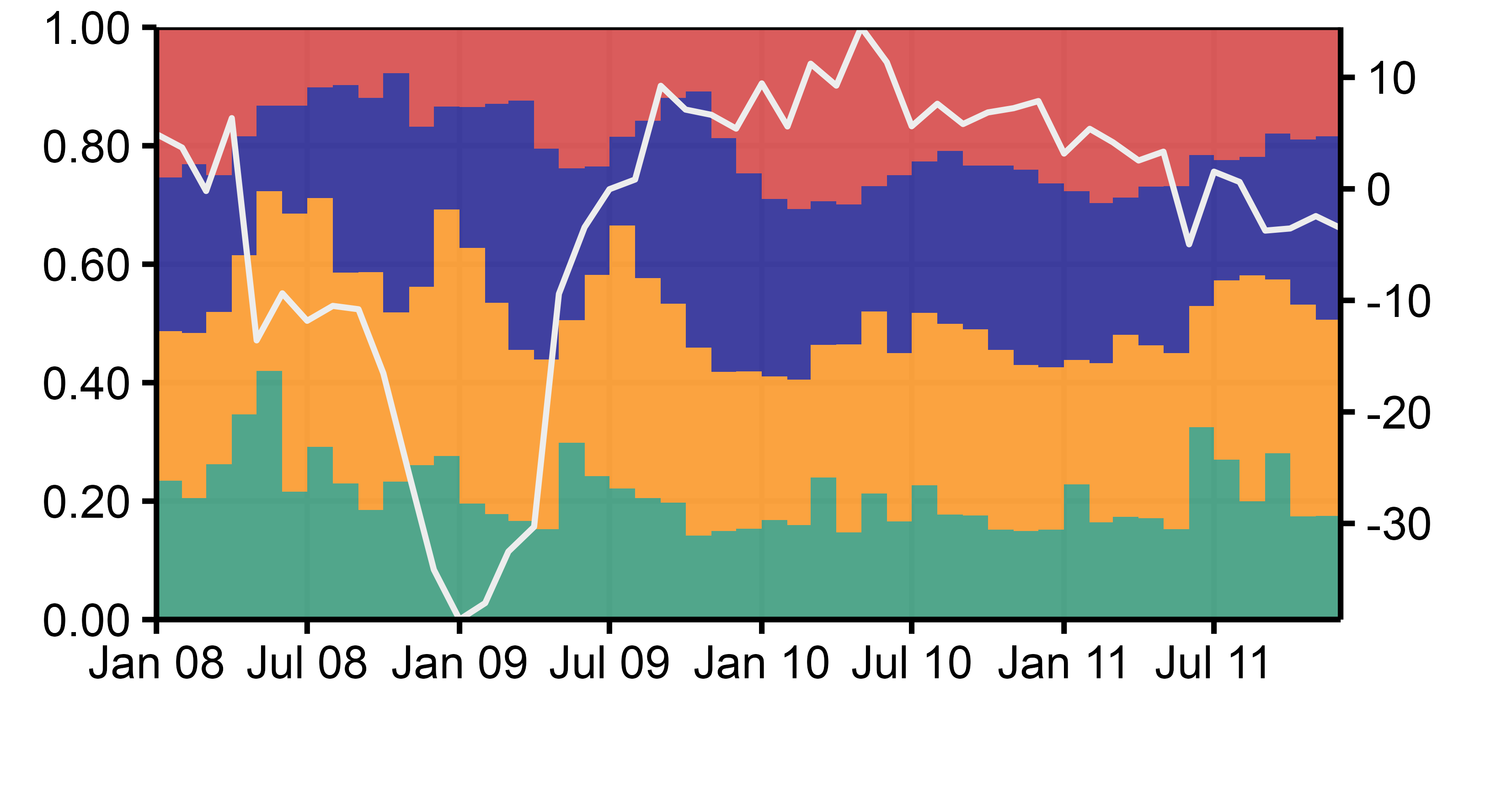}
    \end{minipage}

    \vspace*{-1.2em}
    \begin{minipage}[t]{0.5\textwidth}
      \centering
      \hspace*{-0.8em} \includegraphics[width=1.3\textwidth, trim = 0mm 0mm 0mm 0mm, clip]{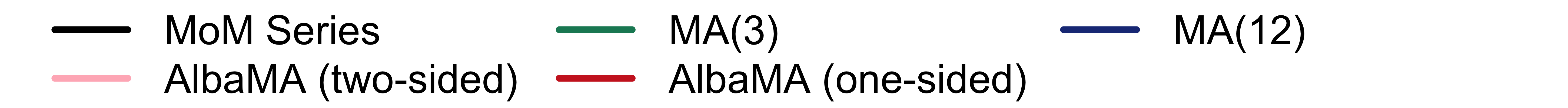}
    \end{minipage}%
    \begin{minipage}[t]{0.5\textwidth}
      \centering
      \vspace*{-2.05em}
     \hspace*{1.1em}  \includegraphics[width=\textwidth, trim = 0mm 0mm 0mm 0mm, clip]{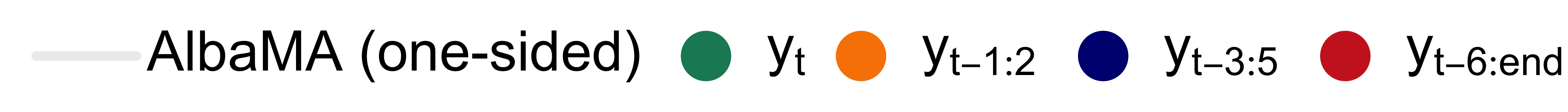}
    \end{minipage}

    \begin{threeparttable}
    \centering
    \begin{minipage}{\textwidth}
      \begin{tablenotes}[para,flushleft]
    \setlength{\lineskip}{0.1ex}
    \notsotiny 
  {\textit{Notes}: The figure presents selected cases to illustrate AlbaMA's adaptability; detailed panels can be found in Figures \ref{fig:EA_CPI} to \ref{fig:US_PMI}. The left panel compares AlbaMA with benchmarks while the right panels give the RF's weights assigned to past observations. Note that we focus on one-sided cases unless stated otherwise.}
    \end{tablenotes}
  \end{minipage}
  \end{threeparttable}
\end{figure}

In this subsection, we present selected results for additional macroeconomic variables, focusing specifically on the Great Recession and the post-Covid inflation surge to illustrate AlbaMA’s adaptability in these contexts. We include inflation series for the Euro Area, along with real activity data for both the Euro Area and the US (see Figure \ref{fig:results_other}). Complete panels, analogous to Figures \ref{fig:US_CPI} and \ref{fig:US_CPIcore}, are provided in Appendix \ref{app:panels}.



\vskip 0.15cm
{\noindent \sc \textbf{EA Inflation Series.}} In all cases, AlbaMA demonstrates the adaptability highlighted in previous sections, responding promptly to economic inflection points by placing greater weight on recent observations. This behavior is especially evident in the inflation data for the Euro Area during the Great Recession. \textcolor{black}{In July 2008, AlbaMA captures a rapid deceleration in inflation from 4.8\% to 1.9\%, marked by strong contributions from recent data points, with $y_t$ almost trippling its weight from 13\% to 33\%.} Unlike the US, where AlbaMA indicated a brief downturn, inflation in the Euro Area shows a slower rebound, similar to an MA(12), driven by the increasing significance of distant lags.

A comparable pattern emerges in AlbaMA’s estimates for Euro Area core inflation during the post-Covid inflation surge. Initially, it rises in line with the MA(3) and transitions into a sustained upward trend until April 2023, resembling the MA(12). Following the peak, AlbaMA captures a relatively rapid disinflation (in contrast to the US), with heightened weight on recent observations beginning in mid-2023. Regarding consistency with the two-sided estimates, the use of both past and future data results in minor forward revisions within the Euro Area inflation subsample, although we note that the adjustments are more pronounced in the top panel.

\newgeometry{left=2 cm, right= 2 cm, top=2.1 cm, bottom=2.3 cm}


\restoregeometry

\vskip 0.15cm
{\noindent \sc \textbf{Real Activity Series.}} Other series, beyond inflation, are often expressed in year-over-year growth rates. This is the case of US PMI and Euro Area industrial production, shown in panels (c) and (d) of Figure \ref{fig:results_other}. These real activity examples emphasize the delayed response of the MA(12) in signaling disruptions. For example, AlbaMA captures a sharp decline in US PMI, reaching -11\% in October 2008, followed by a rapid recovery peaking in August 2009. In contrast, the MA(12) registers only a modest contraction (bottoming out at -3.1\% in December 2009) with a delayed recovery that peaks in early 2010. AlbaMA’s more responsive assessment of PMI stems from (i) assigning higher steady-state weight to recent observations and (ii) doubling the weight on $y_t$ in the fall of 2008.

Similarly, in the Euro Area, the MA(12) lags both the decline and subsequent recovery in industrial production, whereas AlbaMA more closely tracks the MA(3), as evidenced by approximately 60\% of the weighting assigned to the most recent three months. In both cases, the timeliness of AlbaMA is corroborated by its two-sided counterpart, which consistently captures the slowdown and only marginally accelerates the recovery indication.


\subsection{A More Formal Evaluation of One-Sided vs Two-Sided Consistency}\label{sec:consist}

We complement our qualitative assessment with a more systematic examination of the consistency between one-sided and two-sided results. One-sided moving averages and time series filters can experience substantial revisions as new data arrives, even without the data itself being revised. Consequently, real-time estimates may diverge from ex-post optimal two-sided estimates, which incorporate both past and future data points. This issue is known under various names, such as the "boundary problem” in kernel-based approaches and the “filter vs smoother problem” in filtering applications. Thus, it is pertinent to assess how well the ex-post optimal two-sided AlbaMA estimates align with its feasible one-sided counterpart, and to compare AlbaMA’s "performance" with that of traditional moving averages in this context.


\begin{figure}[t]
  \caption{\normalsize{R$^2$ Between Two-Sided and One-Sided Estimates}} \label{fig:R2}
    \centering
    \vspace*{-0.7em}
    \includegraphics[width=\textwidth, trim = 0mm 0mm 0mm 0mm, clip]{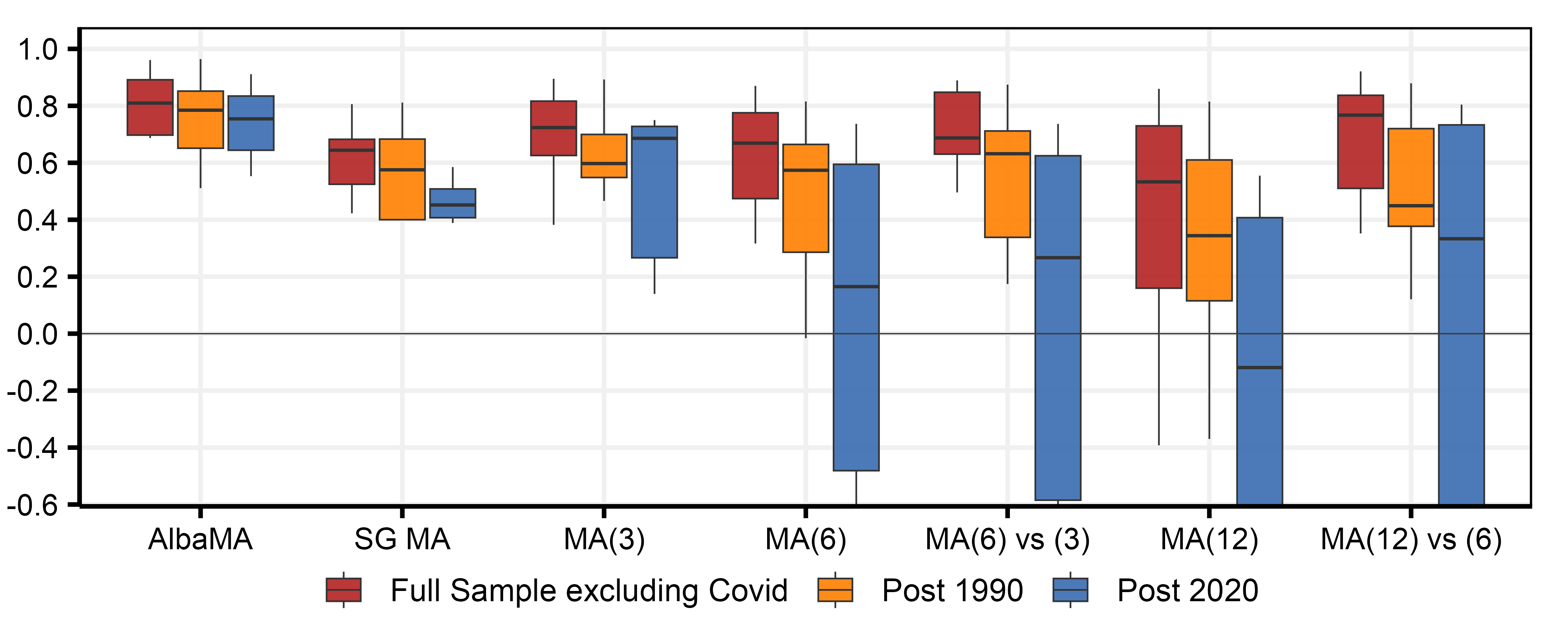}
    \begin{threeparttable}
    \centering
    \begin{minipage}{\textwidth}
      \begin{tablenotes}[para,flushleft]
    \setlength{\lineskip}{0.2ex}
    \notsotiny 
  {\textit{Notes}: The figure presents R$^2$ between two-sided and one-sided estimates of each model. The boxplots summarize the performance across variables (i.e., inflation, core inflation, industrial production, unemployment and the PMI for the US and the EA). SG MA refers to the Savitzgy-Golay filter, while MA(3), MA(6) and MA(12) denote three-, six-, and twelve-months moving averages. MA(6) vs (3) indicates the R$^2$ between the two-sided MA(6) versus the one-sided MA(3), likewise MA(12) vs (6) shows the R$^2$ between the two-sided MA(12) versus the one-sided MA(6). Full sample excluding Covid runs from 1963m1 to 2024m10 excluding the year 2020. Post 1990 runs from 1990m1 to 2024m10 and post 2020 spans 2020m1 to 2024m10.}
    \end{tablenotes}
  \end{minipage}
  \end{threeparttable}
\end{figure}

We use $R^2$ between one-sided and two-sided estimates to assess each MA intertemporal consistency, treating the one-sided estimate as the “prediction” and the two-sided version as the realized value. The choice of $R^2$ over alternatives, such as root mean squared error, reflects the need to account for differences in target series variance; for instance, the variance of MA(12) is inherently lower than that of MA(3). Because the coefficient on the one-sided MA is fixed at 1 and the intercept at 0, meaning they are not estimated, $R^2$ values can sometimes fall below zero indicating that the prediction performs worse than the full-sample average. However, the maximal value for this $R^2$ remains 1.


Figure \ref{fig:R2} displays boxplots for each model and evaluation sample, summarizing performance across nine variables. These include the six previously analyzed variables and three additional ones: US and Euro Area unemployment rate, and US industrial production. Detailed panels for these additional variables are available in Figures \ref{fig:US_INDPRO} to \ref{fig:EA_UNRATE}. Note that the highest-ranked $R^2$ boxplot indicates the best overall model performance across all variables, though it does not imply top performance for each individual variable. For a detailed evaluation, please refer to Tables \ref{tab:results_us} and \ref{tab:results_ea} in the appendix.


\textcolor{black}{Overall, AlbaMA performs well across all three evaluation samples, achieving the highest median $R^2$ in each test set: 0.81 for the full sample excluding Covid, 0.78 for post-1990, and 0.75 for post-2020. Moreover, the interquantile range is notably narrow, with the 25$^{th}$ quantile remaining at 0.64 (for post-2020) or higher (0.65 for post-1990 and 0.7 for the full sample excluding Covid). While benchmarks such as MA(3) and MA(6) vs (3) come in close seconds for the full sample excluding Covid, they fall short for the post-1990 and post-2020 samples.  Overall, \mbox{Figure \ref{fig:R2}} illustrates that AlbaMA offers higher consistency levels that are robust to the choice of target  and evaluation sample, as evidenced by the limited dispersion within and across boxes.}


A closer analysis (see Tables \ref{tab:results_us} and \ref{tab:results_ea}) reveals that standard moving averages are challenging benchmarks for inflation data, which tend to change gradually with the overall price level. For real activity variables, the SG filter provides competitive consistency, followed by shorter-run moving averages, as real activity often exhibits sharper fluctuations that benefit from more responsive measures. AlbaMA adapts effectively to both types of dynamics, consistently ranking among the top performers based on the one-sided to two-sided $R^2$ criterion.





\section{Conclusion}\label{sec:conclusion}

This paper introduced a simple adaptive moving average estimator tailored for macroeconomic monitoring, addressing the challenge of balancing timeliness and stability in tracking noisy series like inflation. By dynamically adjusting the look-back window, our Random Forest-based approach offers a flexible alternative to fixed moving averages. Comparisons with traditional filters reveal that this adaptive method provides unique insights, particularly in capturing shifts such as the 2022 inflation surge and subsequent slowdown in the US and Euro Area.


There are a few avenues for future research. Here are two of them leveraging \cite{MRFjae}'s more sophisticated Macro Random Forest (MRF) algorithm. First, by moving beyond the straightforward moving average interpretation of Random Forest towards a more general time series filter approach, one could implement MRF with a trend as the linear component. This setup would address some of plain RF’s finite-sample limitations in capturing smooth changes. 


A second option would be to incorporate $y_t$ and its lags into the MRF’s linear component, and enforce a sum-to-one constraint on the time-varying coefficients at each point in time. Leveraging large datasets such as FRED-MD \citep{mccrackenng} could enable generalized, time-varying MA weights influenced by external variables. The resulting moving average could adapt as a function of economic fundamentals and potentially exhibit forward-looking behavior in its choice of weights.

 
 \pagebreak
 \clearpage
 \setlength\bibsep{5pt}
        
\bibliographystyle{apalike}

\setstretch{0.75}
\bibliography{AMA}

\clearpage

\appendix
\newcounter{saveeqn}
\setcounter{saveeqn}{\value{section}}
\renewcommand{\theequation}{\mbox{\Alph{saveeqn}.\arabic{equation}}} \setcounter{saveeqn}{1}
\setcounter{equation}{0}
    
\section{Appendix}\label{app}
\setstretch{1.25}

\subsection{Bechnmark Filtering Techniques} \label{app:bench}

We choose a broad set of benchmark filtering techniques. These range from standard moving averages, to adaptive extensions as well as more sophisticated trend filtering techniques.

\vskip 0.15cm 
{\sc \noindent \textbf{Standard and Adpative Moving Averages}.} The most standard ones are moving averages with different window sizes ($k$). In the context of macroeconomic variables, the most commonly used ones are averaging over three, six, and twelve months ($k \in \{3,6,12\}$). For the one-sided case ($MA_{r}$), we compute
$$MA_{r,t} = \frac{1}{k} \sum^{k-1}_{i=0} y_{t-i}$$
and the two-sided case ($MA_{c}$) is obtained by
$$MA_{c,t} = \frac{1}{2k+1} \sum^{k}_{i=-k} y_{t+i}.$$

As an adaptive moving average we use the Savitzky-Golay filter \citep{savitzky1964smoothing}. The smoothing is performed via a local polynomial regression, which ensures that peaks and trends are preserved while noise is reduced. Formally, the resulting measure ($SG$) is given by
$$SG_{t} = \sum_{i=\frac{1-k}{2}}^{\frac{k-1}{2}} c_i y_{t+i}$$ 
where $c_i$ are filter coefficients, which are obtained by fitting a polynomial to the data within a moving window using least squares optimization. For our applications, we choose a window size of $k=11$ and the $3^{rd}$ order polynomial.

We include an exponential moving average (EMA), which gives more weight to recent observations while gradually decreasing weights for past ones. This makes it more responsive to recent changes compared to a simple MA. Note that the EMA is usually applied to the one-sided case and is given by:
$$EMA_t = \alpha y_t + (1-\alpha) EMA_{t-1},$$
where $\alpha$ is the smoothing factor, which satisfies $0 < \alpha < 1$, and determines the weight given to the most recent observation. It is typically calculated as $\alpha = \frac{2}{(k+1)}$, with $k$ determining the window size. In our applications, we choose $k=12$.

\vskip 0.15cm
{\sc \noindent \textbf{$l_1$ Trend Filtering}.} $l_1$ trend filtering is a variation of the HP filter, which estimates trends by minimizing a penalized least squares problem \citep{kim2009ell_1,tibshirani2011solution}. It allows to capture piecewise linear trends as well as sharp changes by imposing an $l_1$ penalty on the trend's discrete derivatives. To obtain the $l_1$ trend filtering estimates, we solve
$$\hat{LT} = \argmin_{\boldsymbol{LT}}  \frac{1}{2} || (y - LT) ||^2_2 + l ||D^{d+1} LT||_1.$$
with $y = (y_1,\dots,y_t)$. We choose the tuning parameter $l$ based on cross-validation and scale the result to vary the degree of smoothness. In particular, we consider the values $0.1l$, $l$, and 4$l$. The polynomial order is set to $d=3$.

\vskip 0.15cm
{\sc \noindent \textbf{Boosted HP Filter}.} As the name suggests, the boosted HP filter combines the well-known Hodrick-Prescott (HP) filter with machine learning techniques, i.e., $l_2$-boosting \citep{phillips2021boosting}. In this setup, the HP filter is repeatedly applied to the residuals from the previous iteration, summarized in:
$$bHP^{(m)} = B_m y, \quad B_m = I - (I - S)^m, \quad S = (I + l DD')^{-1}.$$
$D'$ captures the second differencing vectors $d=(1,-2,1)'$ on the leading tridiagonals and $y = (y_1,\dots,y_t)$. We choose $l \in \{0.1,1,100\}$ to illustrate different degrees of smoothing and a total number of $M=100$ iterations with early stopping based on the BIC criterion.

\vskip 0.15cm
{\sc \noindent \textbf{Unobserved Components Model with Stochastic Volatility (UC-SV)}.} The UC-SV, as proposed by \cite{stock2007has}, is a state-space model used to decompose inflation into a permanent stochastic trend component and a serially uncorrelated transitory component. The model allows the error variance of the shocks to evolve by introducing stochastic volatilities. It can be seen as an integrated moving average, with MA coefficients adapting inversely to the ratio of the variances between the permanent and transitory disturbances.

More formally, the UC-SV is given by:
\begin{align*}
y_t &= \gamma_t + \eta_t, \quad  \eta_t \sim \mathcal{N}(0,\sigma^2_{\eta,t}) \\
\gamma_t &= \gamma_t + \varepsilon_t, \quad  \varepsilon_t \sim \mathcal{N}(0,\sigma^2_{\varepsilon,t}) \\
\text{ln }\sigma^2_{\eta,t} &= \text{ln }\sigma^2_{\eta,t-1} + \nu_{\eta,t}, \quad \nu_{\eta,t} \sim \mathcal{N}(0,\varsigma^2) \\
\text{ln }\sigma^2_{\varepsilon,t} &= \text{ln }\sigma^2_{\varepsilon,t-1} + \nu_{\varepsilon,t}, \quad \nu_{\varepsilon,t} \sim \mathcal{N}(0,\varsigma^2).
\end{align*}
For the one-sided UC-SV estimates for inflation, we run the model recursively, starting in 2000m1. We estimate the model using Kalman filtering techniques and the algorithm proposed in \cite{kastner2014} for stochastic volatility.

\subsection{Details on the Simulation Study} \label{app:sim}

For the illustration of the RF's adaptability in Section \ref{sec:sim}, we simulate data from three different DPGs. They reflect the following scenarios: (i) \textit{gradual change}, (ii) \textit{abrupt change}, and (iii) \textit{combined} scenario. Formally, we define our response variable $y_{j,t}$ as

\begin{equation}
y_{j,t} = a_{j,t} + \epsilon_t, \quad \epsilon_t \sim \mathcal{N}(0, \sigma^2)
\end{equation}
for $t = 1, \dots, T$ and $j \in \{gc,ac,cs\}$, which stands for the three scenarios, i.e., gradual change, abrupt change, combined scenario, respectively. We choose $\sigma = 0.5$ and $T = 300$. $a_{j,t}$ is set in the following way:

\begin{align*}
a_{gc,t} &= \frac{2t - T}{T} \text{ for } t = (1, \dots, T), \\
a_{ac,t} &=
\begin{cases}
-1 & \text{for } t = (1, \dots, \frac{T}{2}),\\
1 & \text{for } t = (\frac{T}{2}+1, \dots, T),
\end{cases} \\
a_{cs,t} &=
\begin{cases}
\frac{2t - T/2}{T/2} & \text{for } t = (1, \dots, \frac{T}{2}),\\
-1 & \text{for } t = (\frac{T}{2}+1, \dots, T).
\end{cases}
\end{align*}

\begin{figure}[h!]
  \caption{\normalsize{Comparing MA and Filtering Techniques on Simulated Data}} \label{fig:Sim_comp}
    \centering
    \vspace*{-1em}
    \includegraphics[width=\textwidth, trim = 0mm 0mm 0mm 0mm, clip]{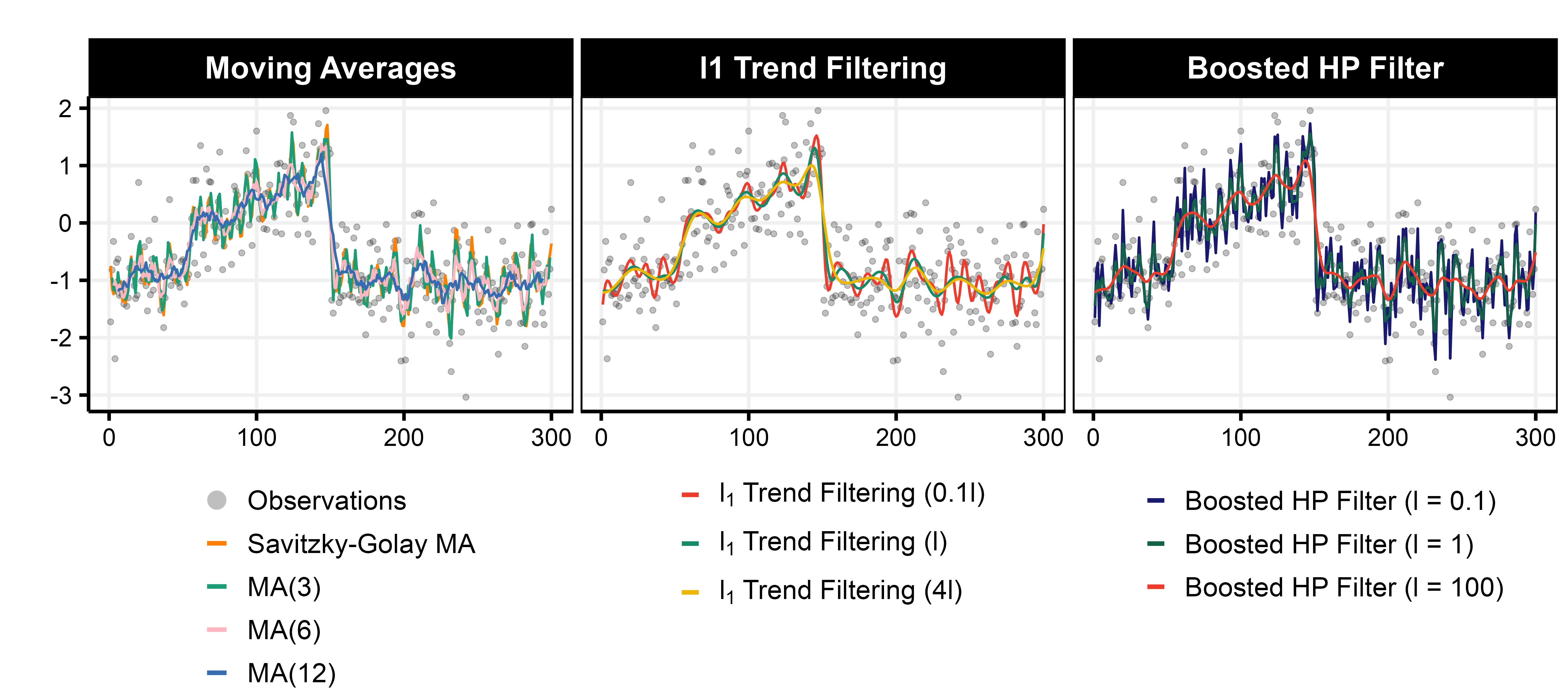}
    \begin{threeparttable}
    \centering
    \begin{minipage}{\textwidth}
      \begin{tablenotes}[para,flushleft]
    \setlength{\lineskip}{0.2ex}
    \notsotiny 
  {\textit{Notes}: The figure compares alternative filtering techniques for the \textit{combined} scenario of the simulation study. The \textbf{first panel} shows the solutions of standard moving averages and the Savitzky-Golay filter. The \textbf{second panel} compares $l_1$ trend filtering with different values for the tuning parameter $l$, which controls the smoothness of the series. In the \textbf{third panel}, we apply a boosted Hodrick-Prescott (HP) filter with increasing values for the tuning paramter $l$. Data is drawn from a normal distribution with mean 0 and standard deviation 0.5 and features a trend from -1 to 1 followed by a sudden shift back to a constant of -1. For details on the formal definition, see \ref{app:sim}. }
    \end{tablenotes}
  \end{minipage}
  \end{threeparttable}
\end{figure}

\clearpage
\subsection{Additional Results} \label{app:panels}


\begin{figure}[h]
  \caption{\normalsize{Two-sided Measures for US CPI Inflation}} \label{fig:US_CPI_2sided}
  
   \begin{minipage}[t]{\textwidth}
      \centering
      \includegraphics[width=\textwidth, trim = -9mm 0mm -19mm 0mm, clip]{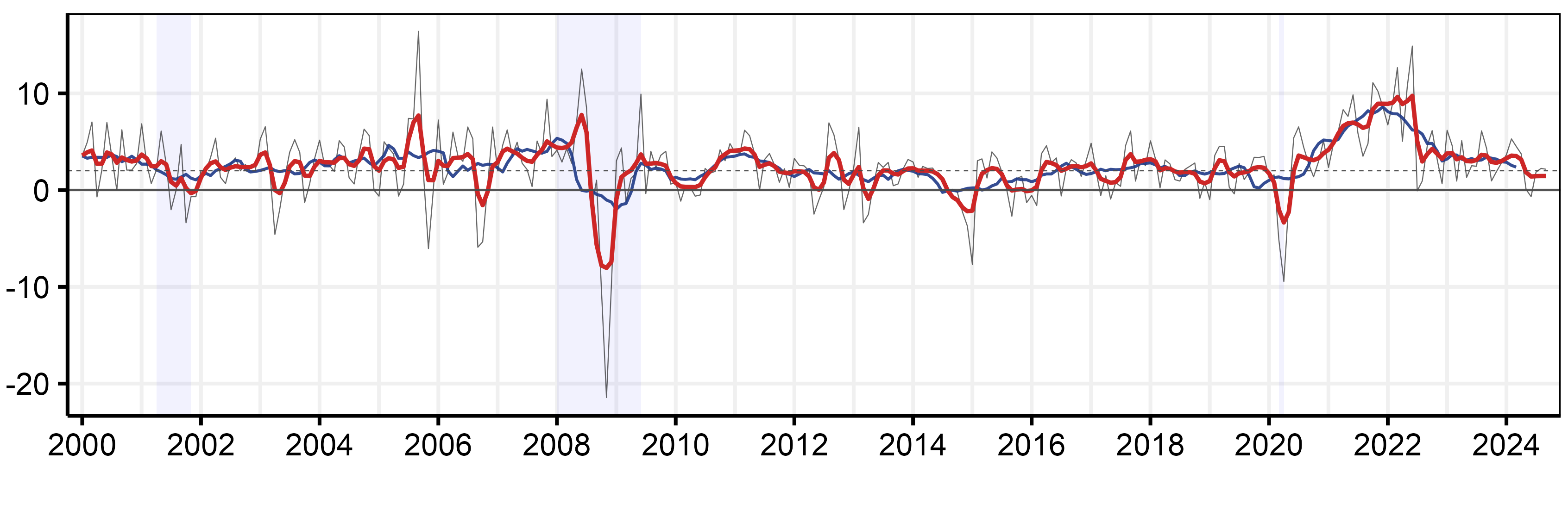}
    \end{minipage}%

\vspace*{-1em}
     \begin{minipage}[t]{0.5\textwidth}
      \centering
      \includegraphics[width=\textwidth, trim = -13mm 0mm -7mm 0mm, clip]{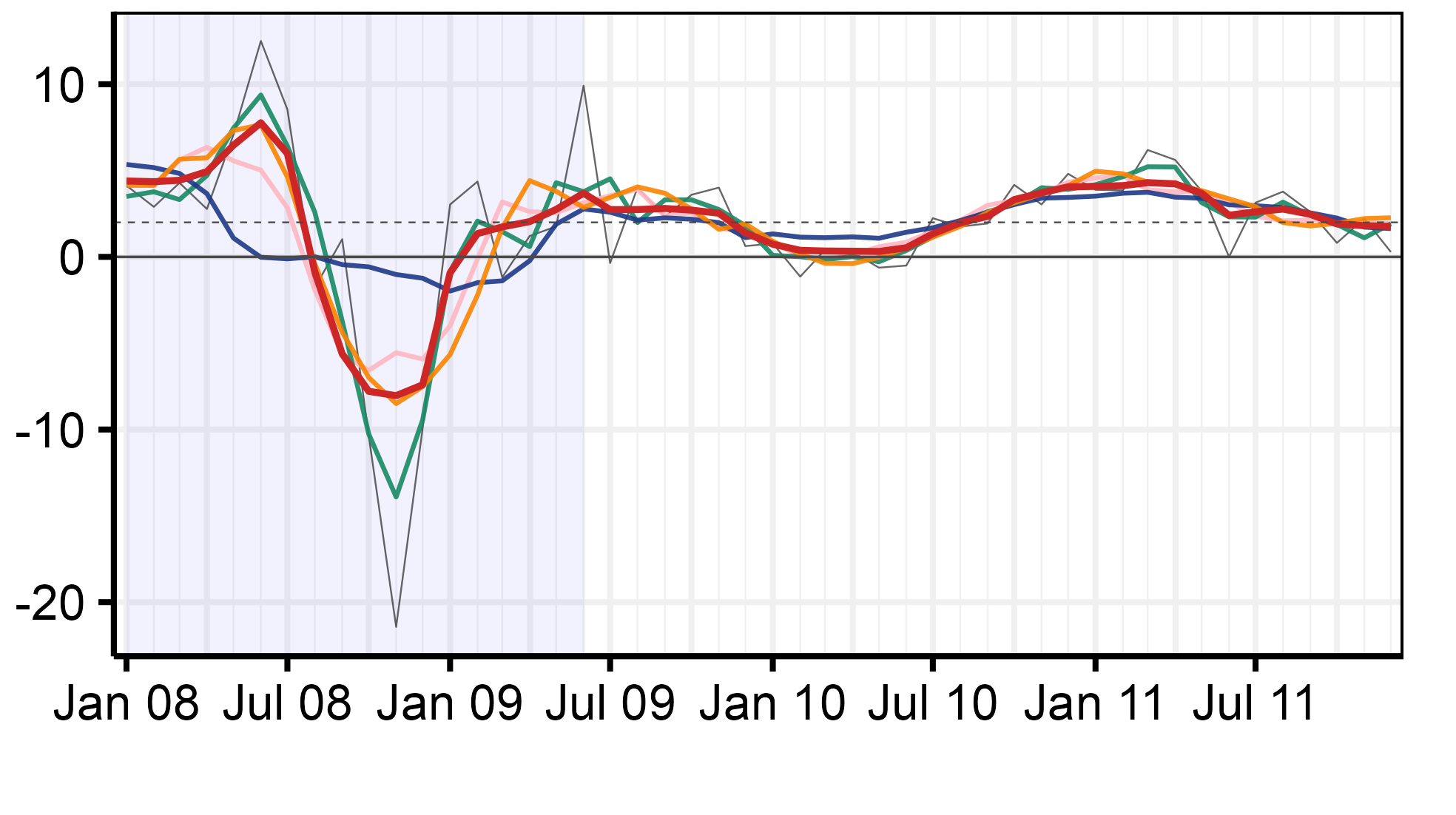}
    \end{minipage}%
    \begin{minipage}[t]{0.5\textwidth}
      \centering
      \includegraphics[width=\textwidth, trim = -3mm 0mm -15mm 0mm, clip]{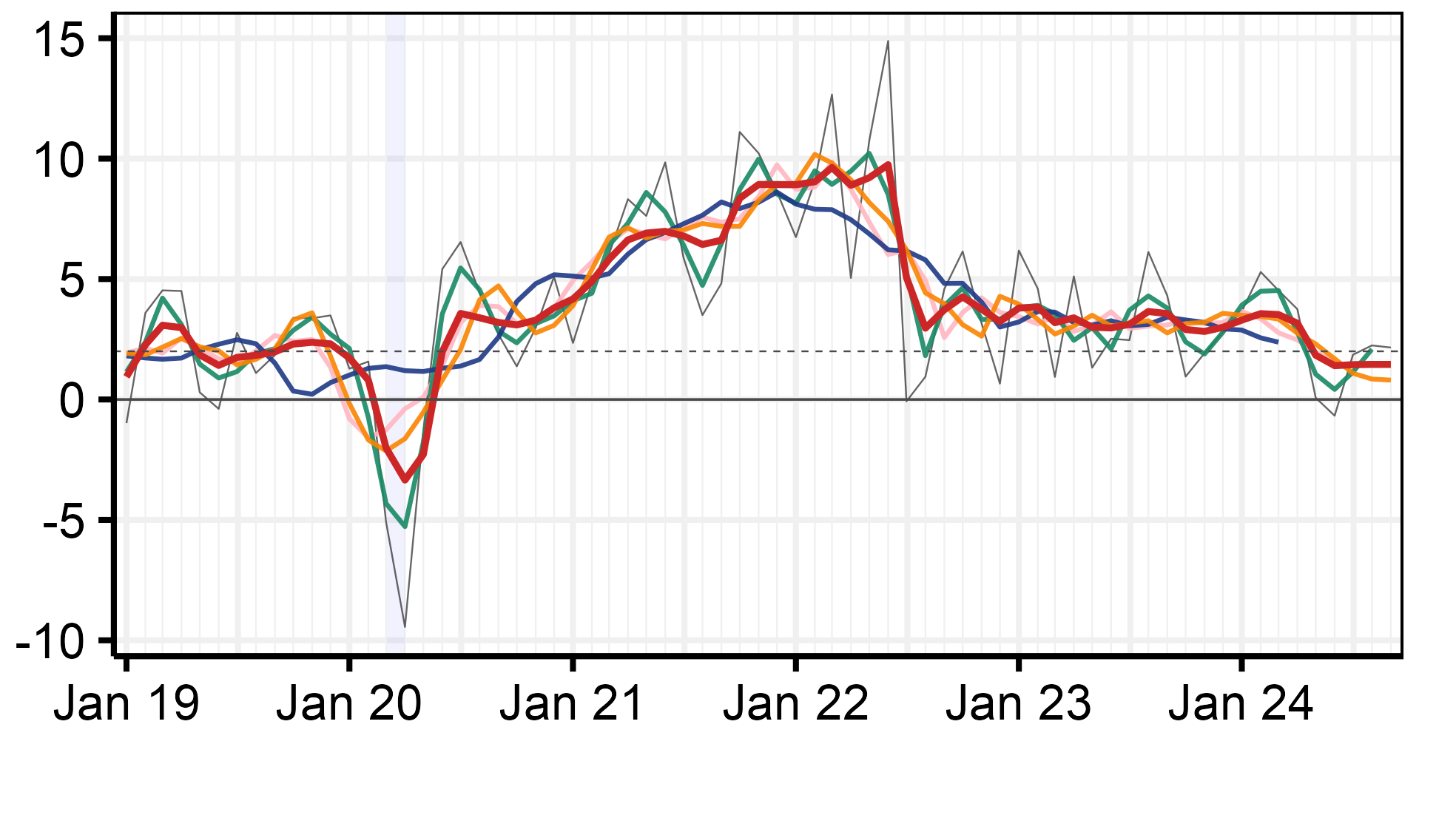}
    \end{minipage}

    \vspace*{-0.8em}
    \begin{minipage}[t]{\textwidth}
      \centering
      \includegraphics[width=0.8\textwidth, trim = 0mm 0mm 0mm 0mm, clip]{AMA_RF_TS_CPIAUCSL__T__legend.png}
    \end{minipage}

    \begin{minipage}[t]{0.5\textwidth}
      \centering
      \includegraphics[width=\textwidth, trim = 8mm 0mm 8mm 0mm, clip]{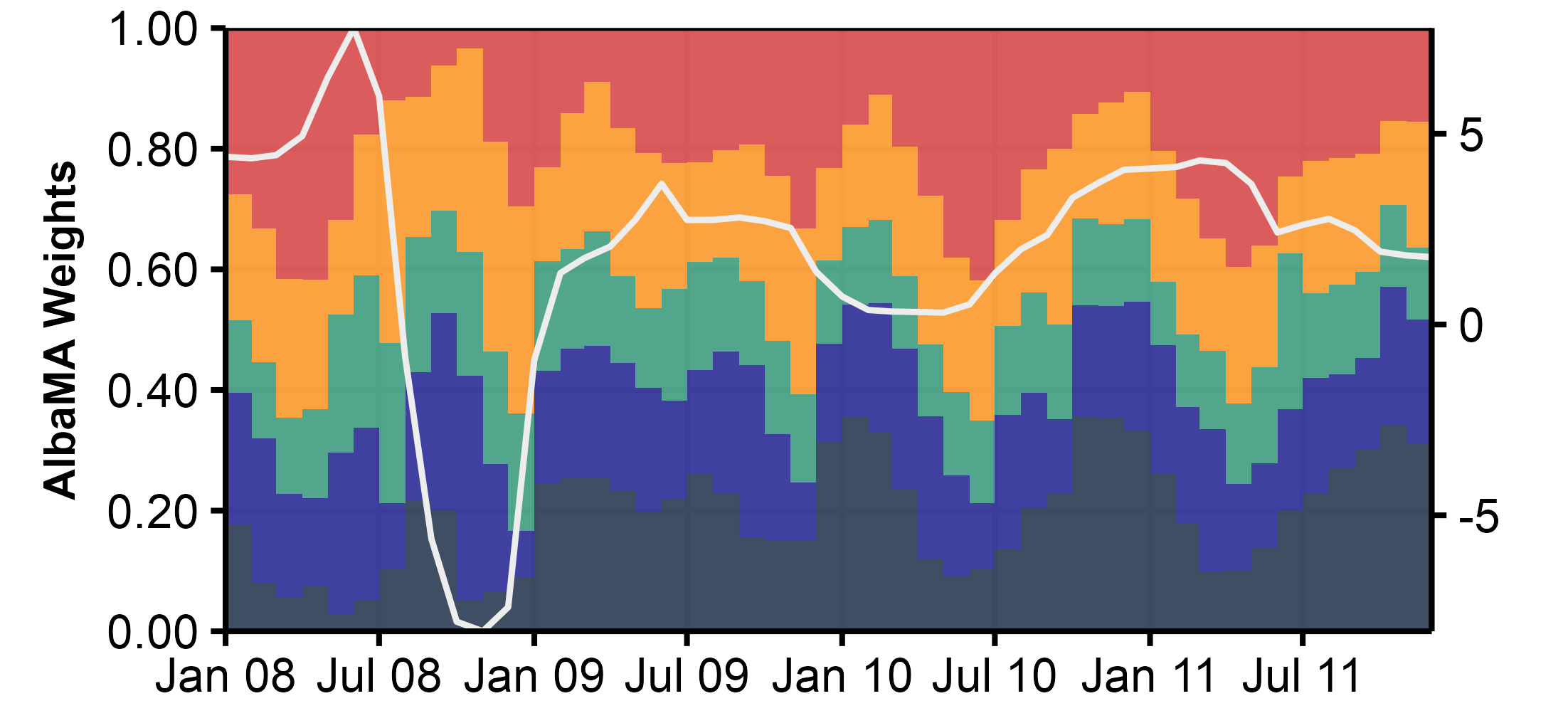}
    \end{minipage}%
    \begin{minipage}[t]{0.5\textwidth}
      \centering
      \includegraphics[width=\textwidth, trim = 10mm 0mm 8mm 0mm, clip]{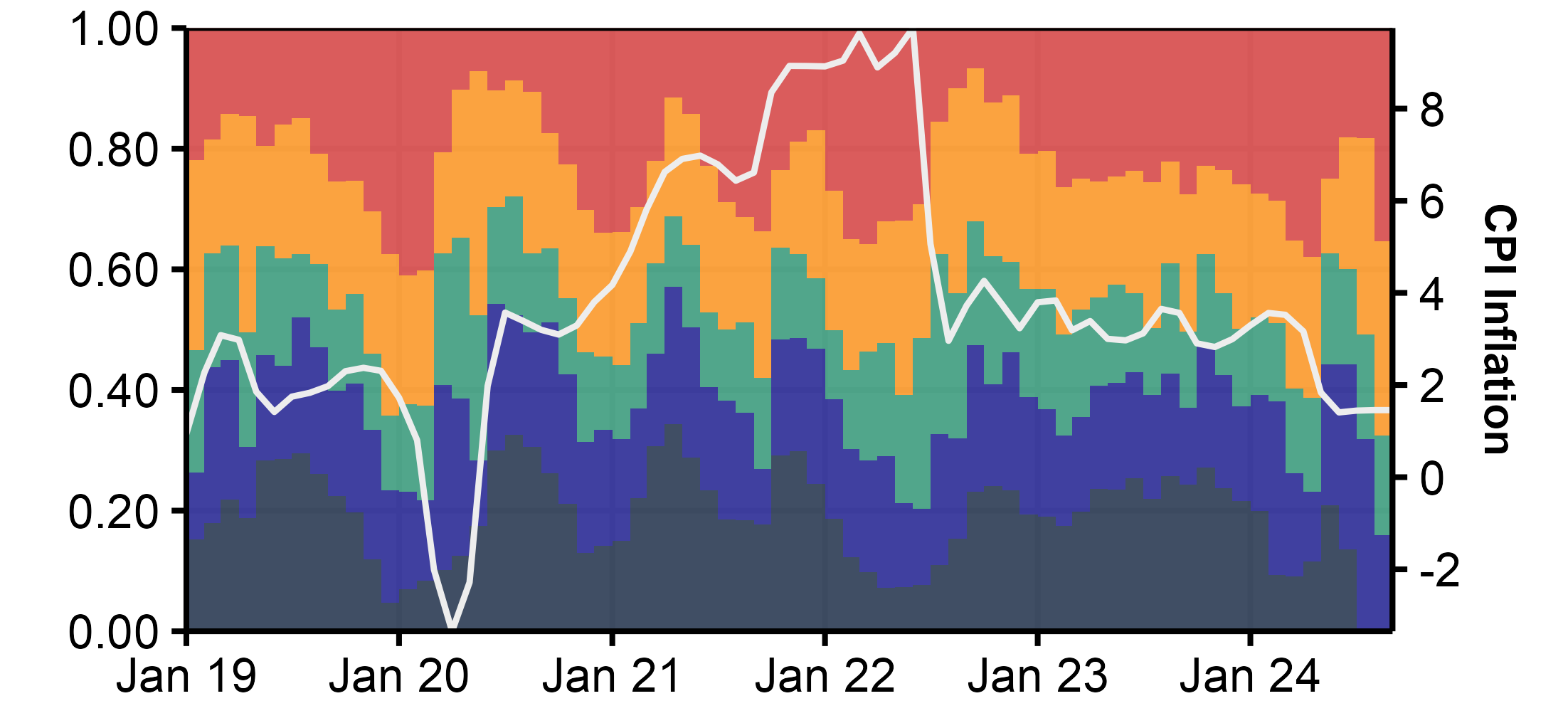}
    \end{minipage}

    \vspace*{-.5em}
    \begin{minipage}[t]{\textwidth}
      \centering
      \hspace*{1.0em} \includegraphics[width=0.95\textwidth, trim = 0mm 0mm 0mm 0mm, clip]{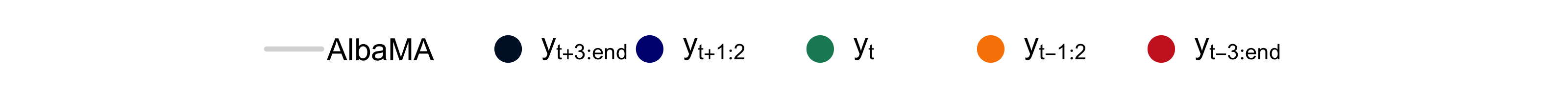}
    \end{minipage}

    \vspace*{-0.3em}
    \begin{minipage}[t]{0.5\textwidth}
      \centering
      (a) Great Recession
    \end{minipage}
    \begin{minipage}[t]{0.5\textwidth}
      \centering
      (b) Post-Covid
    \end{minipage}

    \begin{threeparttable}
    \centering
    \begin{minipage}{\textwidth}
      \begin{tablenotes}[para,flushleft]
    \setlength{\lineskip}{0.2ex}
    \notsotiny 
  {\textit{Notes}: The \textbf{upper panel} shows AlbaMA and the MA(12). The \textbf{middle panels} compare AlbaMA to standard moving averages and the Savitzky-Golay filter for (a) the Great Recession and (b) the post-Covid surge. The \textbf{lower panels} present the weights the RF assigns to past observations. All measures are two-sided.}
    \end{tablenotes}
  \end{minipage}
  \end{threeparttable}
\end{figure}


\begin{figure}[h]
  \caption{\normalsize{Two-sided Measures for US CPI Core Inflation}} \label{fig:US_CPIcore_2sided}
  
   \begin{minipage}[t]{\textwidth}
      \centering
      \includegraphics[width=\textwidth, trim = -12mm 0mm -16mm 0mm, clip]{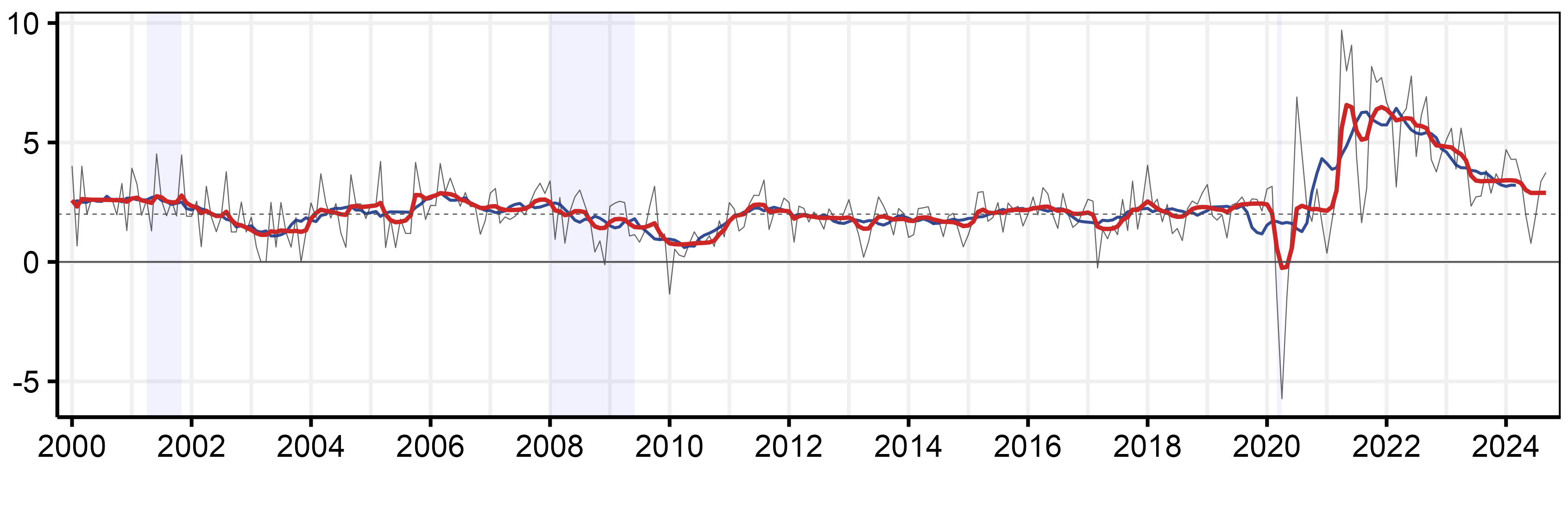}
    \end{minipage}%

\vspace*{-1em}
     \begin{minipage}[t]{0.5\textwidth}
      \centering
      \includegraphics[width=\textwidth, trim = -16mm 0mm -5mm 0mm, clip]{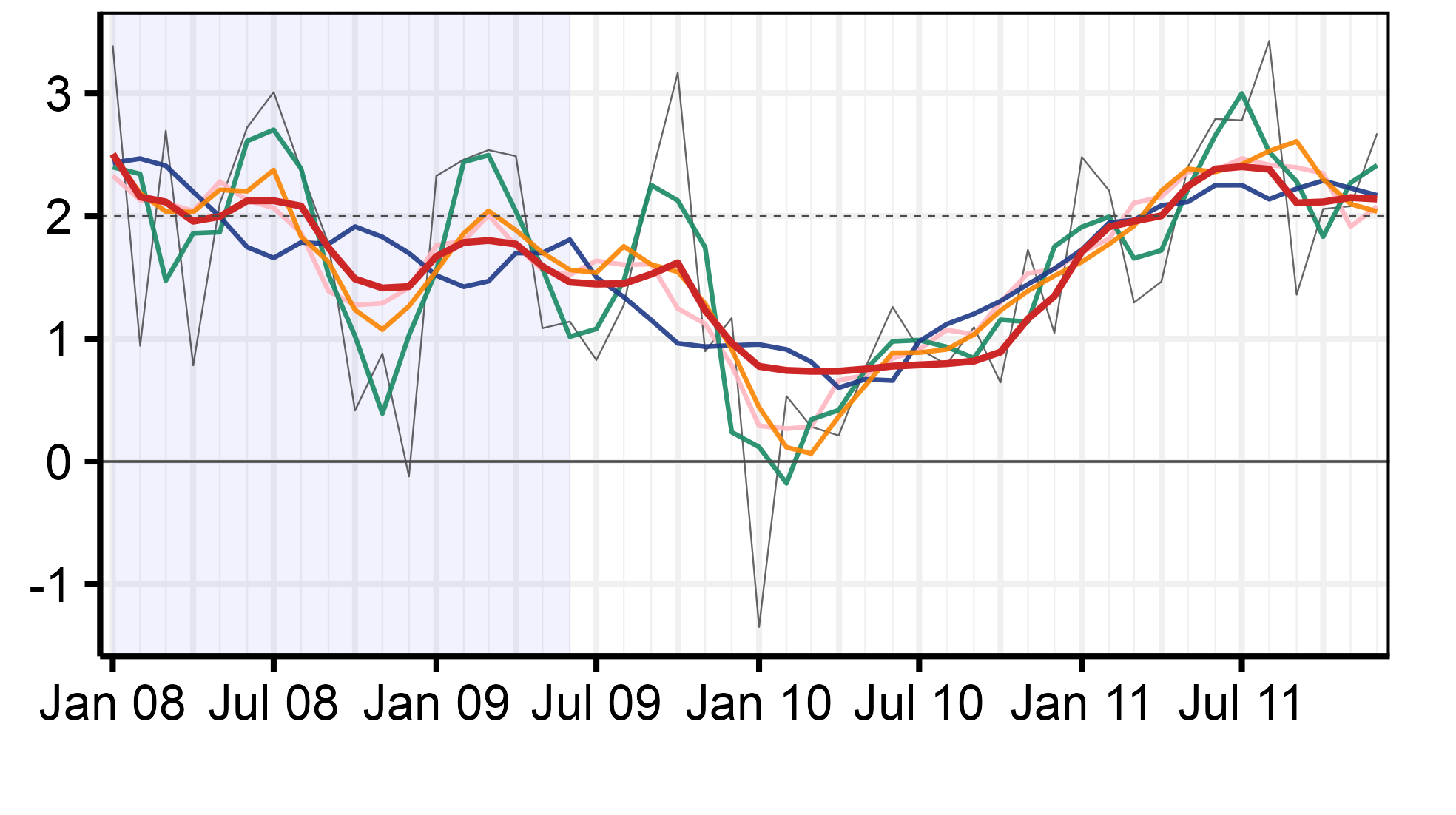}
    \end{minipage}%
    \begin{minipage}[t]{0.5\textwidth}
      \centering
      \includegraphics[width=\textwidth, trim = -5mm 0mm -11mm 0mm, clip]{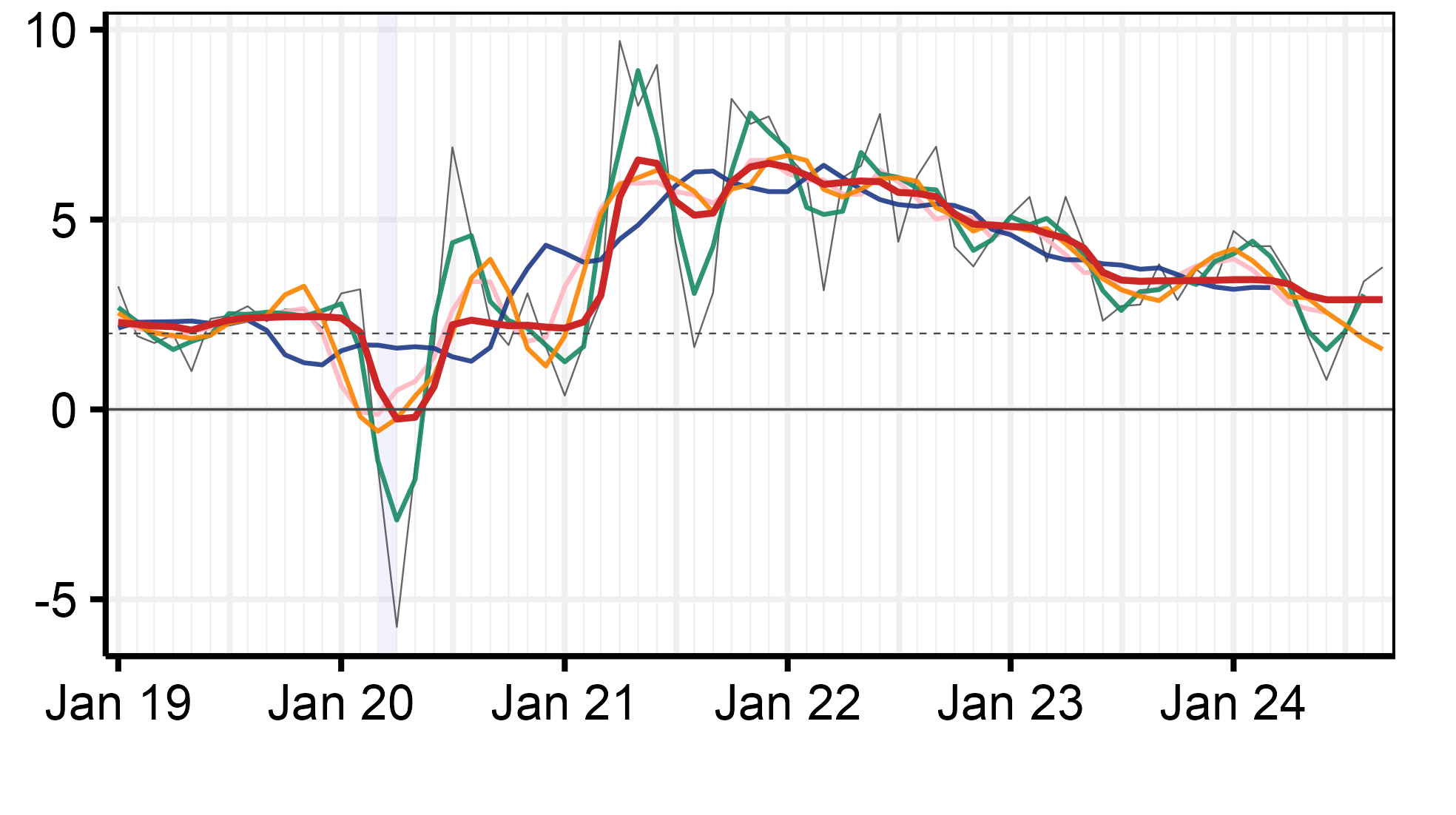}
    \end{minipage}

    \vspace*{-0.8em}
    \begin{minipage}[t]{\textwidth}
      \centering
      \includegraphics[width=0.8\textwidth, trim = 0mm 0mm 0mm 0mm, clip]{AMA_RF_TS_CPILFESL__T__legend.png}
    \end{minipage}

    \begin{minipage}[t]{0.5\textwidth}
      \centering
      \includegraphics[width=\textwidth, trim = 5mm 0mm 10mm 0mm, clip]{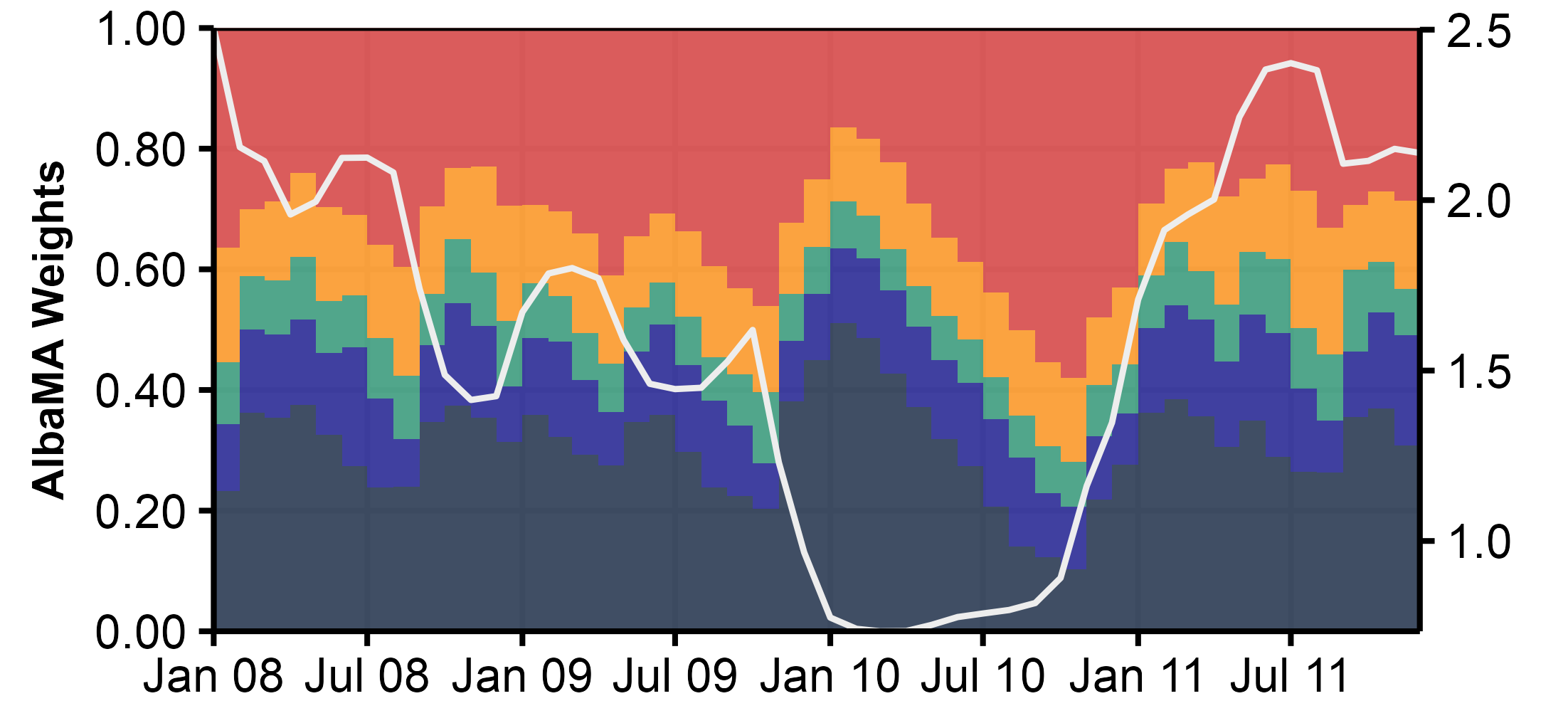}
    \end{minipage}%
    \begin{minipage}[t]{0.5\textwidth}
      \centering
      \includegraphics[width=\textwidth, trim = 12mm 0mm 8mm 0mm, clip]{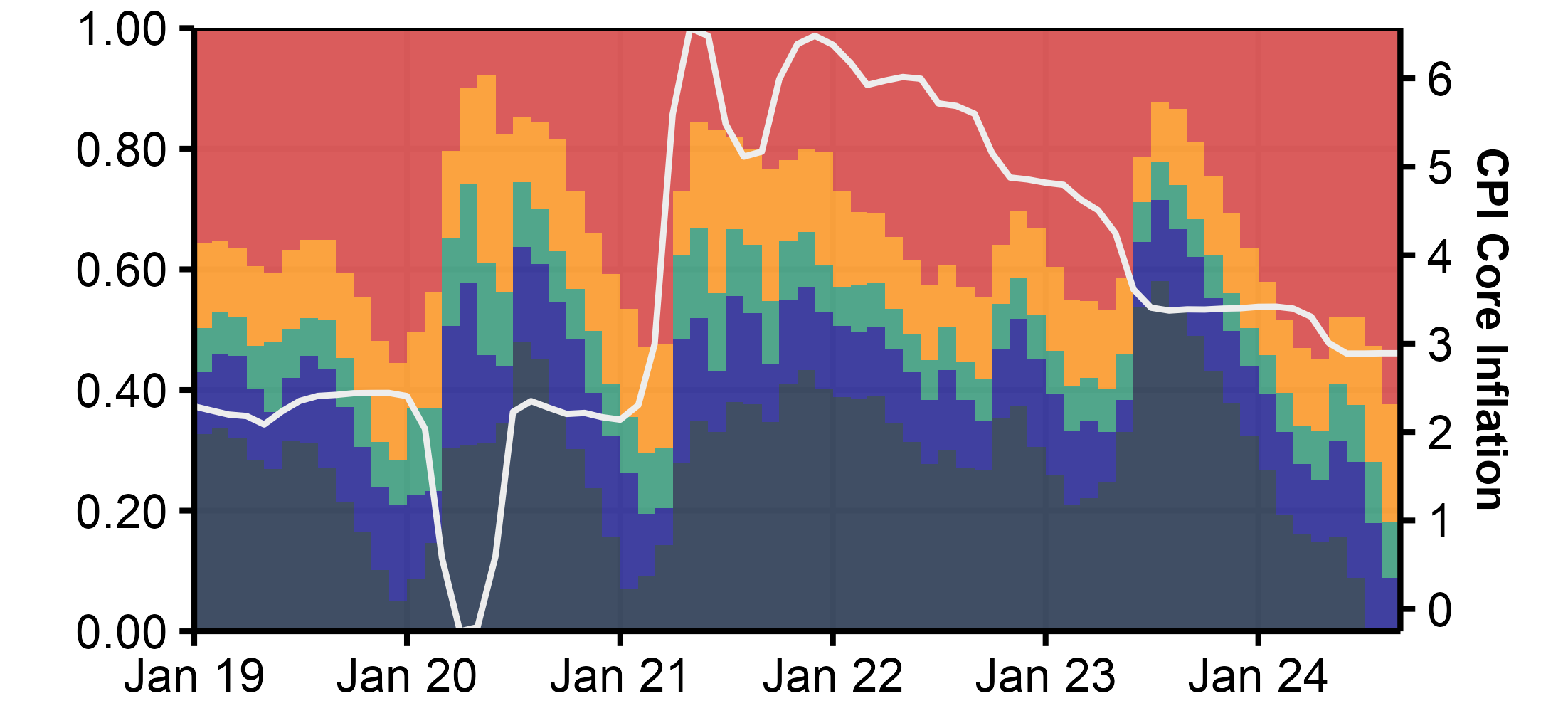}
    \end{minipage}

    \vspace*{-.5em}
    \begin{minipage}[t]{\textwidth}
      \centering
      \hspace*{1.0em} \includegraphics[width=0.95\textwidth, trim = 0mm 0mm 0mm 0mm, clip]{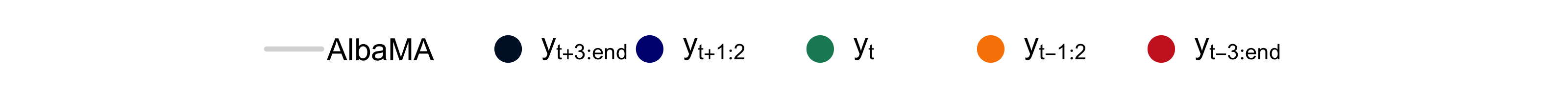}
    \end{minipage}

    \vspace*{-0.3em}
    \begin{minipage}[t]{0.5\textwidth}
      \centering
      (a) Great Recession
    \end{minipage}
    \begin{minipage}[t]{0.5\textwidth}
      \centering
      (b) Post-Covid
    \end{minipage}

    \begin{threeparttable}
    \centering
    \begin{minipage}{\textwidth}
      \begin{tablenotes}[para,flushleft]
    \setlength{\lineskip}{0.2ex}
    \notsotiny 
  {\textit{Notes}: The \textbf{upper panel} shows AlbaMA and the MA(12). The \textbf{middle panels} compare AlbaMA to standard moving averages and the Savitzky-Golay filter for (a) the Great Recession and (b) the post-Covid surge. The \textbf{lower panels} present the weights the RF assigns to past observations. All measures are two-sided.}
    \end{tablenotes}
  \end{minipage}
  \end{threeparttable}
\end{figure}


\begin{figure}[h]
  \caption{\normalsize{EA HICP Inflation}} \label{fig:EA_CPI}
  
  \begin{minipage}[t]{\textwidth}
      \centering
      \includegraphics[width=\textwidth, trim = -14mm 0mm -19mm 0mm, clip]{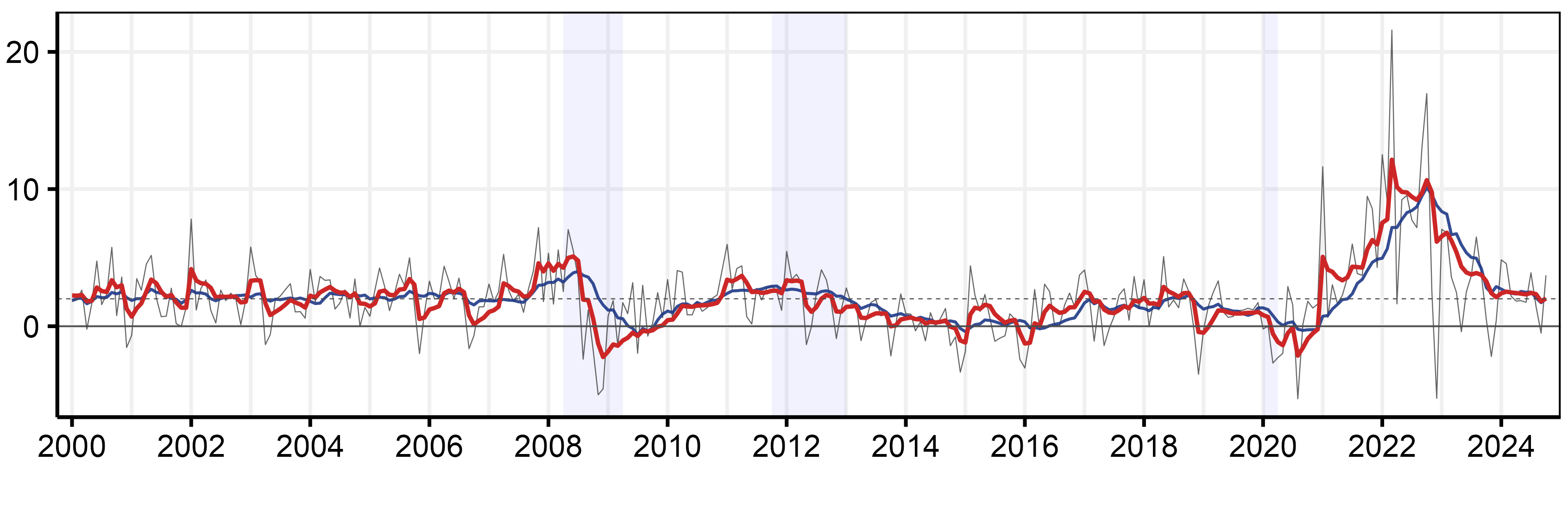}
    \end{minipage}%

\vspace*{-1em}
     \begin{minipage}[t]{0.5\textwidth}
      \centering
      \includegraphics[width=\textwidth, trim = -14mm 0mm -6mm 0mm, clip]{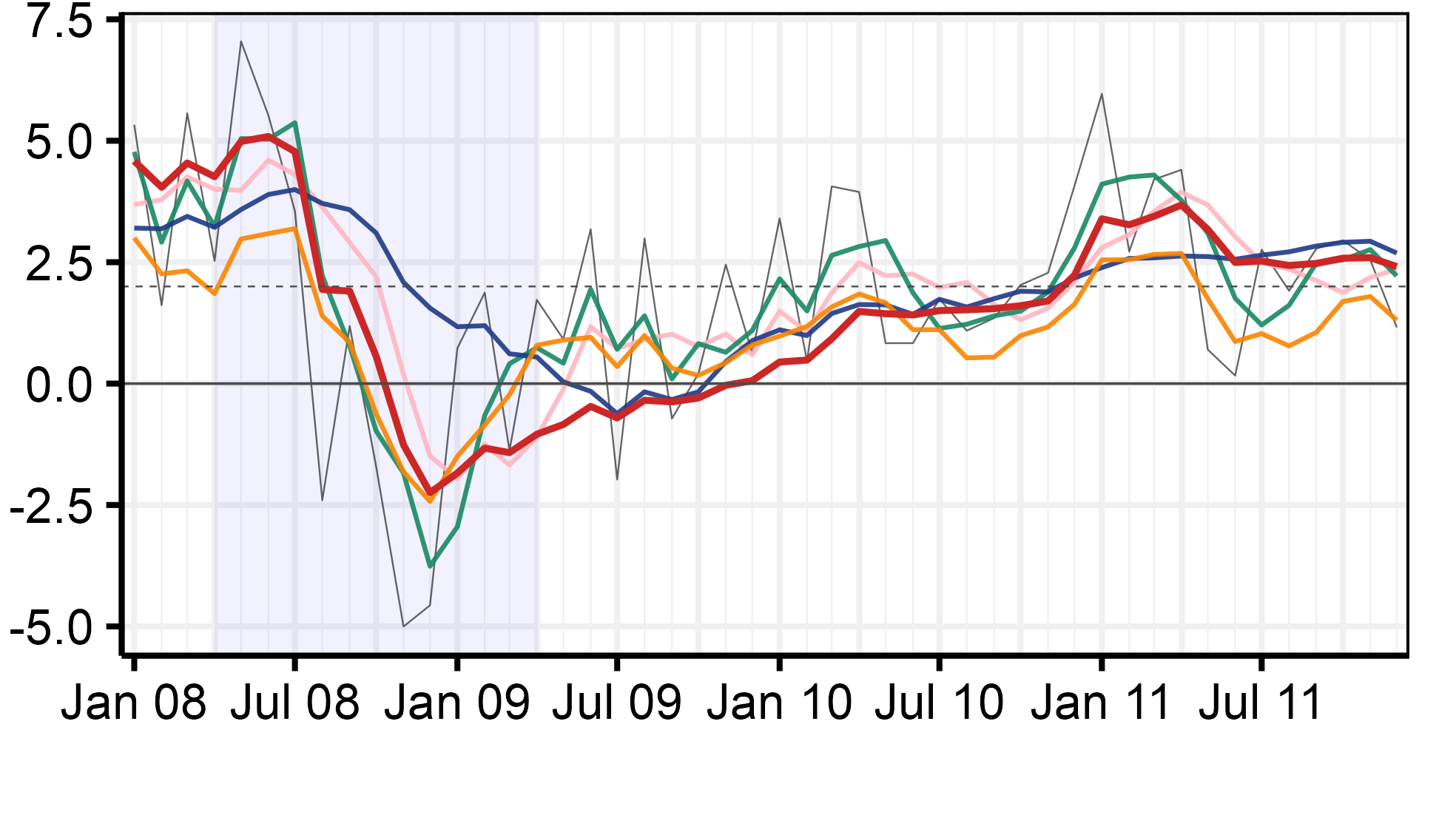}
    \end{minipage}%
    \begin{minipage}[t]{0.5\textwidth}
      \centering
      \includegraphics[width=\textwidth, trim = -4mm 0mm -14mm 0mm, clip]{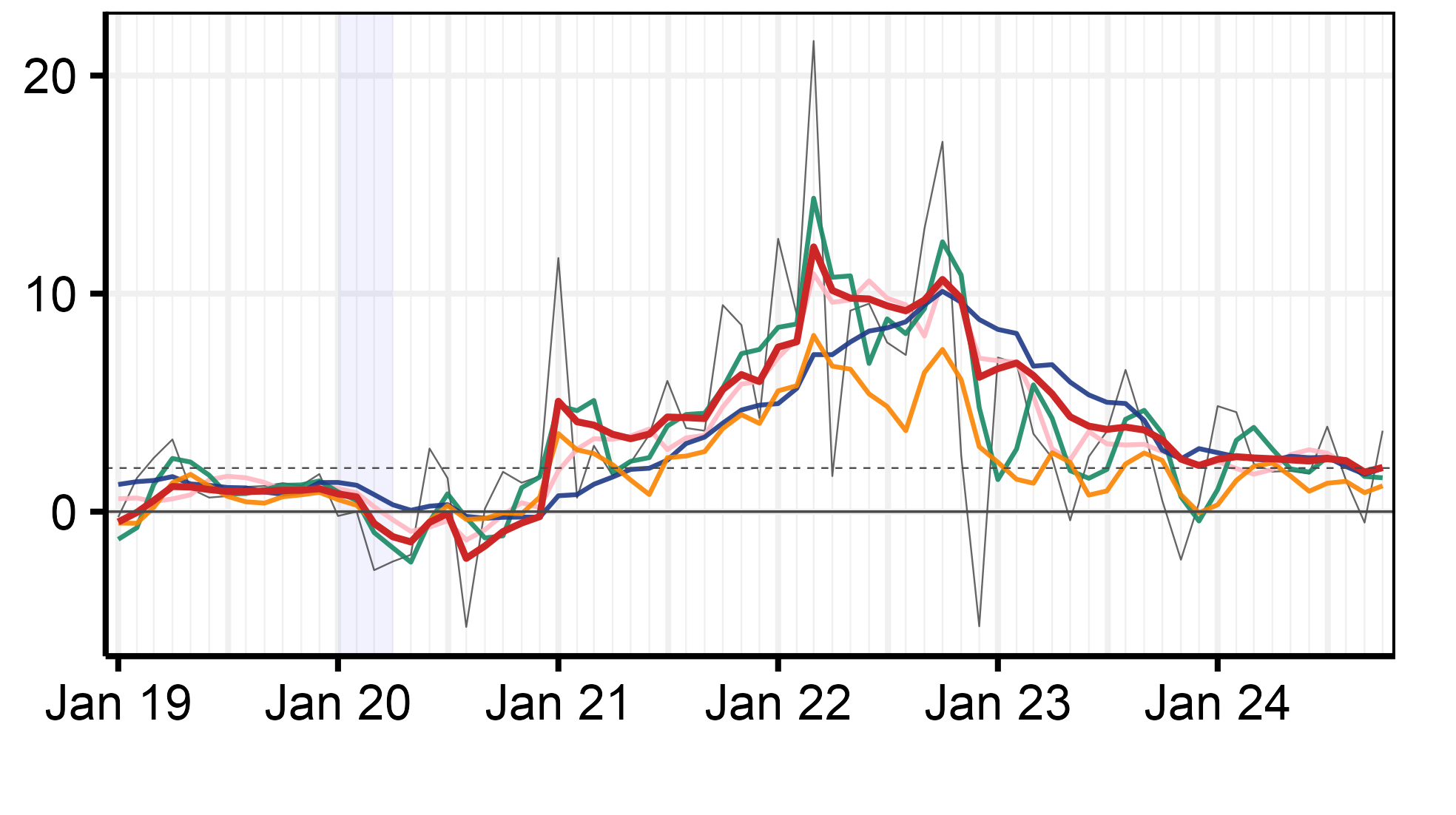}
    \end{minipage}

    \vspace*{-0.8em}
    \begin{minipage}[t]{\textwidth}
      \centering
      \includegraphics[width=0.8\textwidth, trim = 0mm 0mm 0mm 0mm, clip]{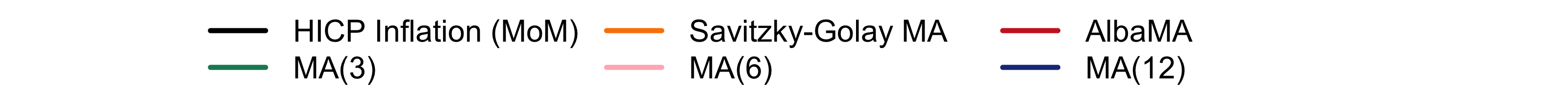}
    \end{minipage}

    \begin{minipage}[t]{0.5\textwidth}
      \centering
      \includegraphics[width=\textwidth, trim = 5mm 0mm 10mm 0mm, clip]{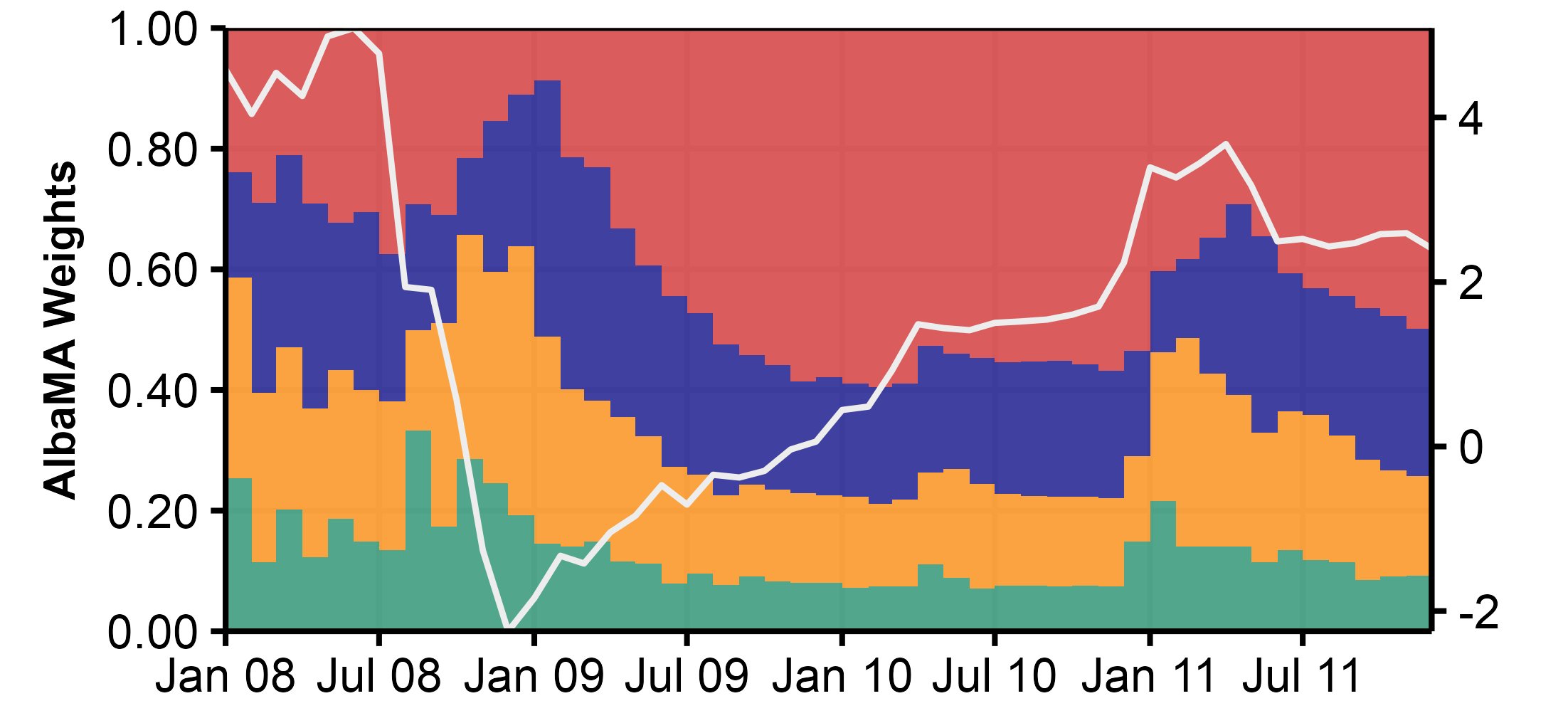}
    \end{minipage}%
    \begin{minipage}[t]{0.5\textwidth}
      \centering
      \includegraphics[width=\textwidth, trim = 10mm 0mm 8mm 0mm, clip]{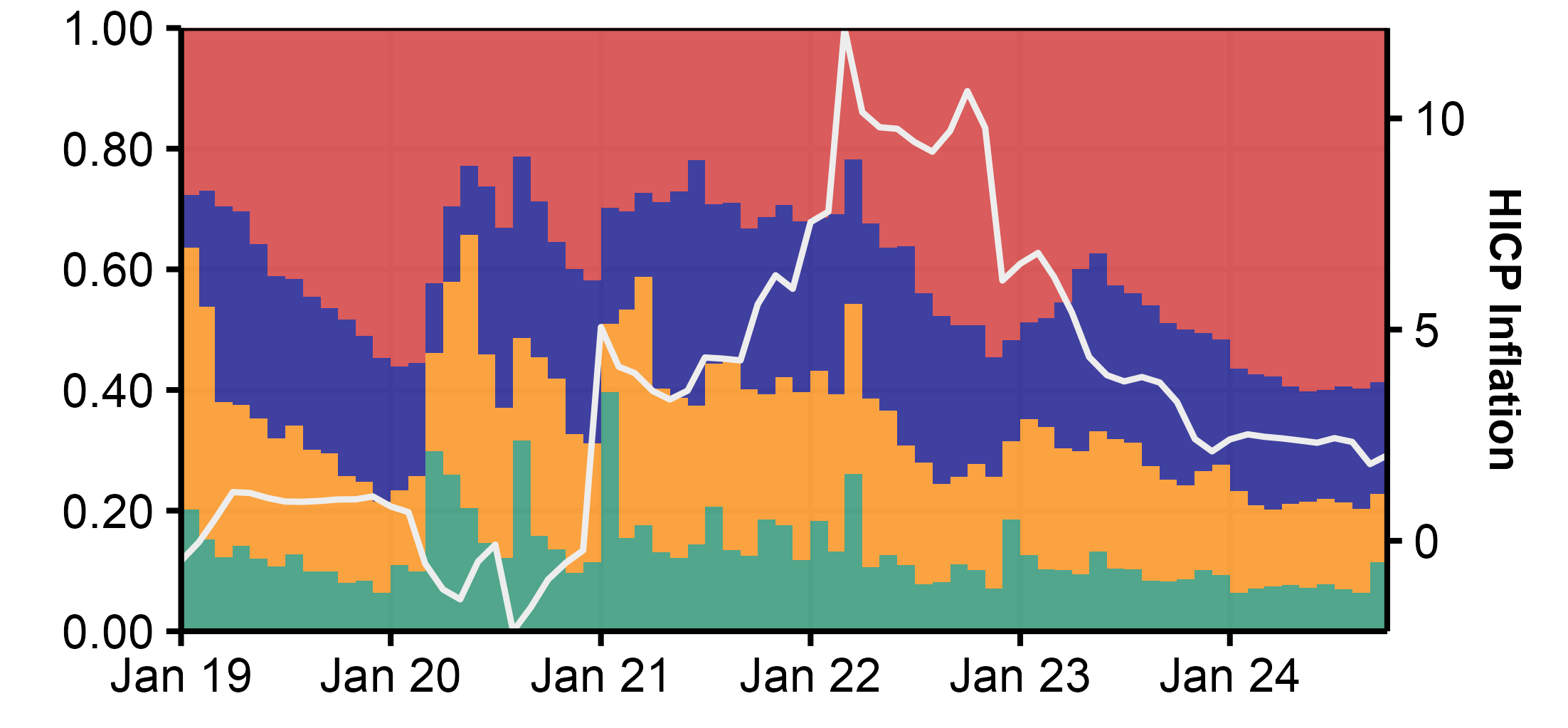}
    \end{minipage}

    \vspace*{-.5em}
    \begin{minipage}[t]{\textwidth}
      \centering
      \hspace*{1.0em} \includegraphics[width=0.95\textwidth, trim = 0mm 0mm 0mm 0mm, clip]{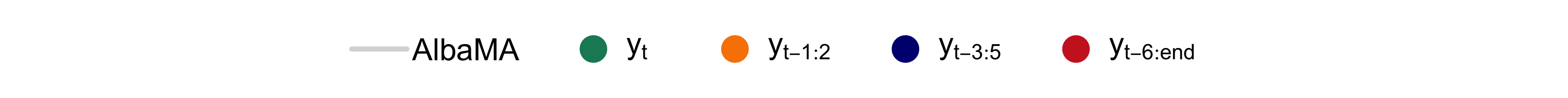}
    \end{minipage}

    \vspace*{-0.3em}
    \begin{minipage}[t]{0.5\textwidth}
      \centering
      (a) Great Recession
    \end{minipage}
    \begin{minipage}[t]{0.5\textwidth}
      \centering
      (b) Post-Covid
    \end{minipage}

    \begin{threeparttable}
    \centering
    \begin{minipage}{\textwidth}
      \begin{tablenotes}[para,flushleft]
    \setlength{\lineskip}{0.2ex}
    \notsotiny 
  {\textit{Notes}: The \textbf{upper panel} shows AlbaMA and the MA(12). The \textbf{middle panels} compare AlbaMA to standard moving averages and the Savitzky-Golay filter for (a) the Great Recession and (b) the post-Covid surge. The \textbf{lower panels} present the weights the RF assigns to past observations. All measures are one-sided.}
    \end{tablenotes}
  \end{minipage}
  \end{threeparttable}
\end{figure}


\begin{figure}[h]
  \caption{\normalsize{EA HICP Core Inflation}} \label{fig:EA_CPIcore}
  
   \begin{minipage}[t]{\textwidth}
      \centering
      \includegraphics[width=\textwidth, trim = -13mm 0mm -18mm 0mm, clip]{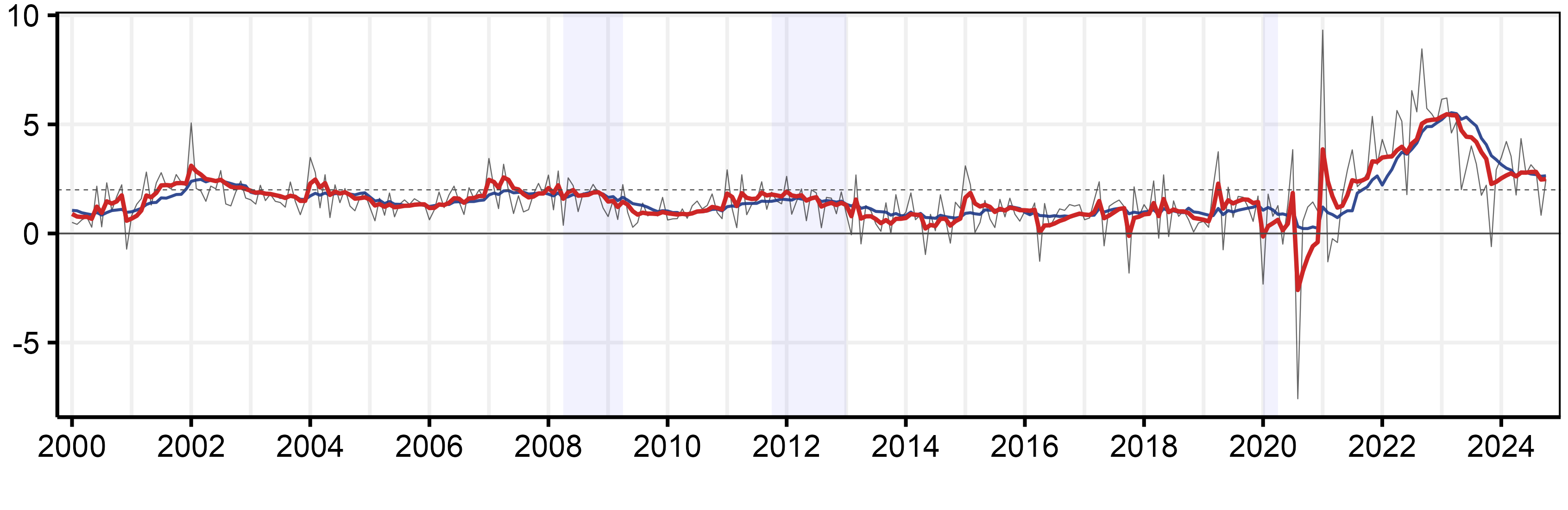}
    \end{minipage}%

\vspace*{-1em}
     \begin{minipage}[t]{0.5\textwidth}
      \centering
      \includegraphics[width=\textwidth, trim = -18mm 0mm -4mm 0mm, clip]{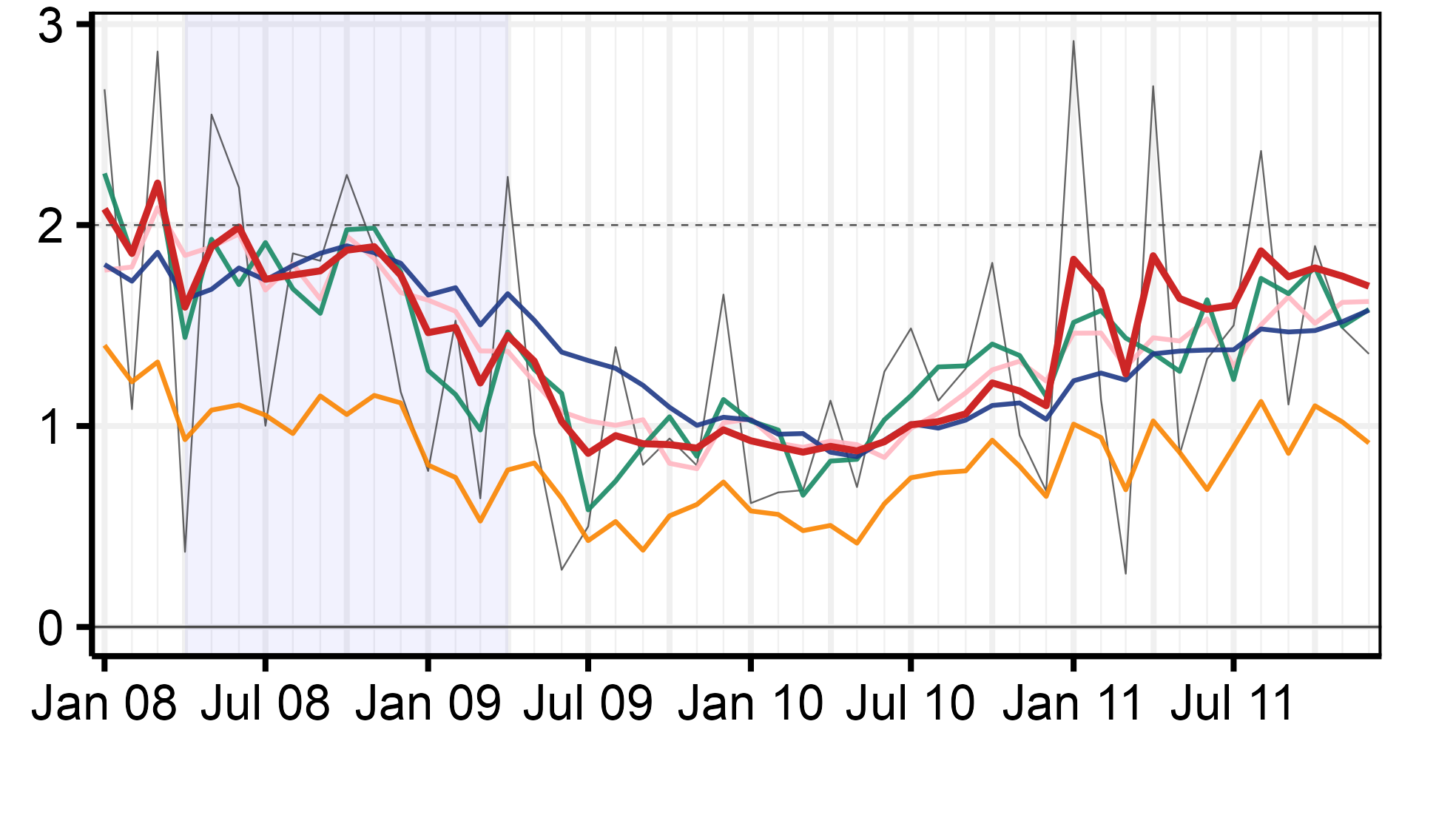}
    \end{minipage}%
    \begin{minipage}[t]{0.5\textwidth}
      \centering
      \includegraphics[width=\textwidth, trim = -5mm 0mm -13mm 0mm, clip]{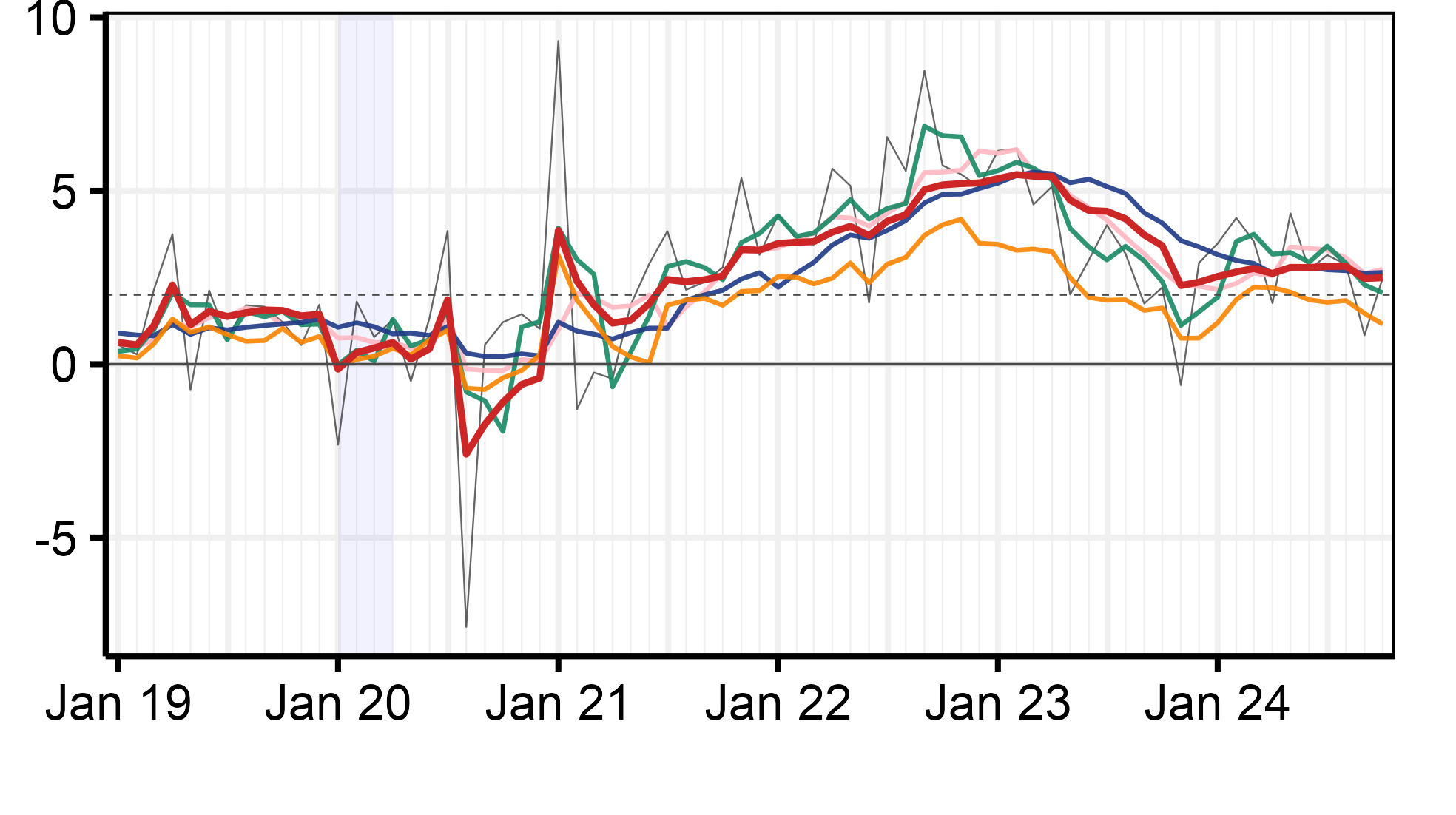}
    \end{minipage}

    \vspace*{-0.8em}
    \begin{minipage}[t]{\textwidth}
      \centering
      \includegraphics[width=0.8\textwidth, trim = 0mm 0mm 0mm 0mm, clip]{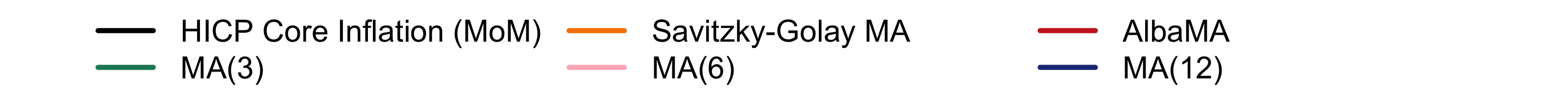}
    \end{minipage}

    \begin{minipage}[t]{0.5\textwidth}
      \centering
      \includegraphics[width=\textwidth, trim = 5mm 0mm 10mm 0mm, clip]{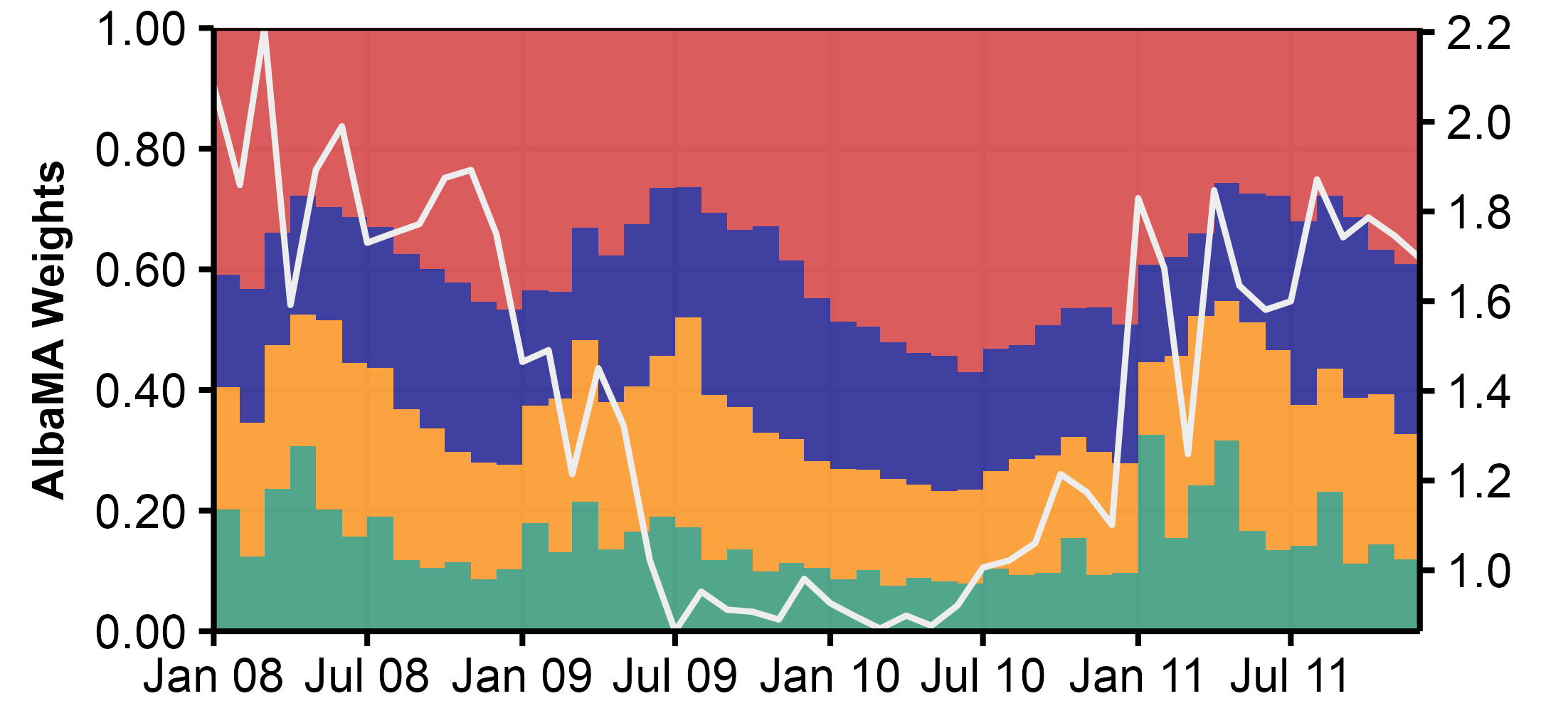}
    \end{minipage}%
    \begin{minipage}[t]{0.5\textwidth}
      \centering
      \includegraphics[width=\textwidth, trim = 10mm 0mm 8mm 0mm, clip]{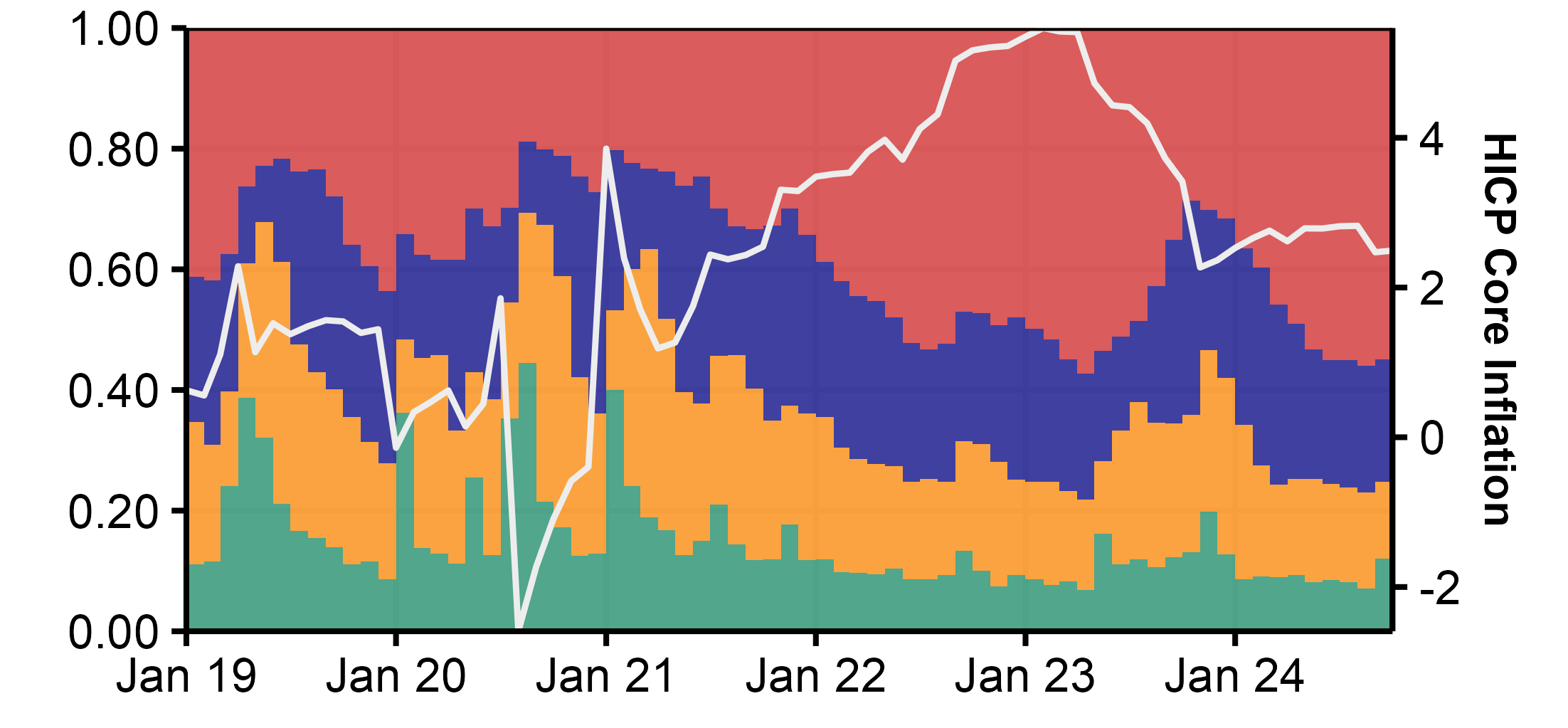}
    \end{minipage}

    \vspace*{-.5em}
    \begin{minipage}[t]{\textwidth}
      \centering
      \hspace*{1.0em} \includegraphics[width=0.95\textwidth, trim = 0mm 0mm 0mm 0mm, clip]{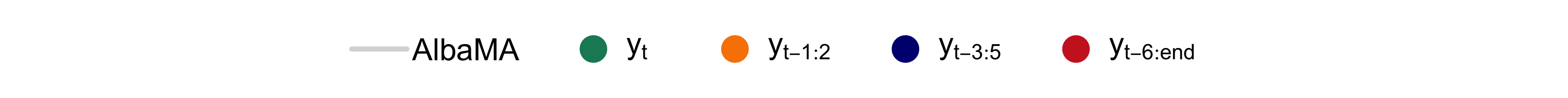}
    \end{minipage}

    \vspace*{-0.3em}
    \begin{minipage}[t]{0.5\textwidth}
      \centering
      (a) Great Recession
    \end{minipage}
    \begin{minipage}[t]{0.5\textwidth}
      \centering
      (b) Post-Covid
    \end{minipage}

    \begin{threeparttable}
    \centering
    \begin{minipage}{\textwidth}
      \begin{tablenotes}[para,flushleft]
    \setlength{\lineskip}{0.2ex}
    \notsotiny 
  {\textit{Notes}: The \textbf{upper panel} shows AlbaMA and the MA(12). The \textbf{middle panels} compare AlbaMA to standard moving averages and the Savitzky-Golay filter for (a) the Great Recession and (b) the post-Covid surge. The \textbf{lower panels} present the weights the RF assigns to past observations. All measures are one-sided.}
    \end{tablenotes}
  \end{minipage}
  \end{threeparttable}
\end{figure}


\begin{figure}[h]
  \caption{\normalsize{US PMI}} \label{fig:US_PMI}
  
     \begin{minipage}[t]{\textwidth}
      \centering
      \includegraphics[width=\textwidth, trim = -10mm 0mm -20mm 0mm, clip]{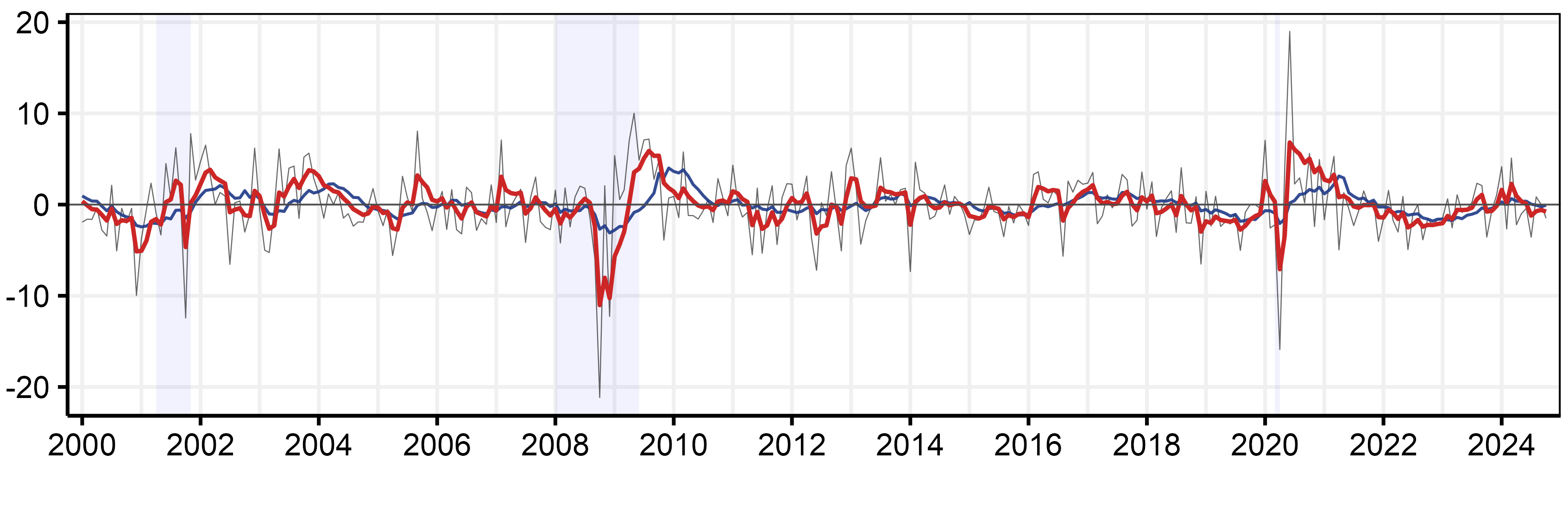}
    \end{minipage}%

\vspace*{-1em}
     \begin{minipage}[t]{0.5\textwidth}
      \centering
      \includegraphics[width=\textwidth, trim = -15mm 0mm -10mm 0mm, clip]{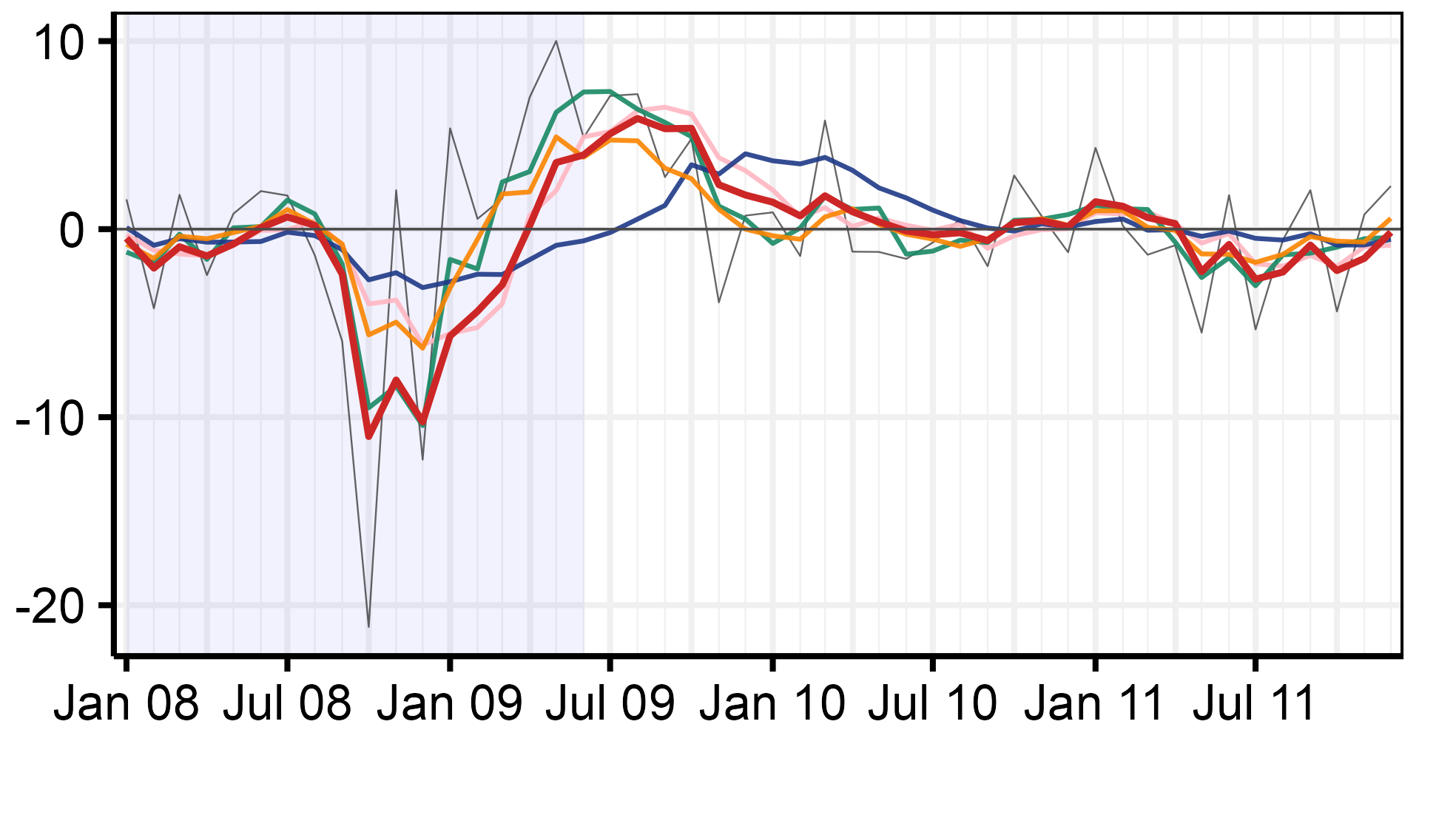}
    \end{minipage}%
    \begin{minipage}[t]{0.5\textwidth}
      \centering
      \includegraphics[width=\textwidth, trim = -5mm 0mm -18mm 0mm, clip]{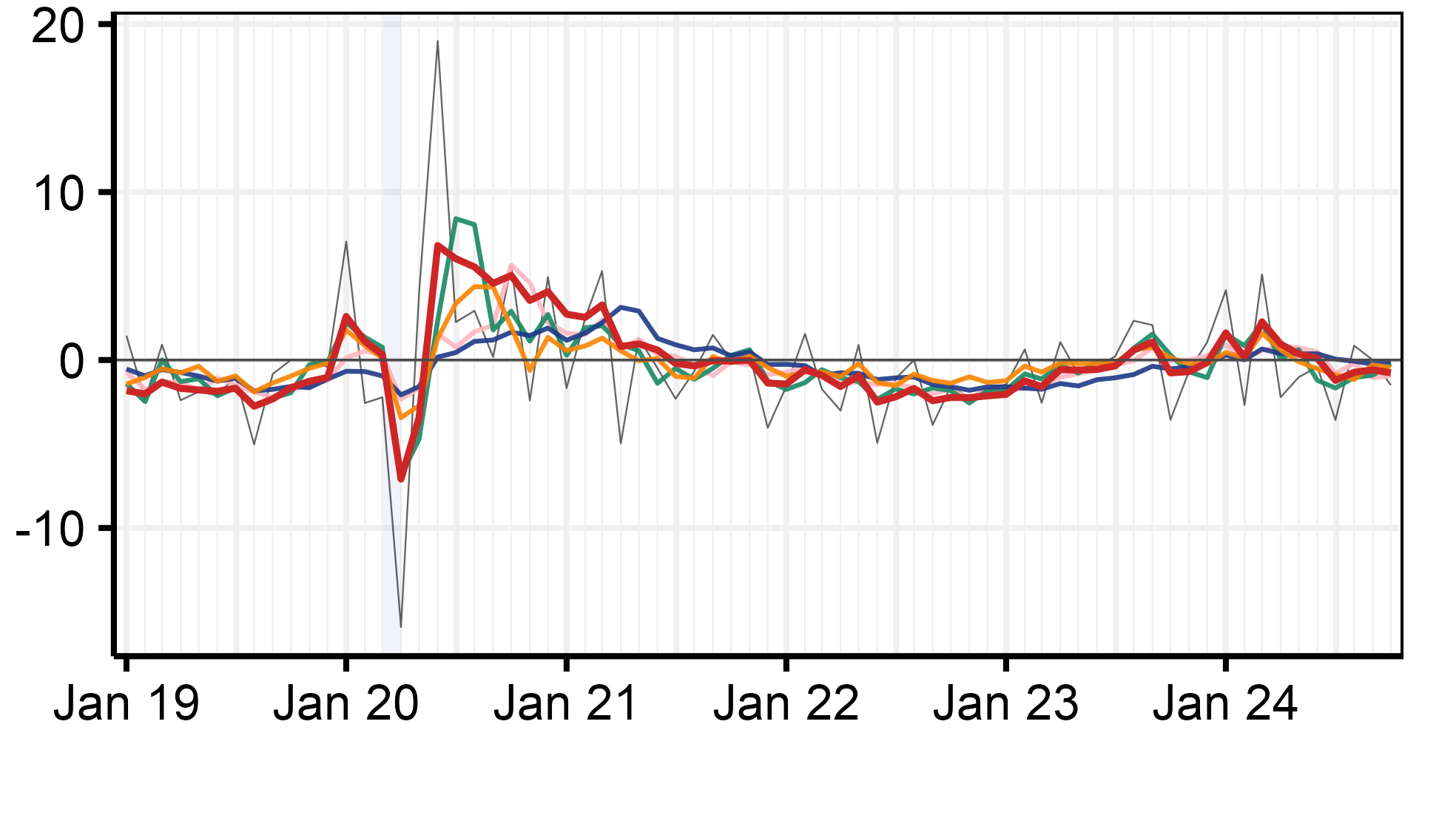}
    \end{minipage}

    \vspace*{-0.8em}
    \begin{minipage}[t]{\textwidth}
      \centering
      \includegraphics[width=0.8\textwidth, trim = 0mm 0mm 0mm 0mm, clip]{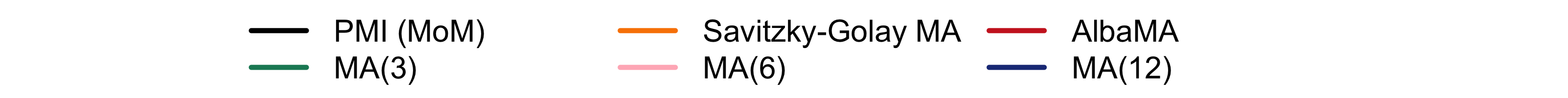}
    \end{minipage}

    \begin{minipage}[t]{0.5\textwidth}
      \centering
      \includegraphics[width=\textwidth, trim = 4mm -2mm 8mm 0mm, clip]{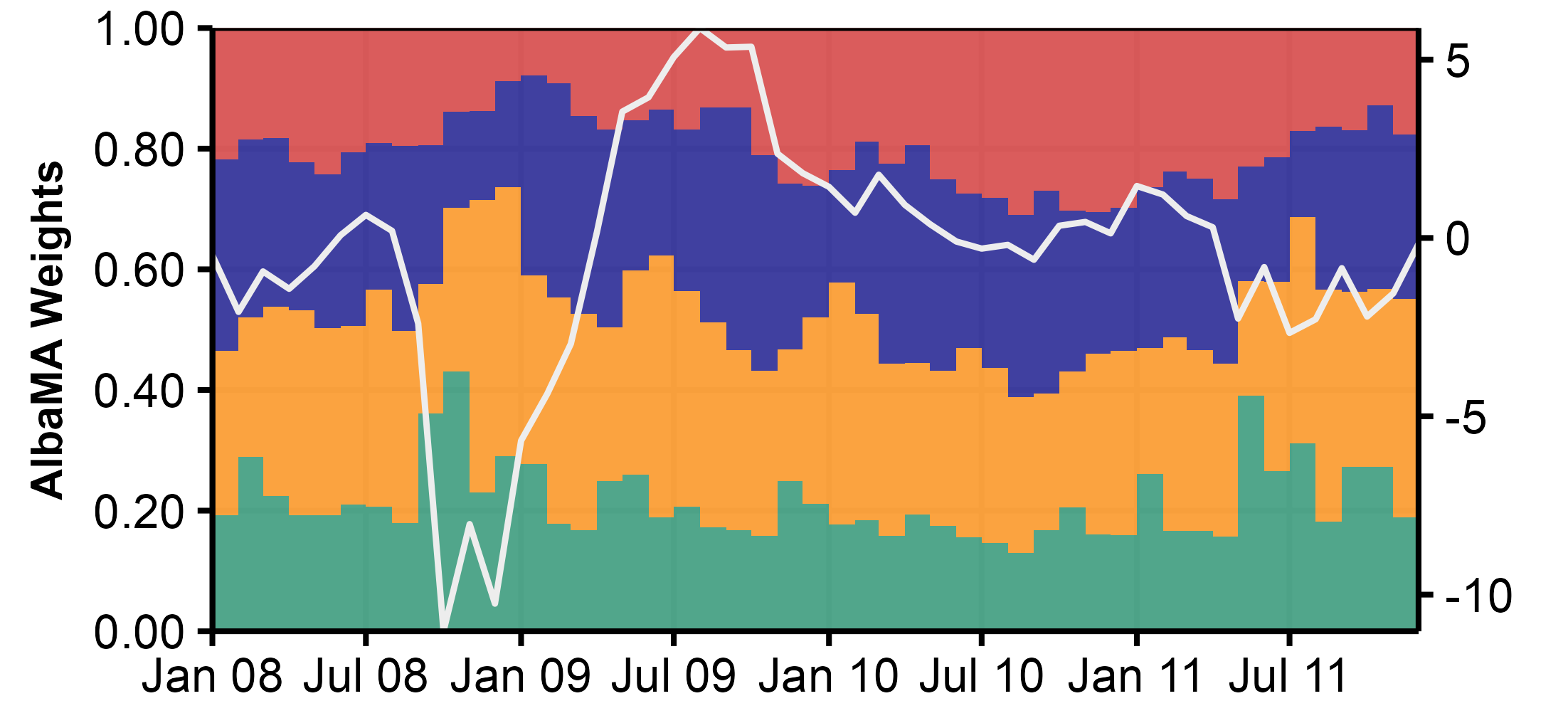}
    \end{minipage}%
    \begin{minipage}[t]{0.5\textwidth}
      \centering
      \includegraphics[width=\textwidth, trim = 10mm 0mm 5mm 0mm, clip]{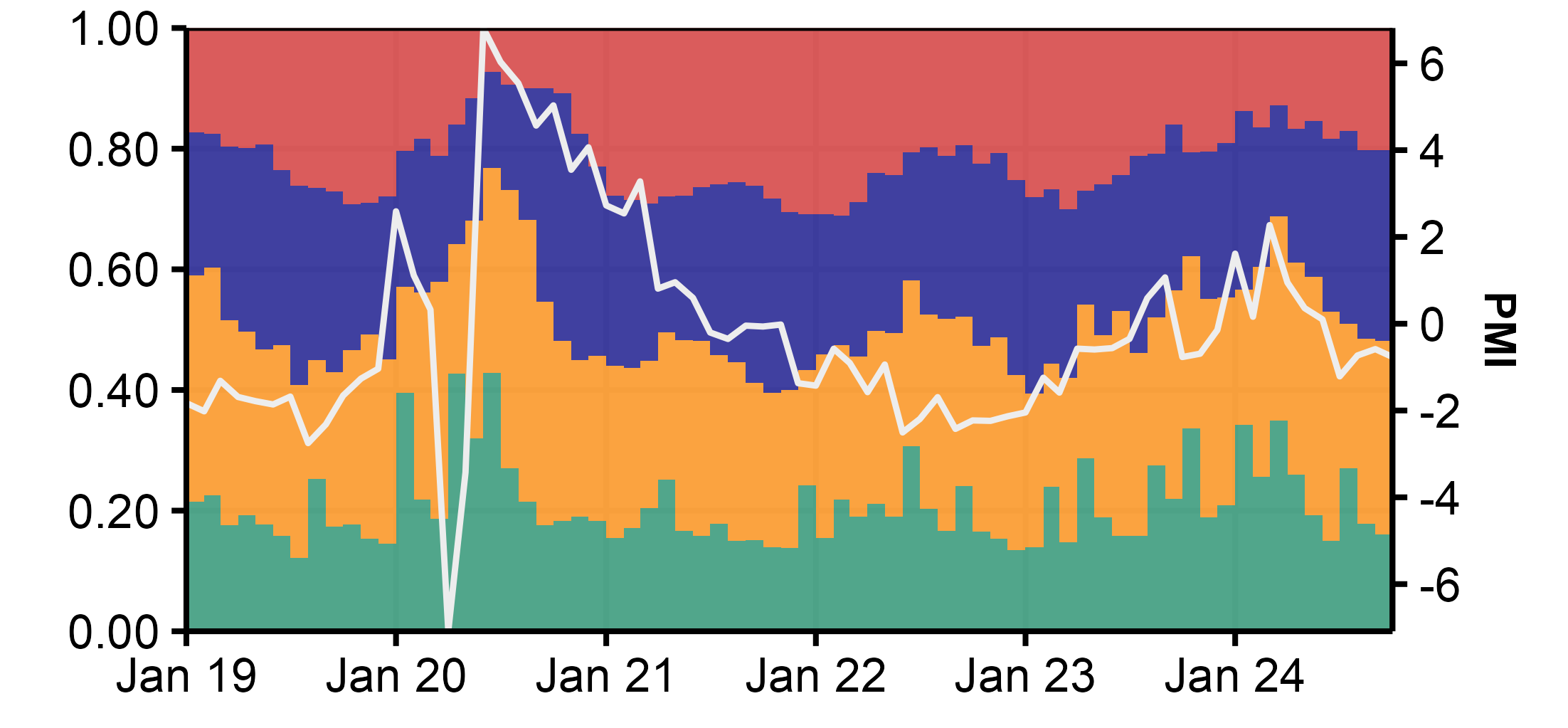}
    \end{minipage}

    \vspace*{-.5em}
    \begin{minipage}[t]{\textwidth}
      \centering
      \hspace*{1.0em} \includegraphics[width=0.95\textwidth, trim = 0mm 0mm 0mm 0mm, clip]{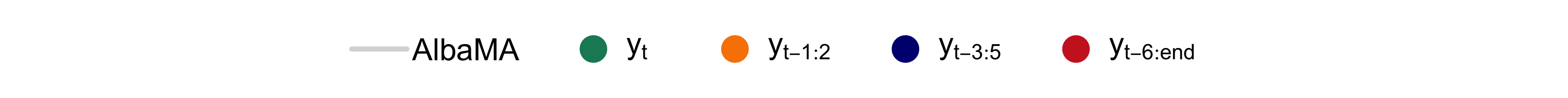}
    \end{minipage}

    \vspace*{-0.3em}
    \begin{minipage}[t]{0.5\textwidth}
      \centering
      (a) Great Recession
    \end{minipage}
    \begin{minipage}[t]{0.5\textwidth}
      \centering
      (b) Post-Covid
    \end{minipage}

    \begin{threeparttable}
    \centering
    \begin{minipage}{\textwidth}
      \begin{tablenotes}[para,flushleft]
    \setlength{\lineskip}{0.2ex}
    \notsotiny 
  {\textit{Notes}: The \textbf{upper panel} shows AlbaMA and the MA(12). The \textbf{middle panels} compare AlbaMA to standard moving averages and the Savitzky-Golay filter for (a) the Great Recession and (b) the post-Covid surge. The \textbf{lower panels} present the weights the RF assigns to past observations. All measures are one-sided.}
    \end{tablenotes}
  \end{minipage}
  \end{threeparttable}
\end{figure}


\begin{figure}[h]
  \caption{\normalsize{EA Industrial Production}} \label{fig:EA_INDPRO}
  
    \begin{minipage}[t]{\textwidth}
      \centering
      \includegraphics[width=\textwidth, trim = -5mm 0mm -25mm 0mm, clip]{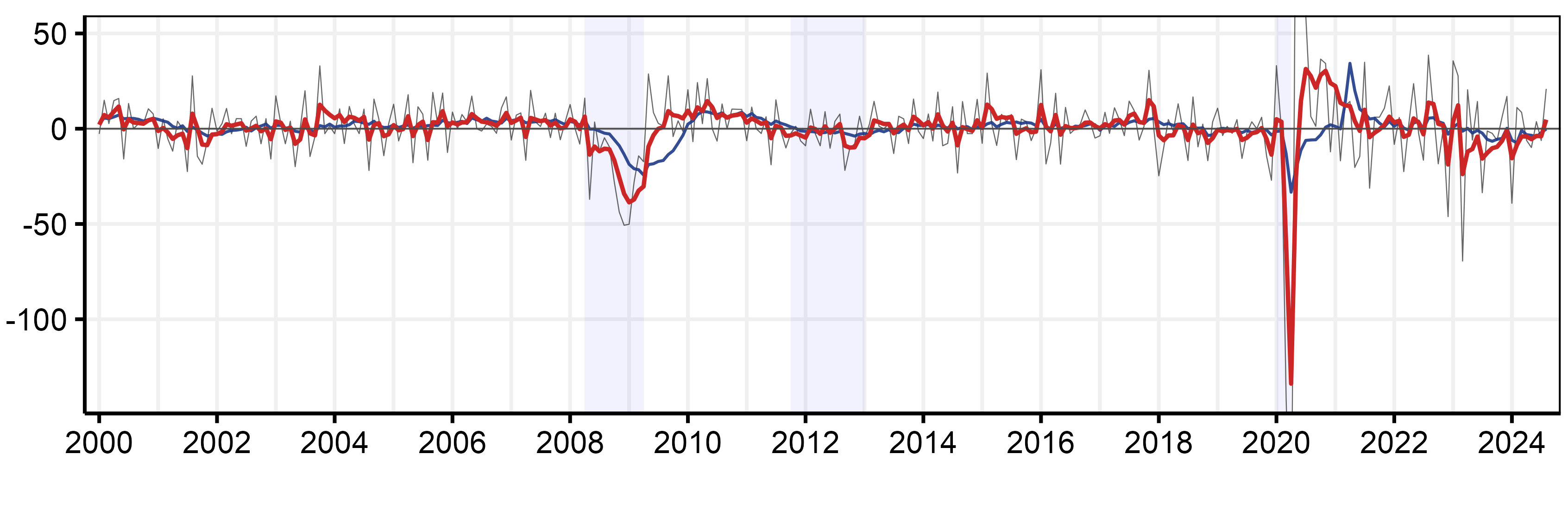}
    \end{minipage}%

\vspace*{-1em}
     \begin{minipage}[t]{0.5\textwidth}
      \centering
      \includegraphics[width=\textwidth, trim = -15mm 0mm -8mm 0mm, clip]{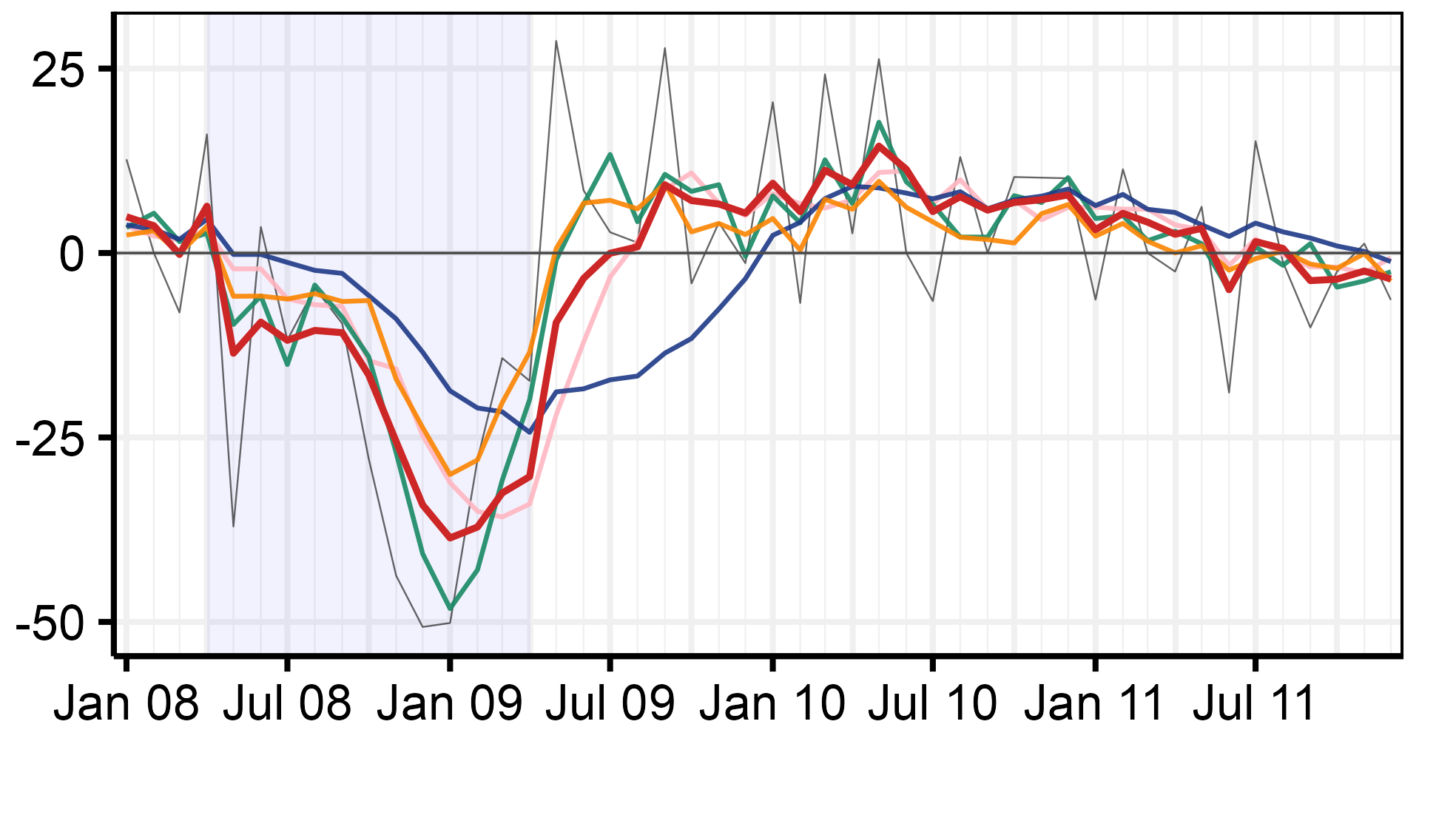}
    \end{minipage}%
    \begin{minipage}[t]{0.5\textwidth}
      \centering
      \includegraphics[width=\textwidth, trim = 0mm 0mm -24mm 0mm, clip]{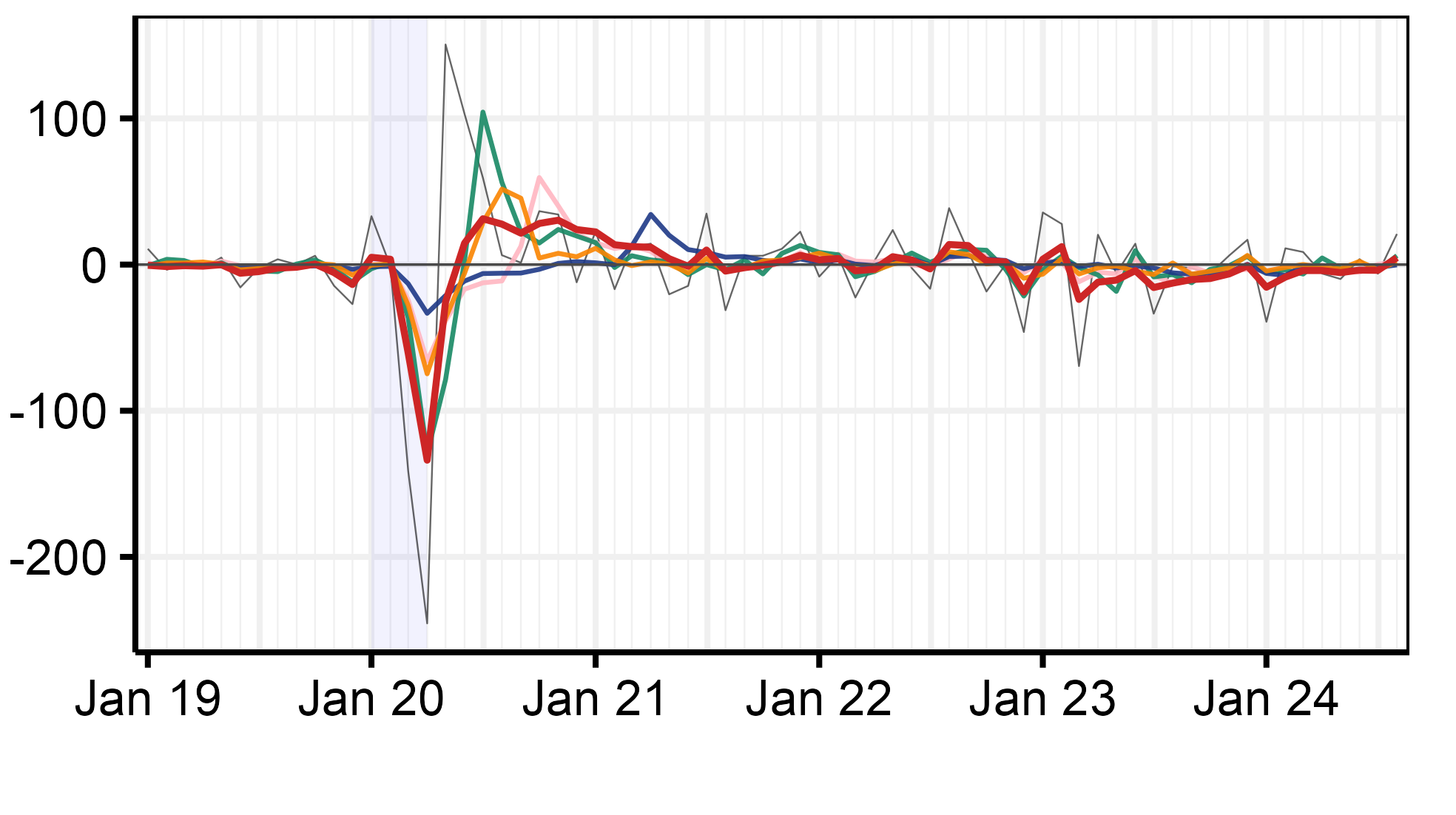}
    \end{minipage}

    \vspace*{-0.8em}
    \begin{minipage}[t]{\textwidth}
      \centering
      \includegraphics[width=0.8\textwidth, trim = 0mm 0mm 0mm 0mm, clip]{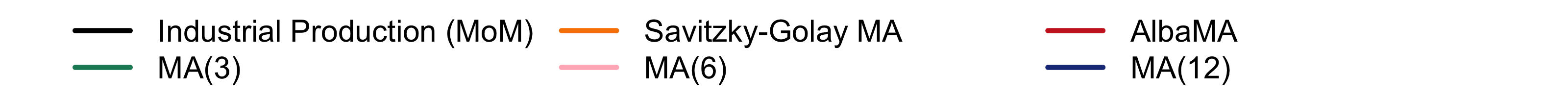}
    \end{minipage}

    \begin{minipage}[t]{0.5\textwidth}
      \centering
      \includegraphics[width=\textwidth, trim = 4mm 0mm 10mm 0mm, clip]{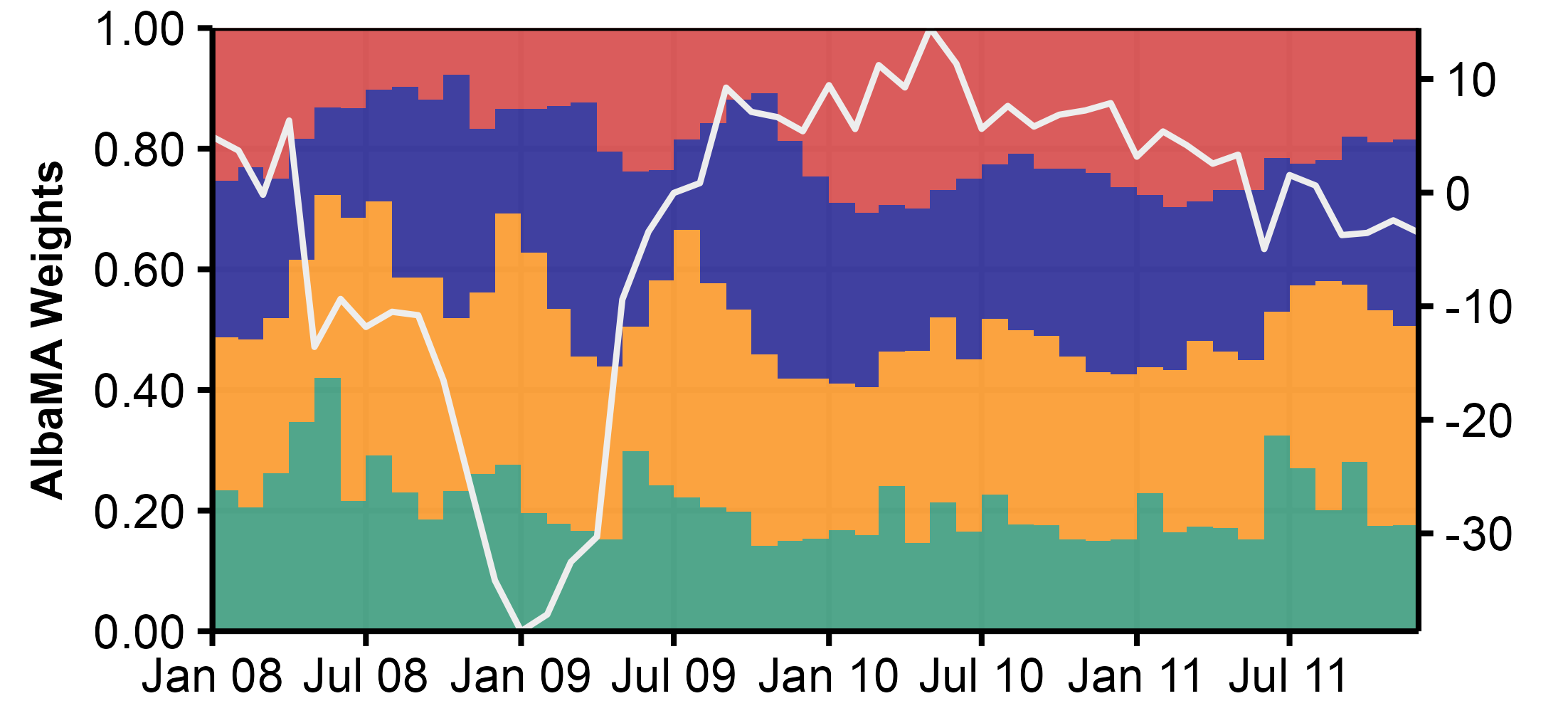}
    \end{minipage}%
    \begin{minipage}[t]{0.5\textwidth}
      \centering
      \includegraphics[width=\textwidth, trim = 6mm 0mm 5mm 0mm, clip]{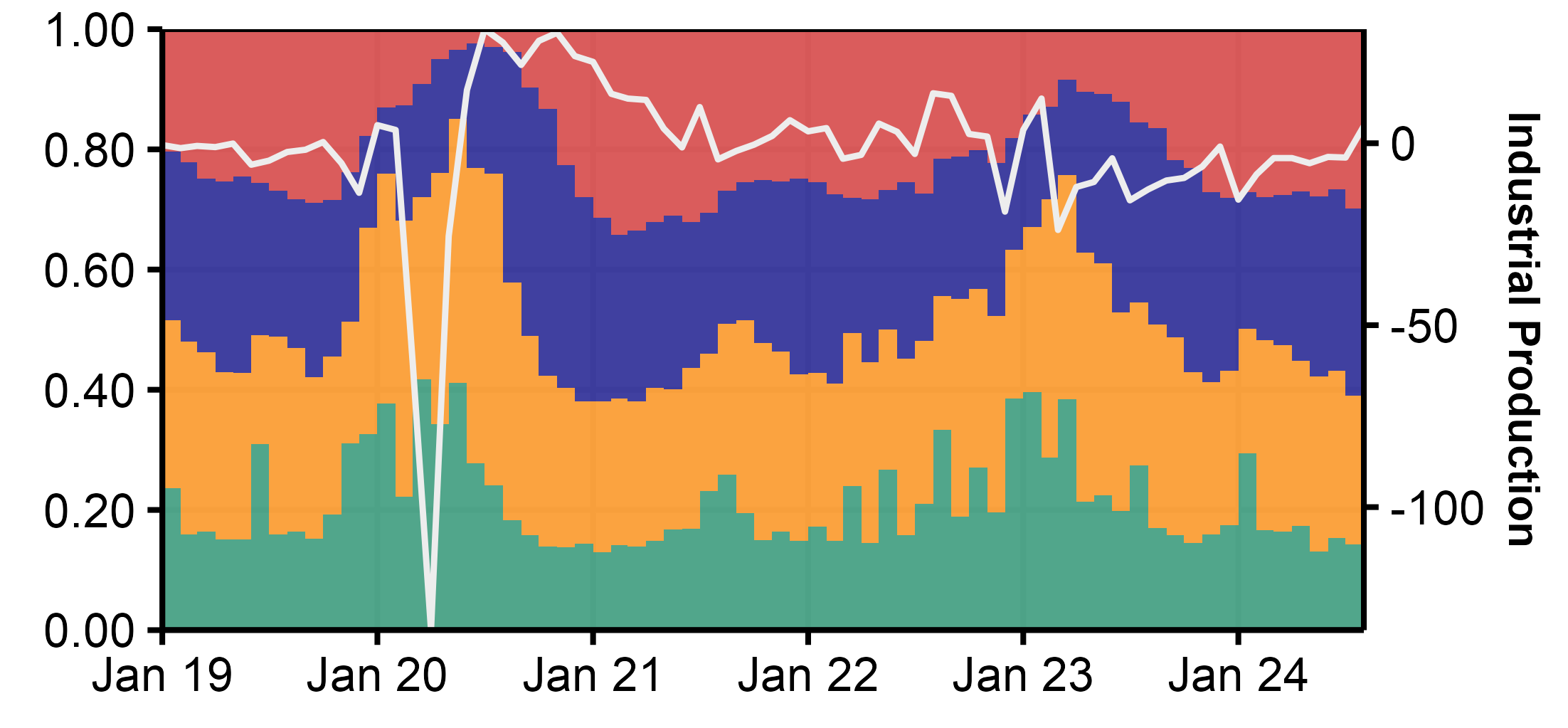}
    \end{minipage}

    \vspace*{-.5em}
    \begin{minipage}[t]{\textwidth}
      \centering
      \hspace*{1.0em} \includegraphics[width=0.95\textwidth, trim = 0mm 0mm 0mm 0mm, clip]{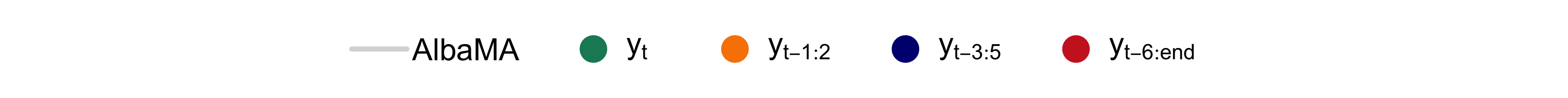}
    \end{minipage}

    \vspace*{-0.3em}
    \begin{minipage}[t]{0.5\textwidth}
      \centering
      (a) Great Recession
    \end{minipage}
    \begin{minipage}[t]{0.5\textwidth}
      \centering
      (b) Post-Covid
    \end{minipage}

    \begin{threeparttable}
    \centering
    \begin{minipage}{\textwidth}
      \begin{tablenotes}[para,flushleft]
    \setlength{\lineskip}{0.2ex}
    \notsotiny 
  {\textit{Notes}: The \textbf{upper panel} shows AlbaMA and the MA(12). The \textbf{middle panels} compare AlbaMA to standard moving averages and the Savitzky-Golay filter for (a) the Great Recession and (b) the post-Covid surge. The \textbf{lower panels} present the weights the RF assigns to past observations. All measures are one-sided.}
    \end{tablenotes}
  \end{minipage}
  \end{threeparttable}
\end{figure}


\begin{figure}[h]
  \caption{\normalsize{US Industrial Production}} \label{fig:US_INDPRO}
  
   \begin{minipage}[t]{\textwidth}
      \centering
      \includegraphics[width=\textwidth, trim = -8mm 0mm -20mm 0mm, clip]{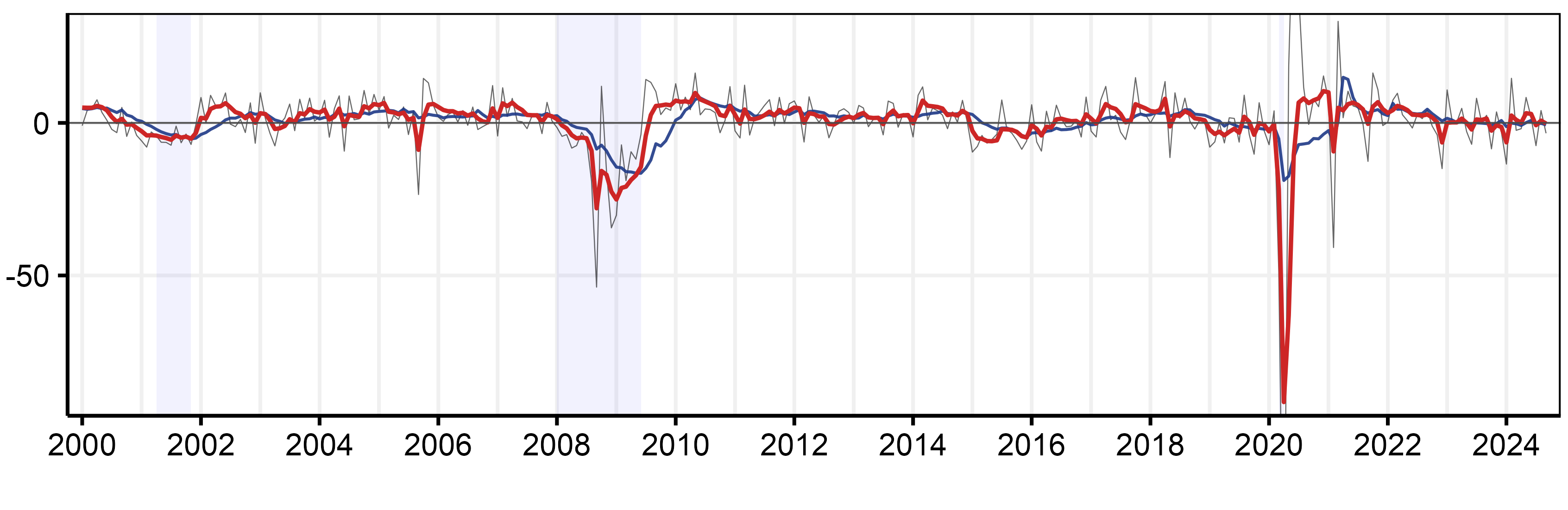}
    \end{minipage}%

\vspace*{-1em}
     \begin{minipage}[t]{0.5\textwidth}
      \centering
      \includegraphics[width=\textwidth, trim = -13mm 0mm -7mm 0mm, clip]{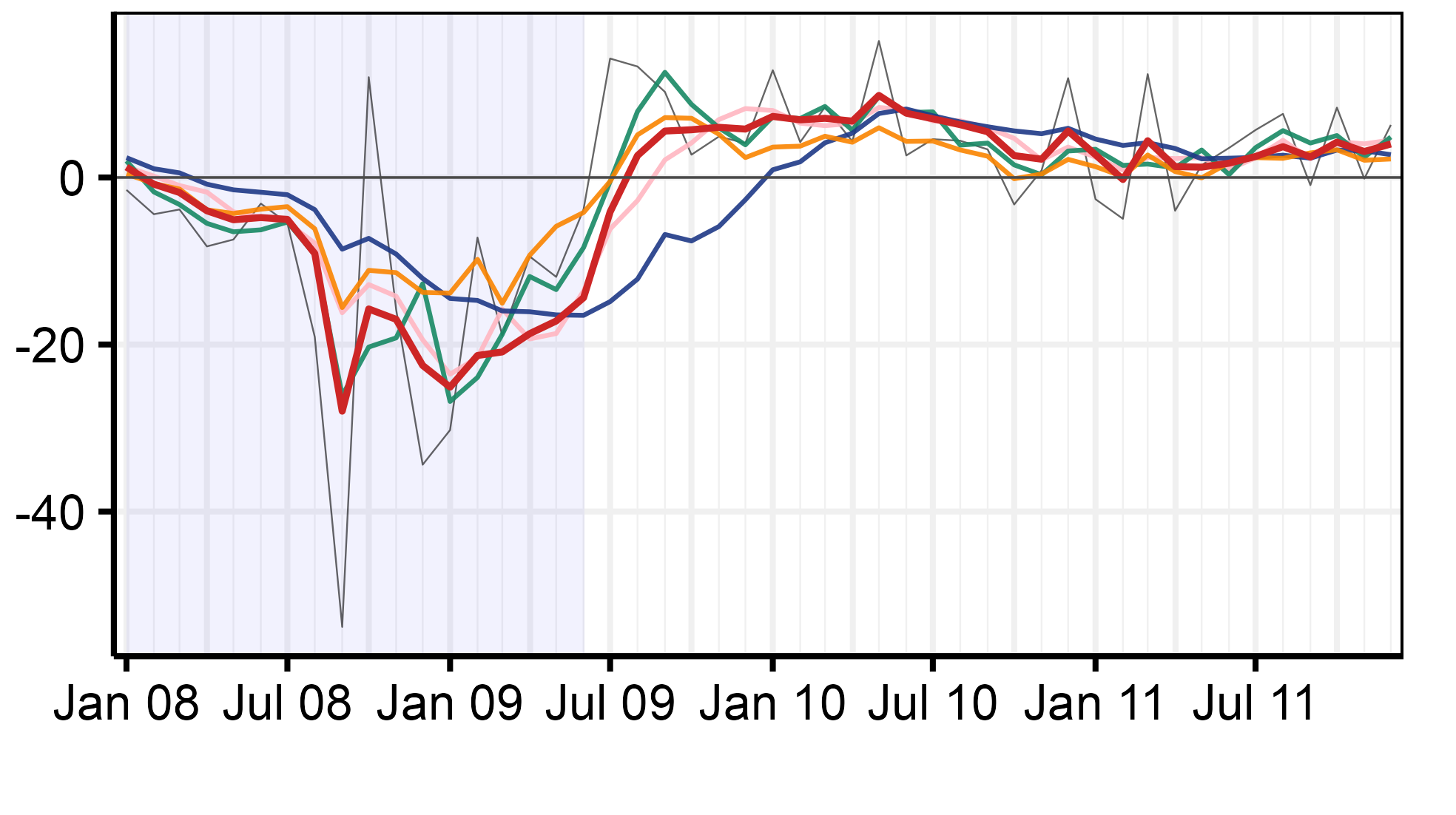}
    \end{minipage}%
    \begin{minipage}[t]{0.5\textwidth}
      \centering
      \includegraphics[width=\textwidth, trim = -1mm 0mm -18mm 0mm, clip]{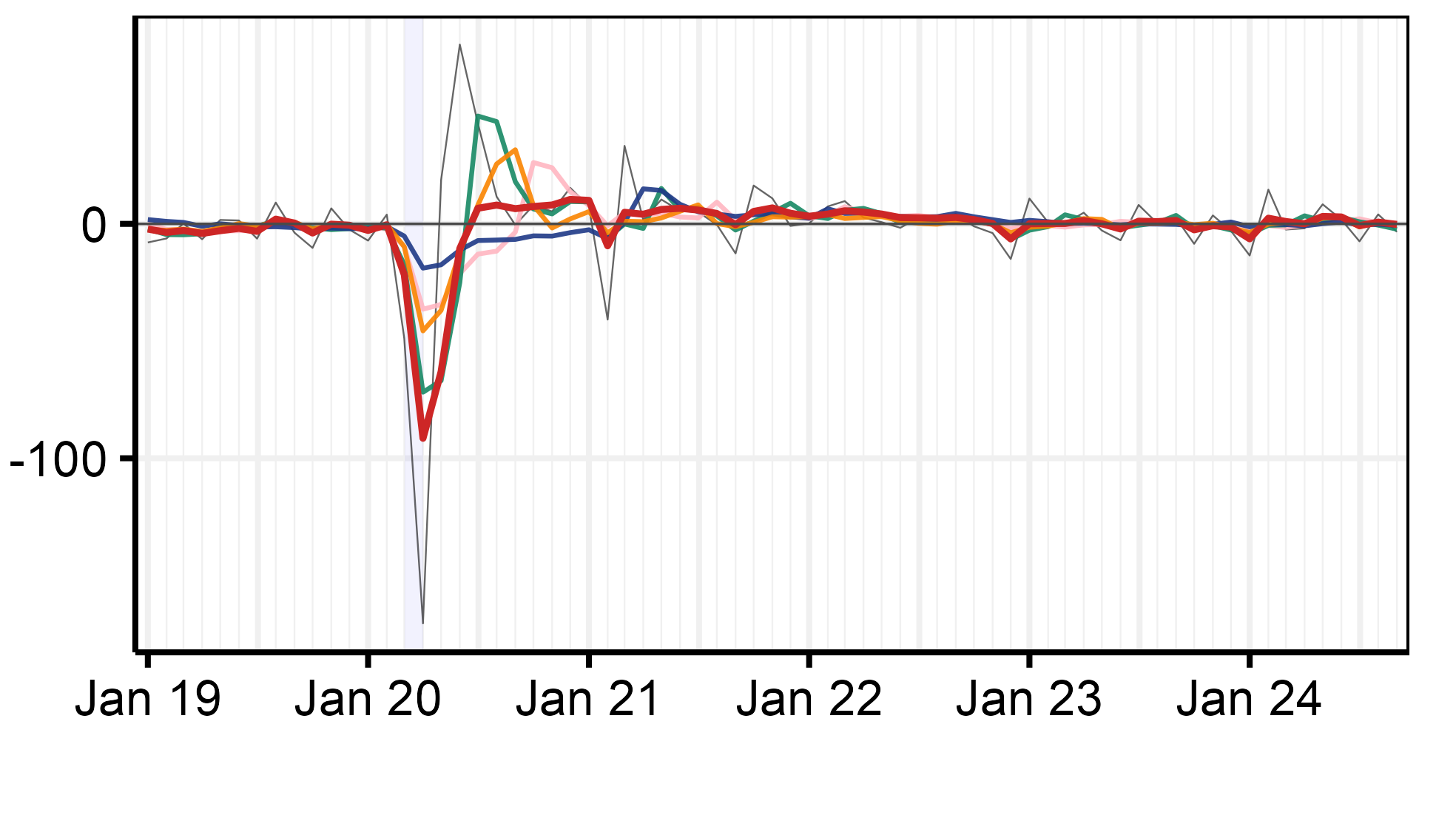}
    \end{minipage}

    \vspace*{-0.8em}
    \begin{minipage}[t]{\textwidth}
      \centering
      \includegraphics[width=0.8\textwidth, trim = 0mm 0mm 0mm 0mm, clip]{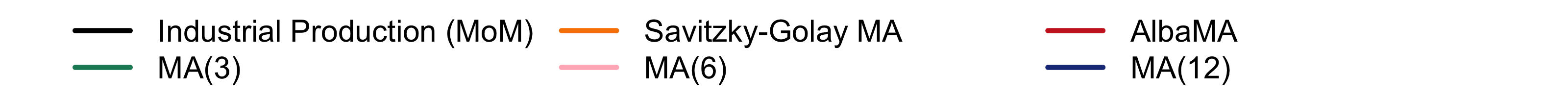}
    \end{minipage}

    \begin{minipage}[t]{0.5\textwidth}
      \centering
      \includegraphics[width=\textwidth, trim = 5mm 0mm 10mm 0mm, clip]{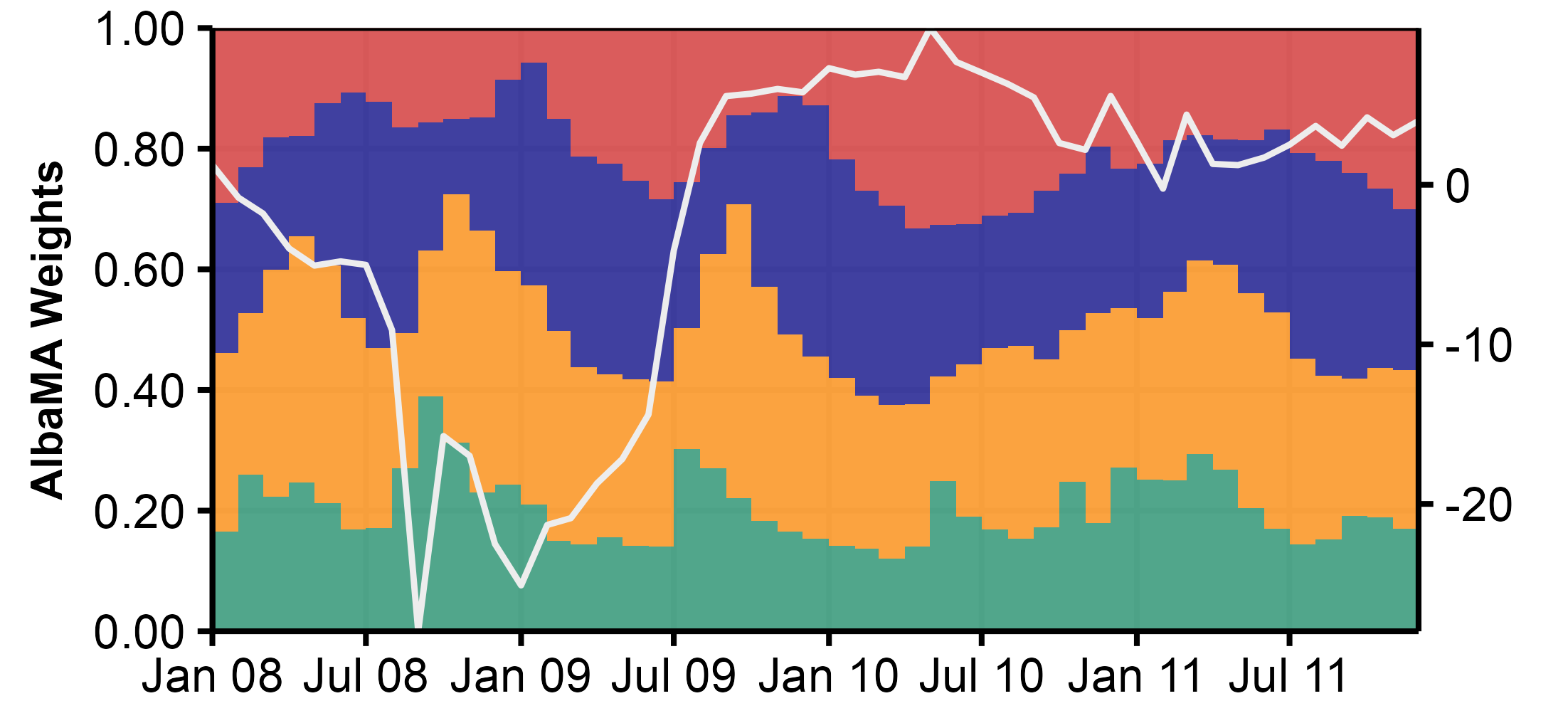}
    \end{minipage}%
    \begin{minipage}[t]{0.5\textwidth}
      \centering
      \includegraphics[width=\textwidth, trim = 6mm 0mm 7mm 0mm, clip]{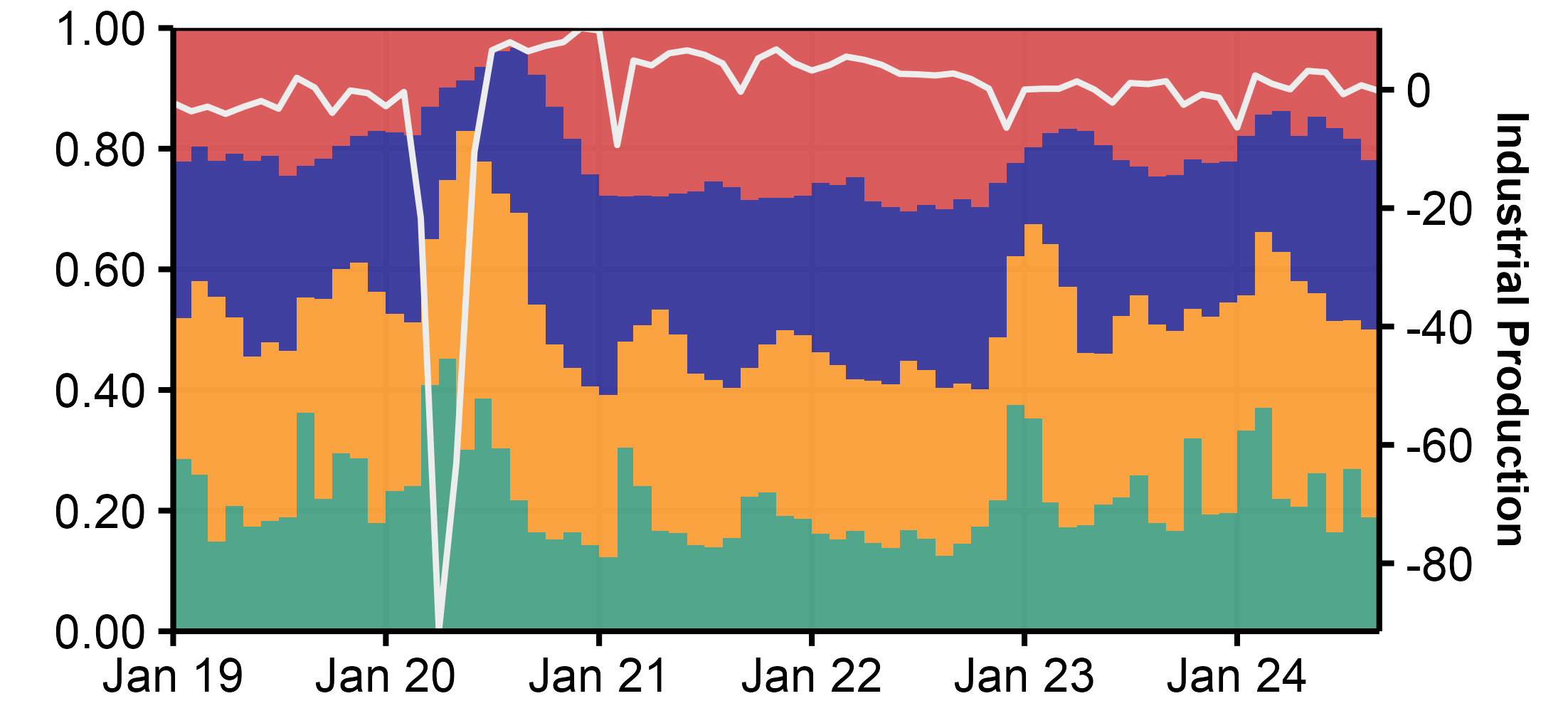}
    \end{minipage}

    \vspace*{-.5em}
    \begin{minipage}[t]{\textwidth}
      \centering
      \hspace*{1.0em} \includegraphics[width=0.95\textwidth, trim = 0mm 0mm 0mm 0mm, clip]{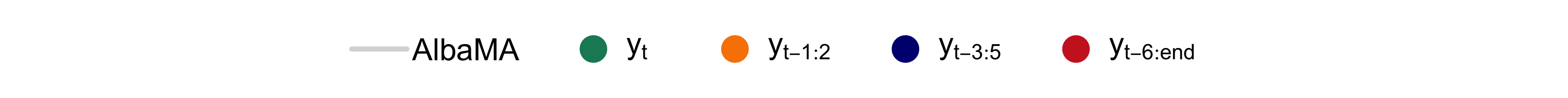}
    \end{minipage}

    \vspace*{-0.3em}
    \begin{minipage}[t]{0.5\textwidth}
      \centering
      (a) Great Recession
    \end{minipage}
    \begin{minipage}[t]{0.5\textwidth}
      \centering
      (b) Post-Covid
    \end{minipage}

    \begin{threeparttable}
    \centering
    \begin{minipage}{\textwidth}
      \begin{tablenotes}[para,flushleft]
    \setlength{\lineskip}{0.2ex}
    \notsotiny 
  {\textit{Notes}: The \textbf{upper panel} shows AlbaMA and the MA(12). The \textbf{middle panels} compare AlbaMA to standard moving averages and the Savitzky-Golay filter for (a) the Great Recession and (b) the post-Covid surge. The \textbf{lower panels} present the weights the RF assigns to past observations. All measures are one-sided.}
    \end{tablenotes}
  \end{minipage}
  \end{threeparttable}
\end{figure}


\begin{figure}[h]
  \caption{\normalsize{US Unemployment Rate}} \label{fig:US_UNRATE}
  
     \begin{minipage}[t]{\textwidth}
      \centering
      \includegraphics[width=\textwidth, trim = -14mm 0mm -20mm 0mm, clip]{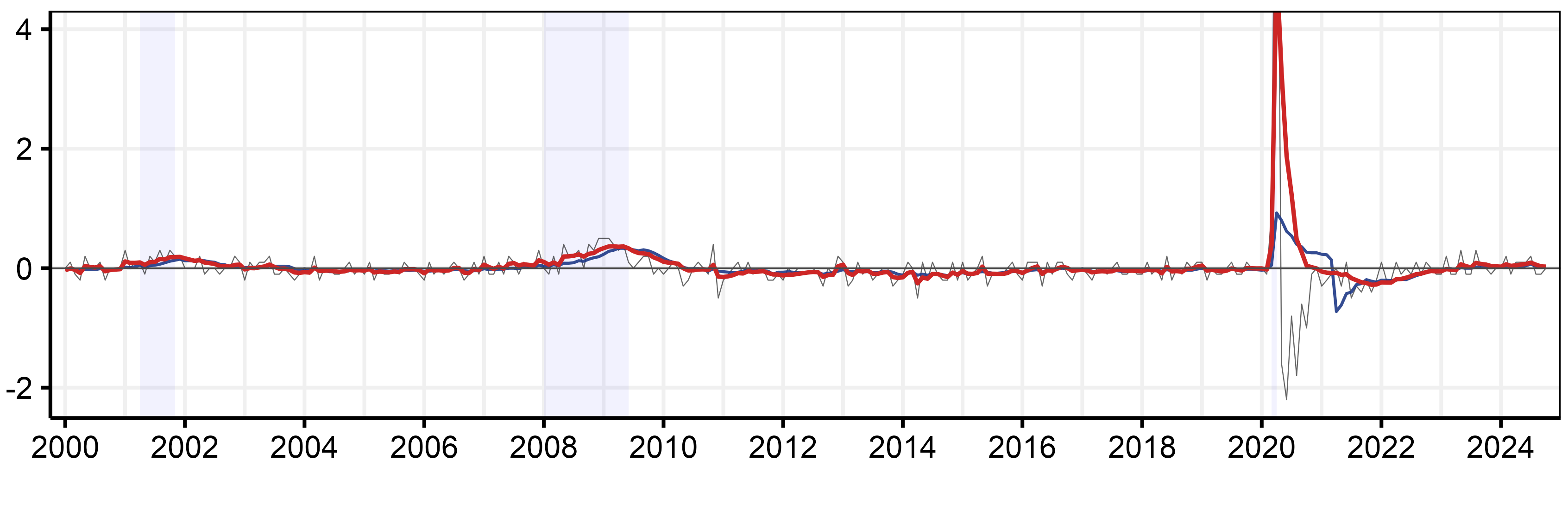}
    \end{minipage}%

\vspace*{-1em}
     \begin{minipage}[t]{0.5\textwidth}
      \centering
      \includegraphics[width=\textwidth, trim = -5mm 0mm -9mm 0mm, clip]{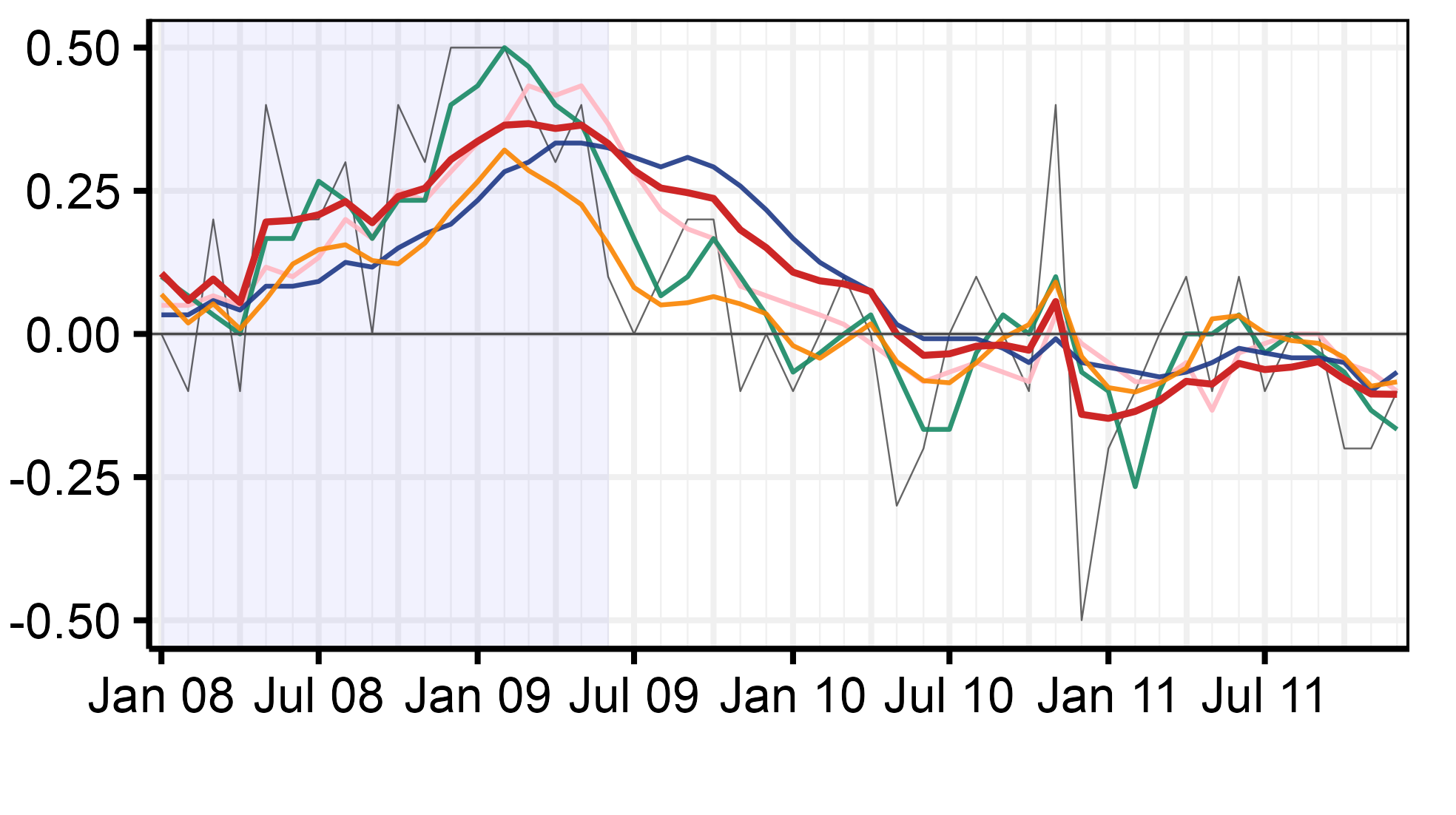}
    \end{minipage}%
    \begin{minipage}[t]{0.5\textwidth}
      \centering
      \includegraphics[width=\textwidth, trim = -3mm 0mm -14mm 0mm, clip]{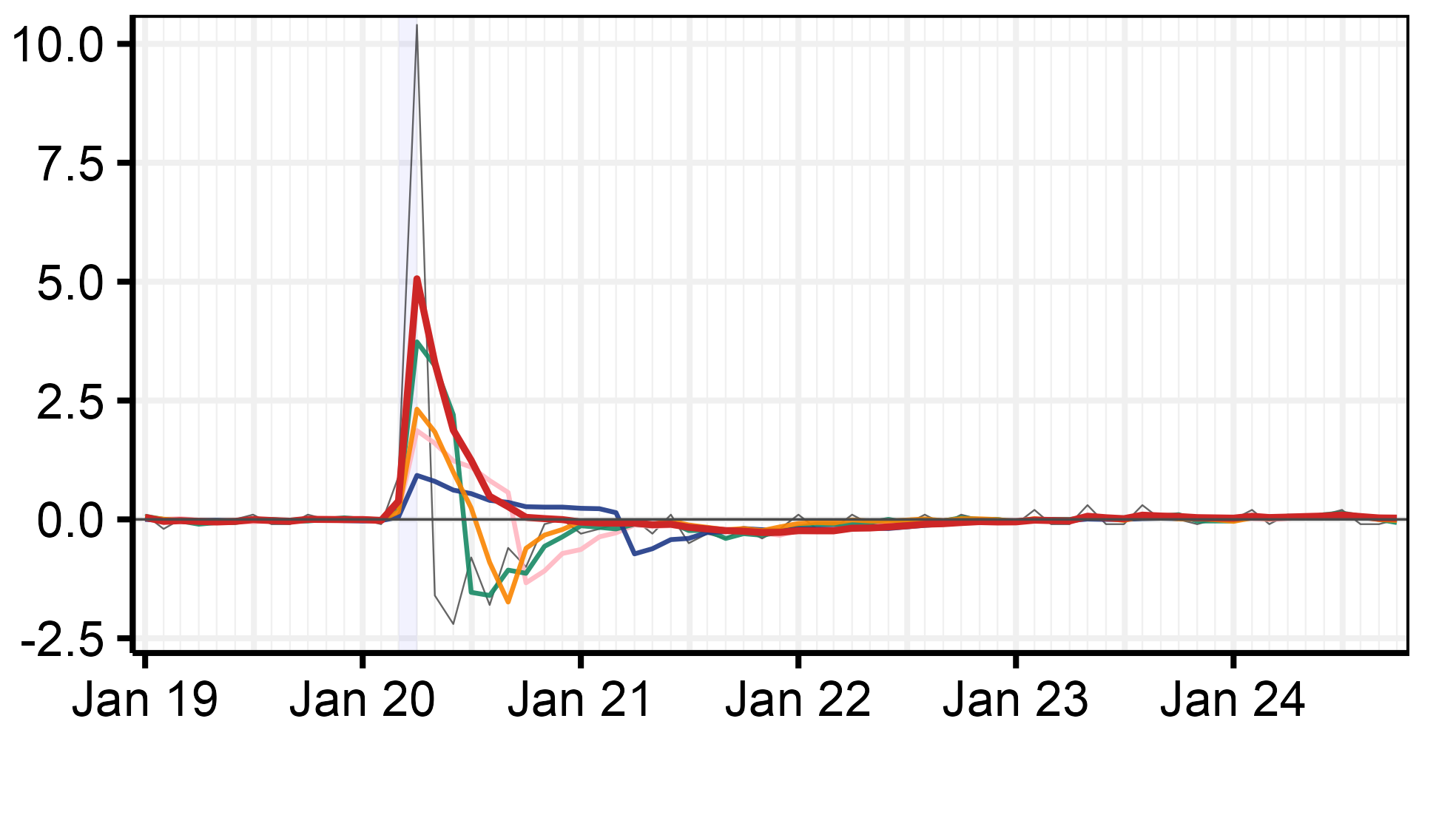}
    \end{minipage}

    \vspace*{-0.8em}
    \begin{minipage}[t]{\textwidth}
      \centering
      \includegraphics[width=0.8\textwidth, trim = 0mm 0mm 0mm 0mm, clip]{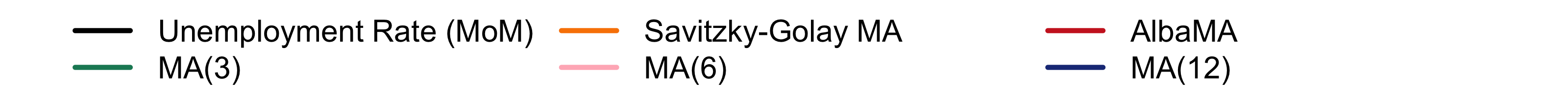}
    \end{minipage}

    \begin{minipage}[t]{0.5\textwidth}
      \centering
      \includegraphics[width=\textwidth, trim = 4mm 0mm 10mm 0mm, clip]{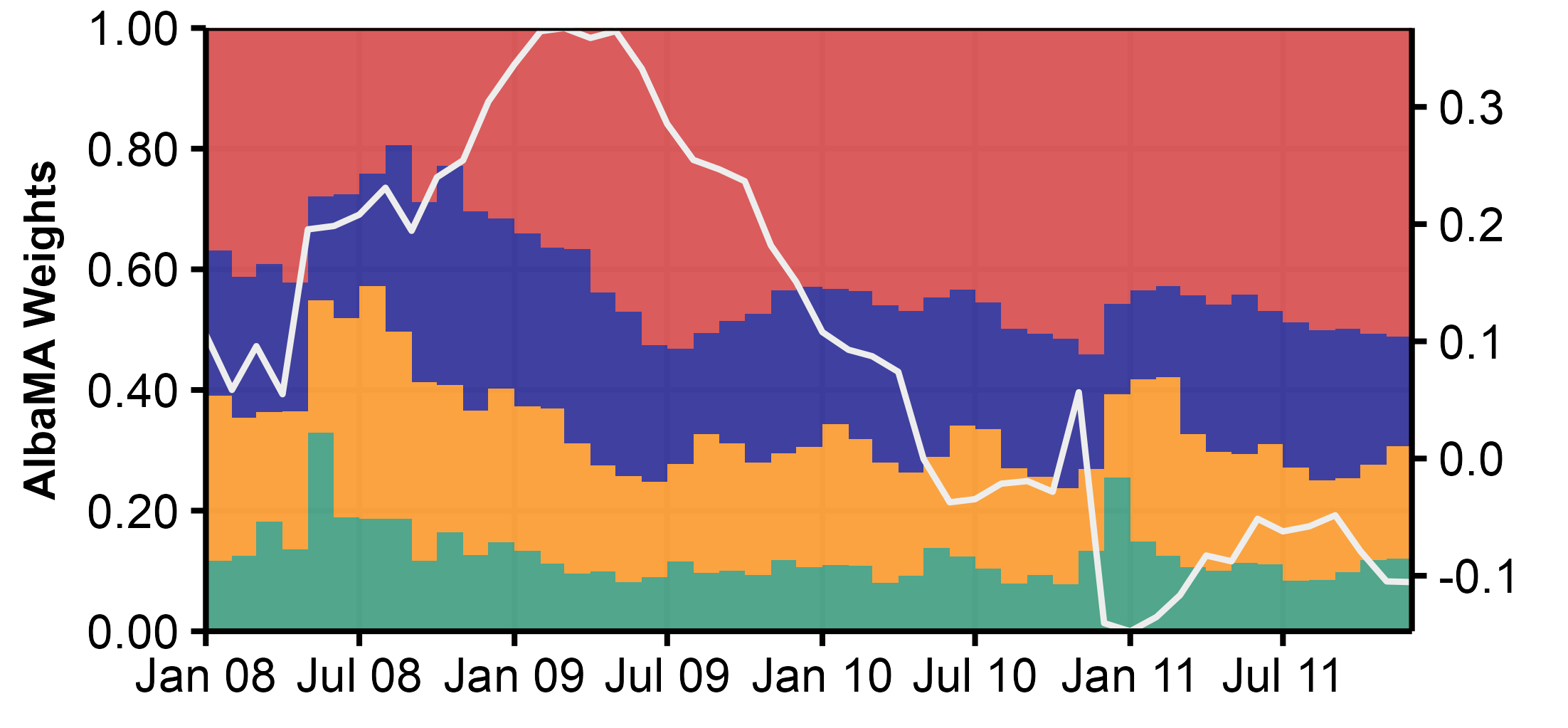}
    \end{minipage}%
    \begin{minipage}[t]{0.5\textwidth}
      \centering
      \includegraphics[width=\textwidth, trim = 10mm 0mm 7mm 0mm, clip]{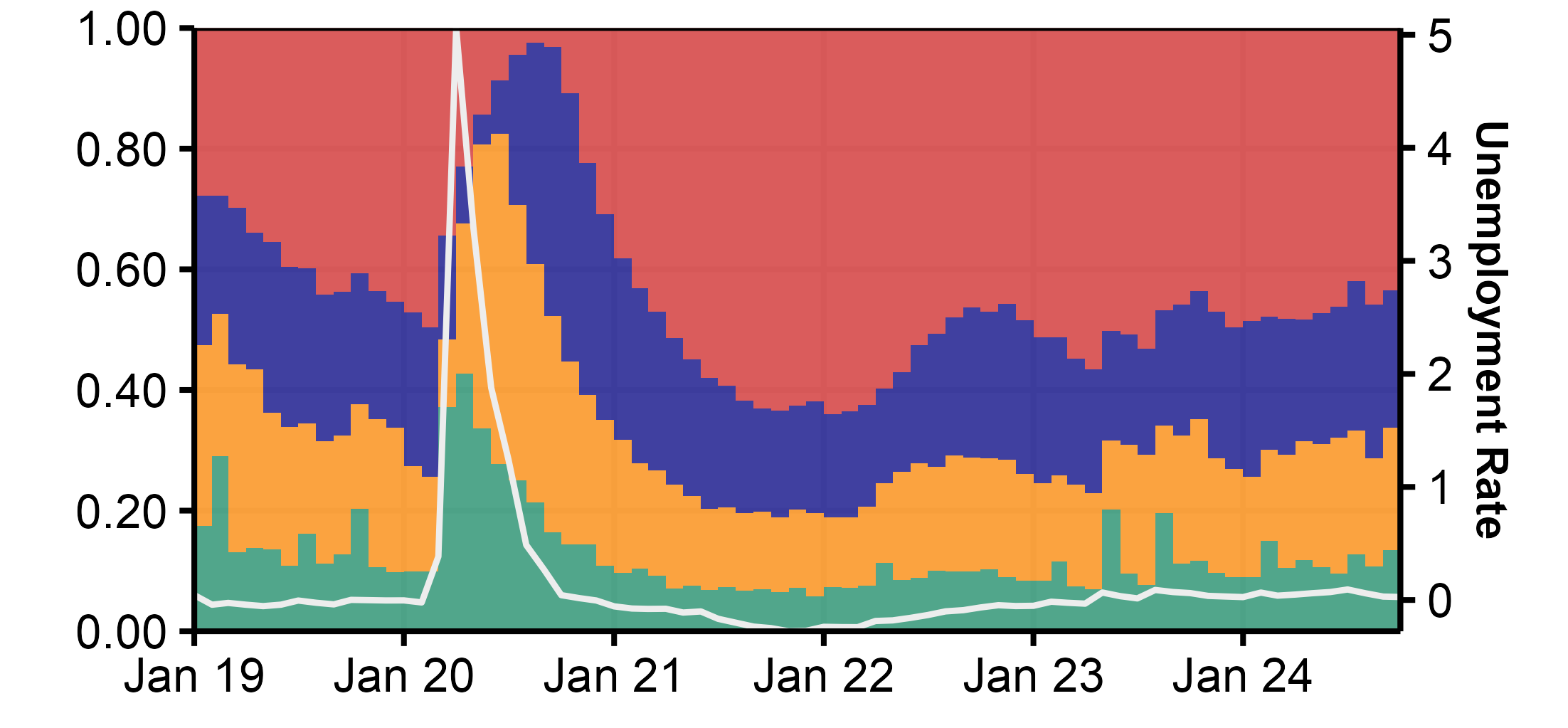}
    \end{minipage}

    \vspace*{-.5em}
    \begin{minipage}[t]{\textwidth}
      \centering
      \hspace*{1.0em} \includegraphics[width=0.95\textwidth, trim = 0mm 0mm 0mm 0mm, clip]{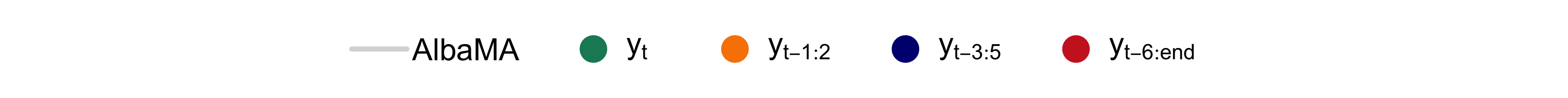}
    \end{minipage}

    \vspace*{-0.3em}
    \begin{minipage}[t]{0.5\textwidth}
      \centering
      (a) Great Recession
    \end{minipage}
    \begin{minipage}[t]{0.5\textwidth}
      \centering
      (b) Post-Covid
    \end{minipage}

    \begin{threeparttable}
    \centering
    \begin{minipage}{\textwidth}
      \begin{tablenotes}[para,flushleft]
    \setlength{\lineskip}{0.2ex}
    \notsotiny 
  {\textit{Notes}: The \textbf{upper panel} shows AlbaMA and the MA(12). The \textbf{middle panels} compare AlbaMA to standard moving averages and the Savitzky-Golay filter for (a) the Great Recession and (b) the post-Covid surge. The \textbf{lower panels} present the weights the RF assigns to past observations. All measures are one-sided.}
    \end{tablenotes}
  \end{minipage}
  \end{threeparttable}
\end{figure}


\begin{figure}[h]
  \caption{\normalsize{EA Unemployment Rate}} \label{fig:EA_UNRATE}
  
  \begin{minipage}[t]{\textwidth}
      \centering
      \includegraphics[width=\textwidth, trim = 0mm 0mm -22mm 0mm, clip]{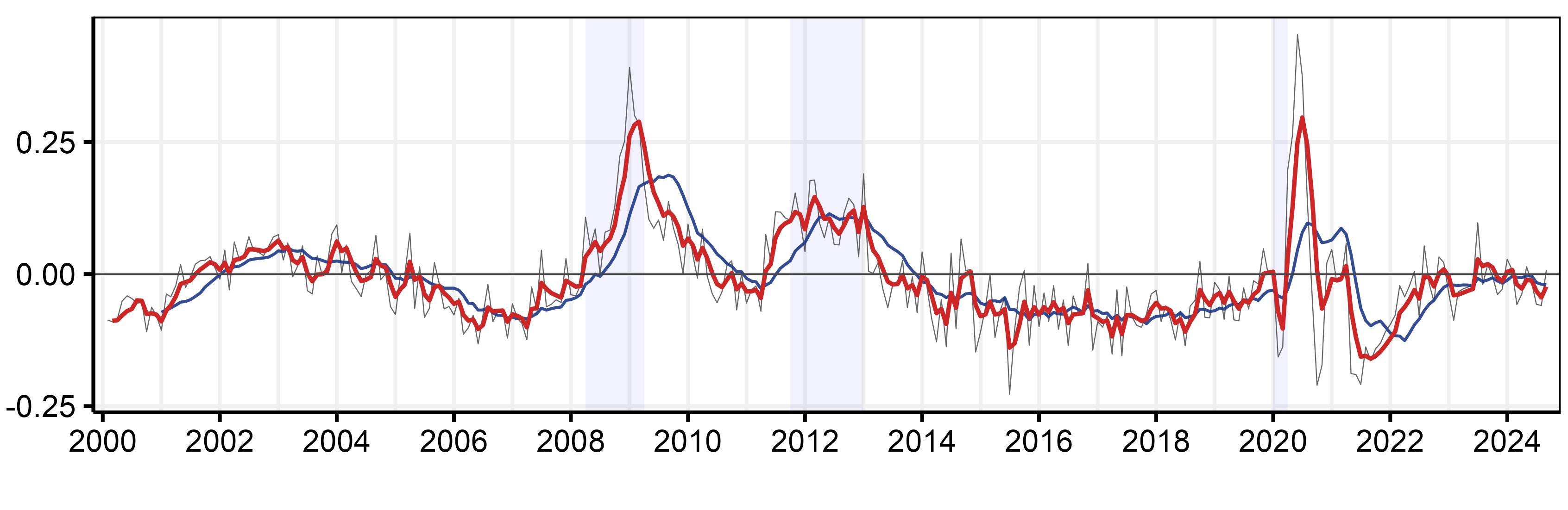}
    \end{minipage}%

\vspace*{-1em}
     \begin{minipage}[t]{0.5\textwidth}
      \centering
      \includegraphics[width=\textwidth, trim = -12mm 0mm -13mm 0mm, clip]{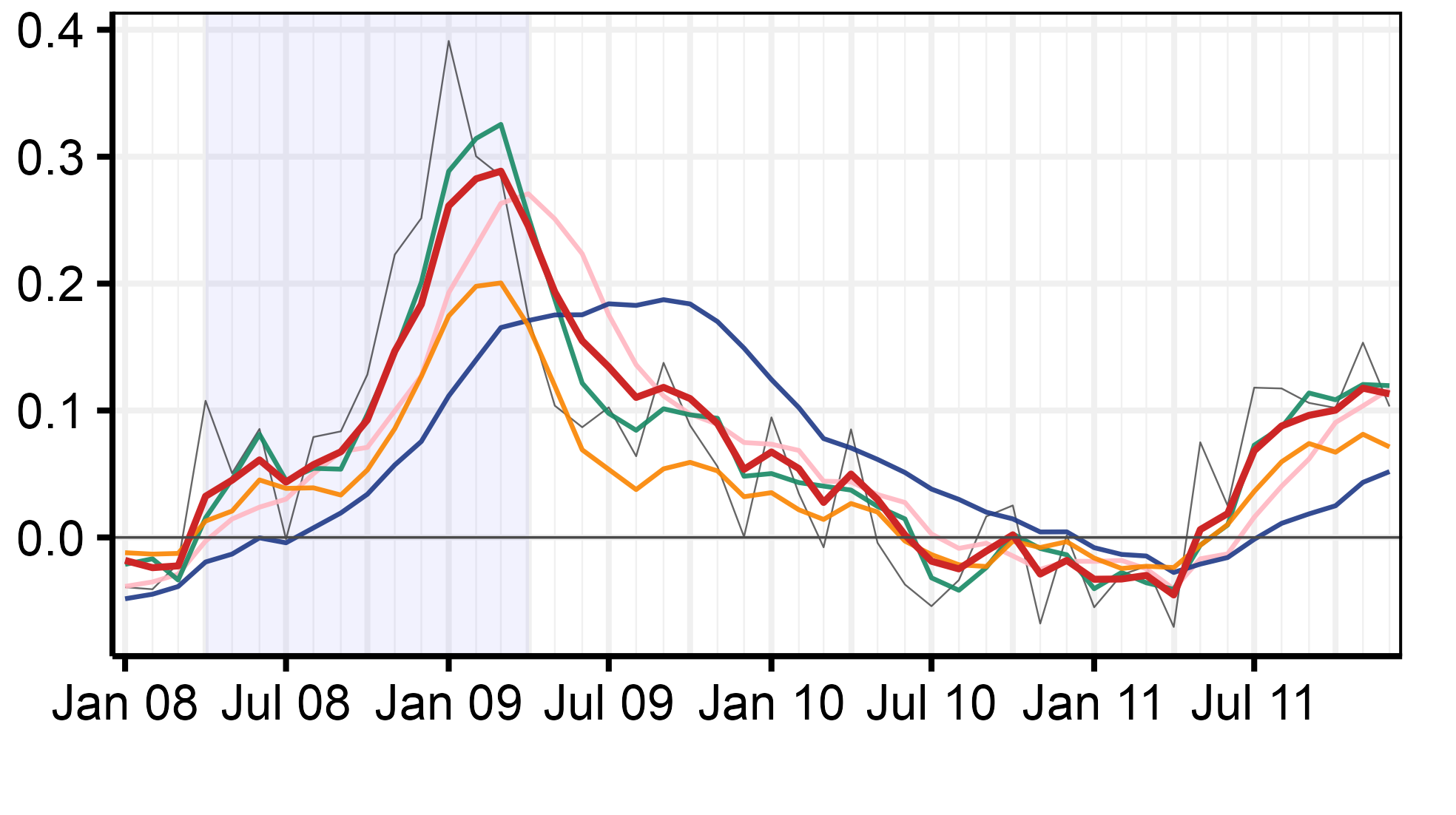}
    \end{minipage}%
    \begin{minipage}[t]{0.5\textwidth}
      \centering
      \includegraphics[width=\textwidth, trim = -3mm 0mm -22mm 0mm, clip]{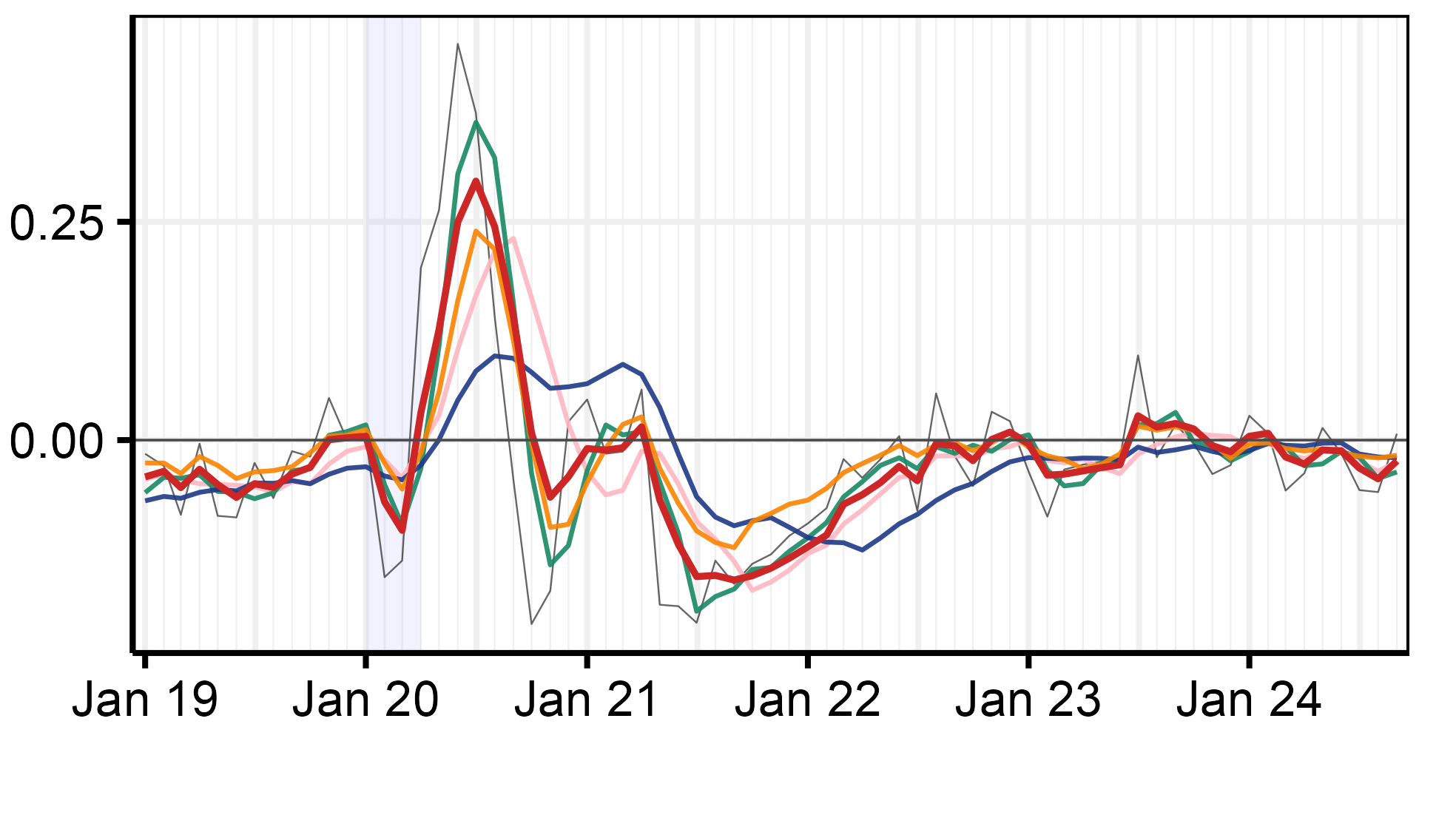}
    \end{minipage}

    \vspace*{-0.8em}
    \begin{minipage}[t]{\textwidth}
      \centering
      \includegraphics[width=0.8\textwidth, trim = 0mm 0mm 0mm 0mm, clip]{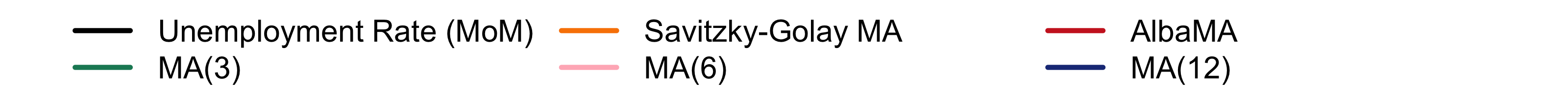}
    \end{minipage}

    \begin{minipage}[t]{0.5\textwidth}
      \centering
      \includegraphics[width=\textwidth, trim = 4mm 0mm 6mm 0mm, clip]{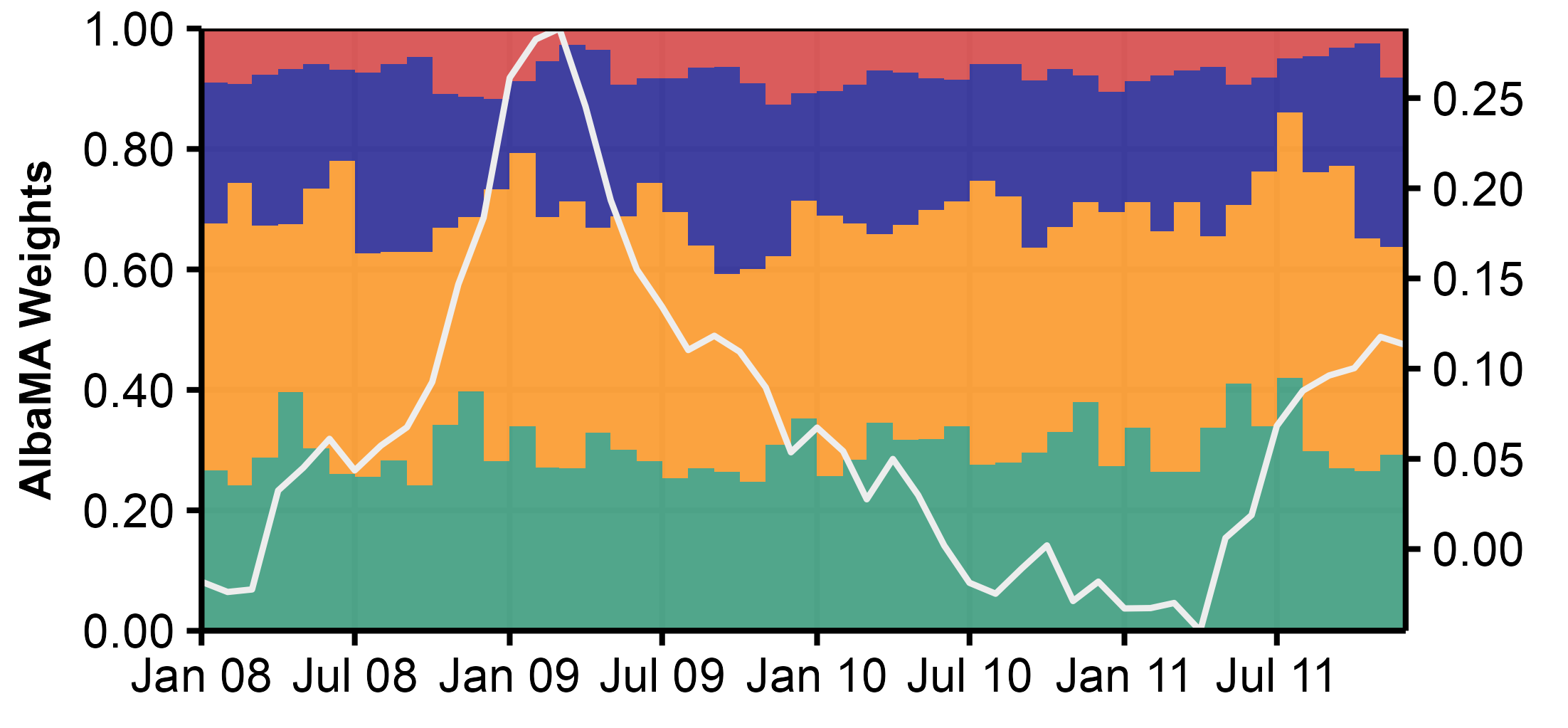}
    \end{minipage}%
    \begin{minipage}[t]{0.5\textwidth}
      \centering
      \includegraphics[width=\textwidth, trim = 8mm 0mm 5mm 0mm, clip]{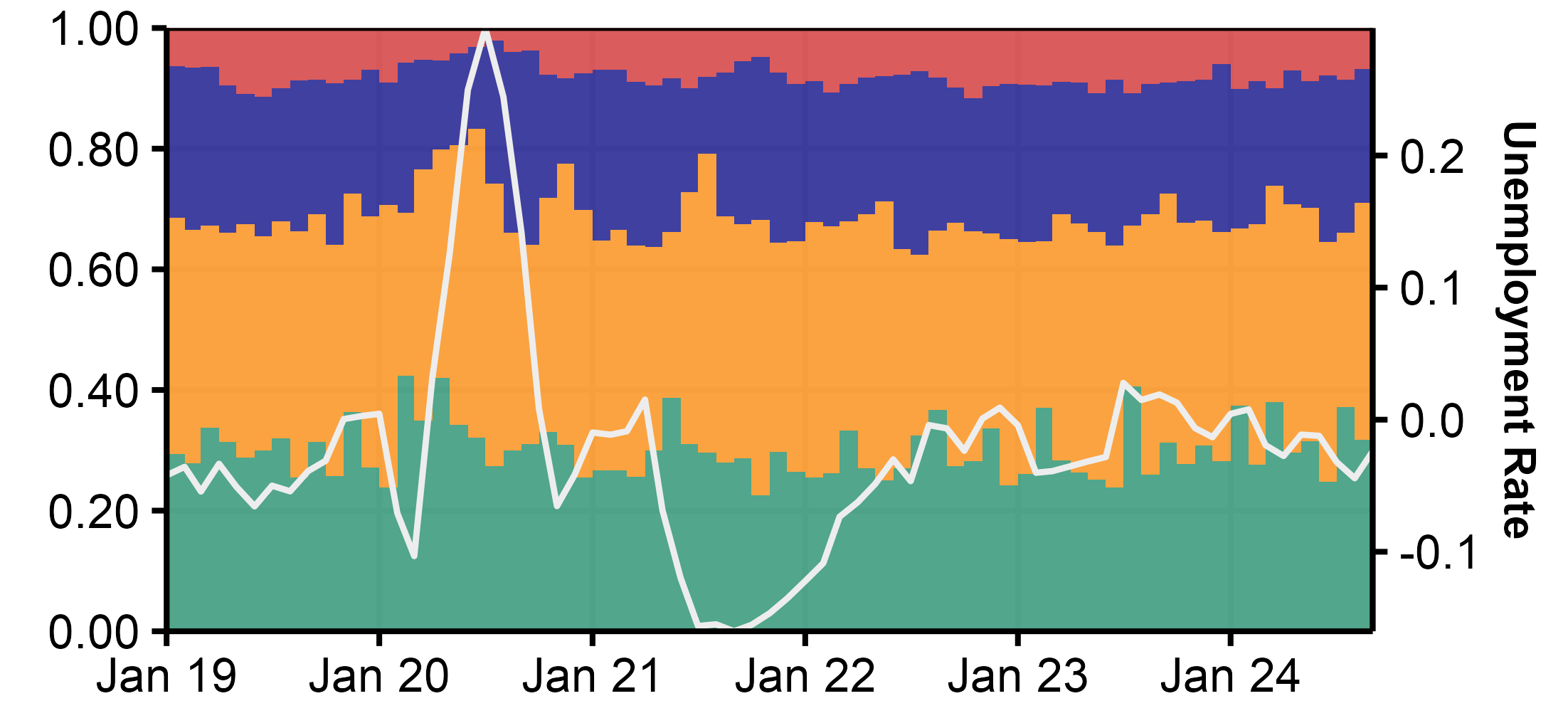}
    \end{minipage}

    \vspace*{-.5em}
    \begin{minipage}[t]{\textwidth}
      \centering
      \hspace*{1.0em} \includegraphics[width=0.95\textwidth, trim = 0mm 0mm 0mm 0mm, clip]{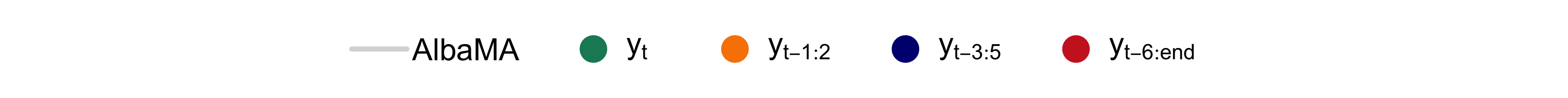}
    \end{minipage}

    \vspace*{-0.3em}
    \begin{minipage}[t]{0.5\textwidth}
      \centering
      (a) Great Recession
    \end{minipage}
    \begin{minipage}[t]{0.5\textwidth}
      \centering
      (b) Post-Covid
    \end{minipage}

    \begin{threeparttable}
    \centering
    \begin{minipage}{\textwidth}
      \begin{tablenotes}[para,flushleft]
    \setlength{\lineskip}{0.2ex}
    \notsotiny 
  {\textit{Notes}: The \textbf{upper panel} shows AlbaMA and the MA(12). The \textbf{middle panels} compare AlbaMA to standard moving averages and the Savitzky-Golay filter for (a) the Great Recession and (b) the post-Covid surge. The \textbf{lower panels} present the weights the RF assigns to past observations. All measures are one-sided.}
    \end{tablenotes}
  \end{minipage}
  \end{threeparttable}
\end{figure}


\begin{figure}[t!]
  \caption{\normalsize{Additional Benchmarks for Inflation (one-sided)}} \label{fig:infl_ucsv_right}
  
  \vspace*{-0.4em}
    \begin{minipage}[t]{\textwidth}
      \centering
      (a) US CPI Inflation
    \end{minipage}

    \begin{minipage}[t]{\textwidth}
      \centering
       \includegraphics[width=\textwidth, trim = 2mm -5mm 0mm 0mm, clip]{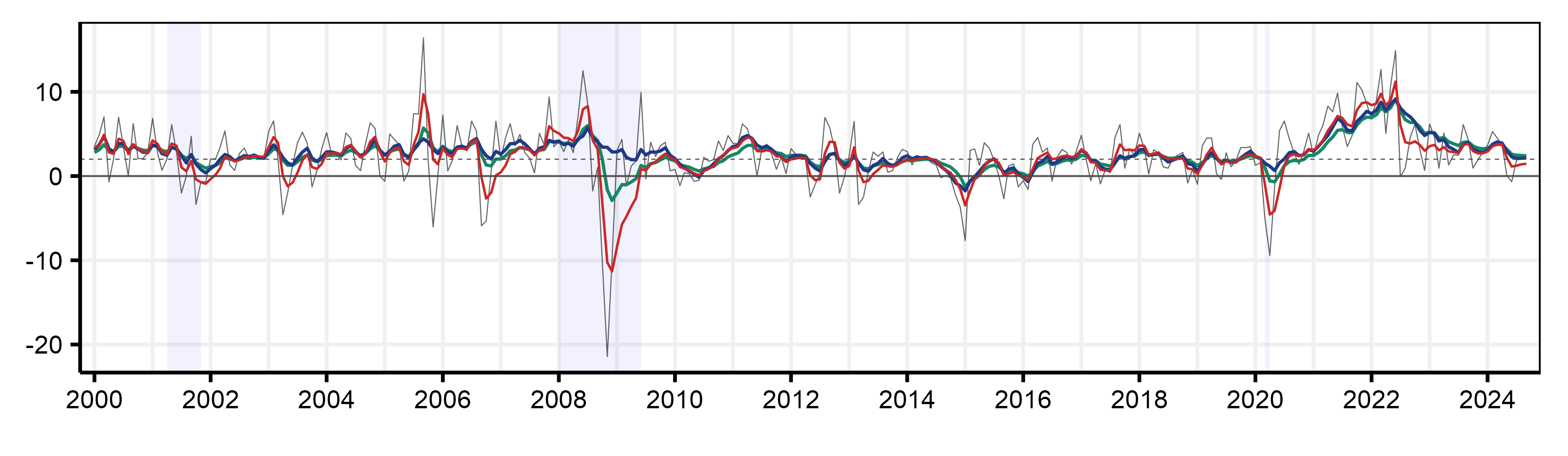}
    \end{minipage}%
   
    \vspace*{-1.1em}
    \begin{minipage}[t]{\textwidth}
      \centering 
      (b) US CPI Core Inflation
    \end{minipage}

    \begin{minipage}[t]{\textwidth}
      \centering
       \includegraphics[width=\textwidth, trim = -1mm -6mm 3mm 0mm, clip]{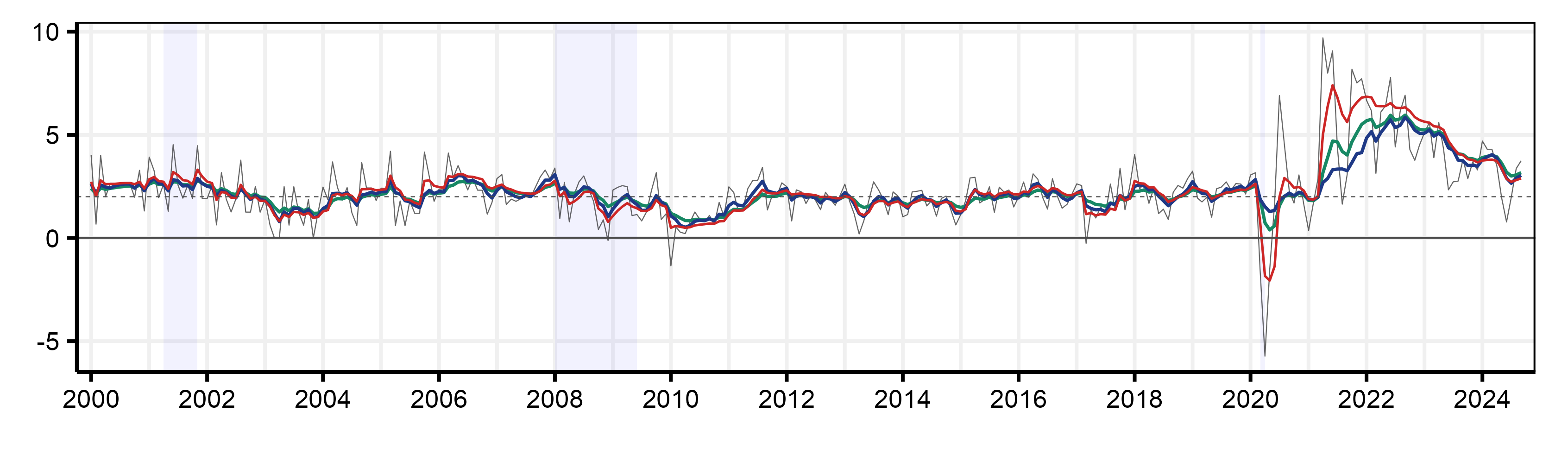}
    \end{minipage}%

    \vspace*{-1.1em}
    \begin{minipage}[t]{\textwidth}
      \centering
      (c) EA HICP Inflation
    \end{minipage}

     \begin{minipage}[t]{\textwidth}
      \centering
       \includegraphics[width=\textwidth, trim = 0mm -6mm 0mm 0mm, clip]{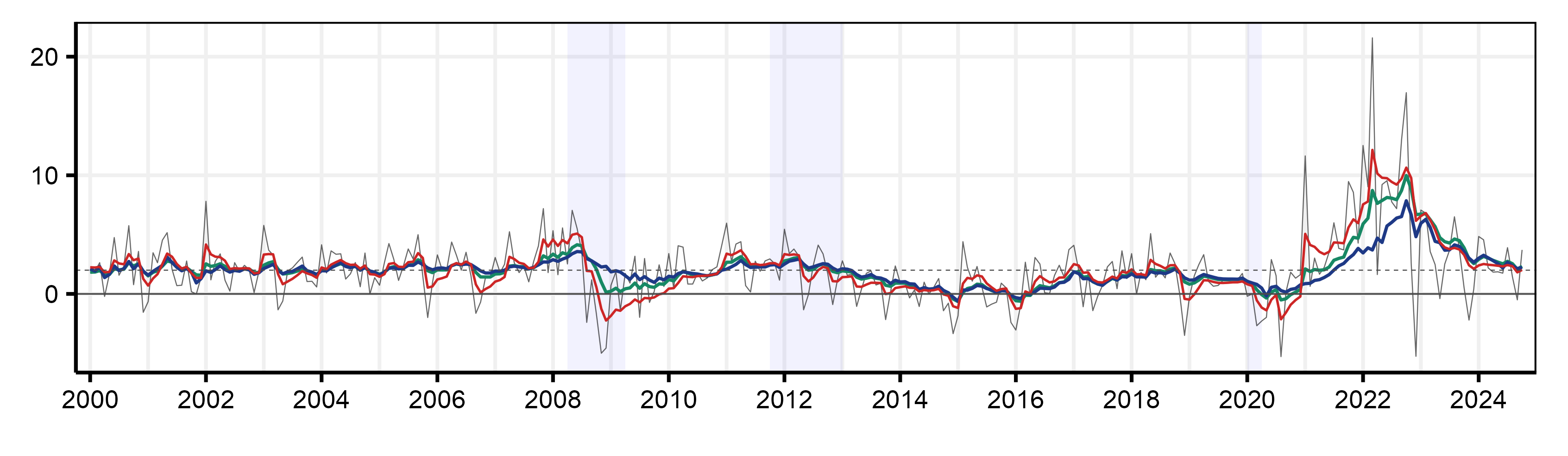}
    \end{minipage}%
 
    \vspace*{-1.1em}
     \begin{minipage}[t]{\textwidth}
      \centering
      (d) EA HICP Core Inflation
    \end{minipage}    

    \begin{minipage}[t]{\textwidth}
      \centering
       \includegraphics[width=\textwidth, trim = 1mm -5mm 2mm 0mm, clip]{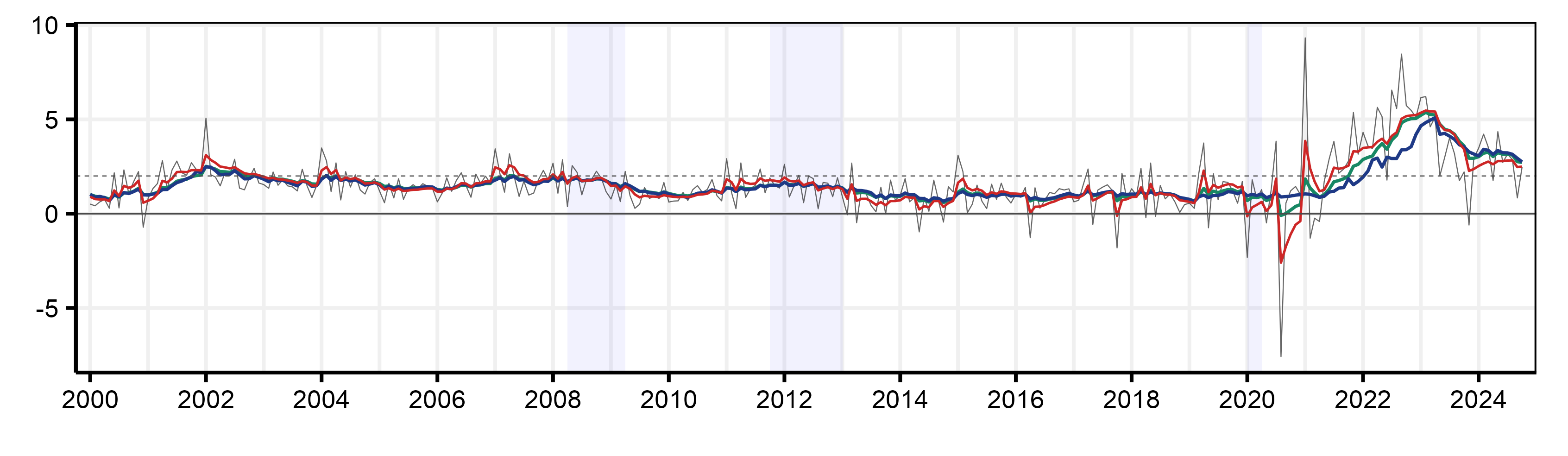}
    \end{minipage}%

    \vspace*{-1.8em}
    \begin{minipage}[t]{\textwidth}
      \centering
      \includegraphics[width=0.8\textwidth, trim = -100mm 0mm 0mm 0mm, clip]{AMA_RF_TS_right_UCSV_legend.png}
    \end{minipage}%

    \begin{threeparttable}
    \centering
    \begin{minipage}{\textwidth}
      \begin{tablenotes}[para,flushleft]
    \setlength{\lineskip}{0.1ex}
    \notsotiny 
  {\textit{Notes}: The figure compares AlbaMA to \cite{stock2007has}'s unobserved components model with stochastic volatility (UC-SV) and an exponential moving average with a smoothing factor over twelve months (EMA(12)). The panels refer to inflation and core inflation series in the US and the Euro Area.}
    \end{tablenotes}
  \end{minipage}
  \end{threeparttable}
\end{figure}


\begin{figure}[t!]
  \caption{\normalsize{Additional Benchmarks for Inflation (two-sided)}} \label{fig:infl_ucsv_center}
  
  \vspace*{-0.4em}
    \begin{minipage}[t]{\textwidth}
      \centering
      (a) US CPI Inflation
    \end{minipage}

    \begin{minipage}[t]{\textwidth}
      \centering
       \includegraphics[width=\textwidth, trim = 2mm -5mm 0mm 0mm, clip]{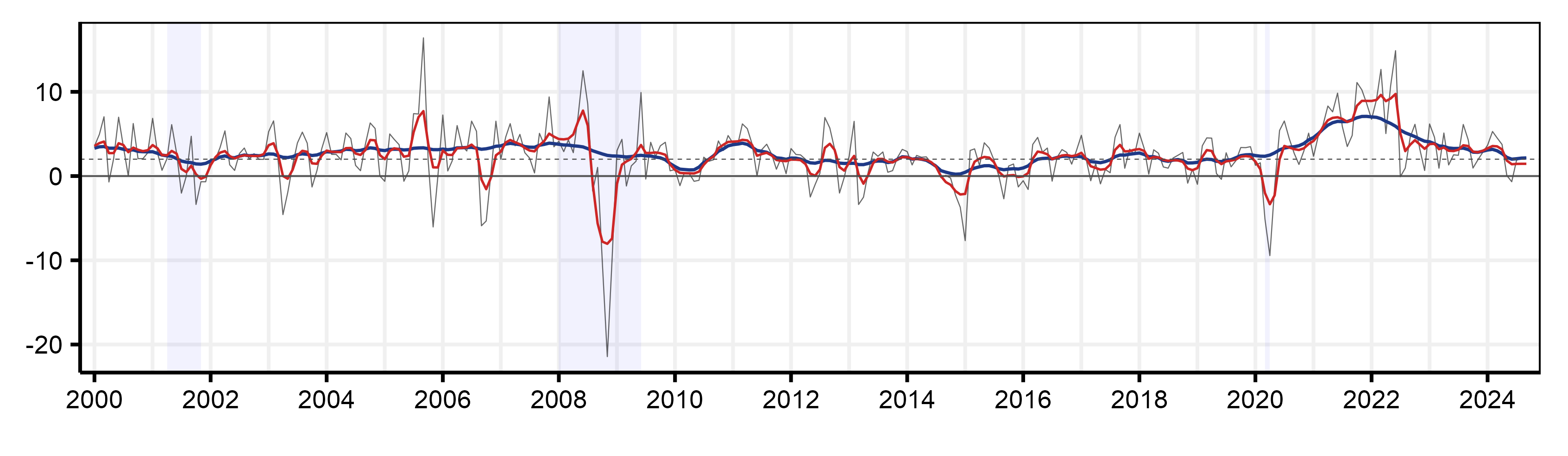}
    \end{minipage}%
   
    \vspace*{-1.1em}
    \begin{minipage}[t]{\textwidth}
      \centering 
      (b) US CPI Core Inflation
    \end{minipage}

    \begin{minipage}[t]{\textwidth}
      \centering
       \includegraphics[width=\textwidth, trim = -1mm -6mm 3mm 0mm, clip]{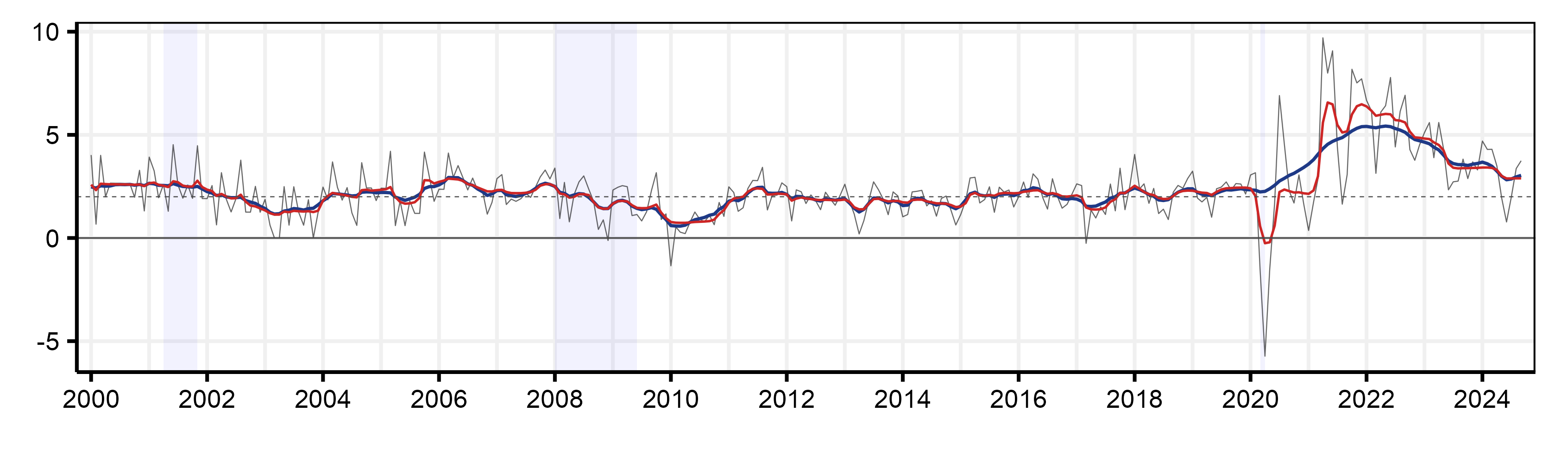}
    \end{minipage}%

    \vspace*{-1.1em}
    \begin{minipage}[t]{\textwidth}
      \centering
      (c) EA HICP Inflation
    \end{minipage}

     \begin{minipage}[t]{\textwidth}
      \centering
       \includegraphics[width=\textwidth, trim = 0mm -6mm 0mm 0mm, clip]{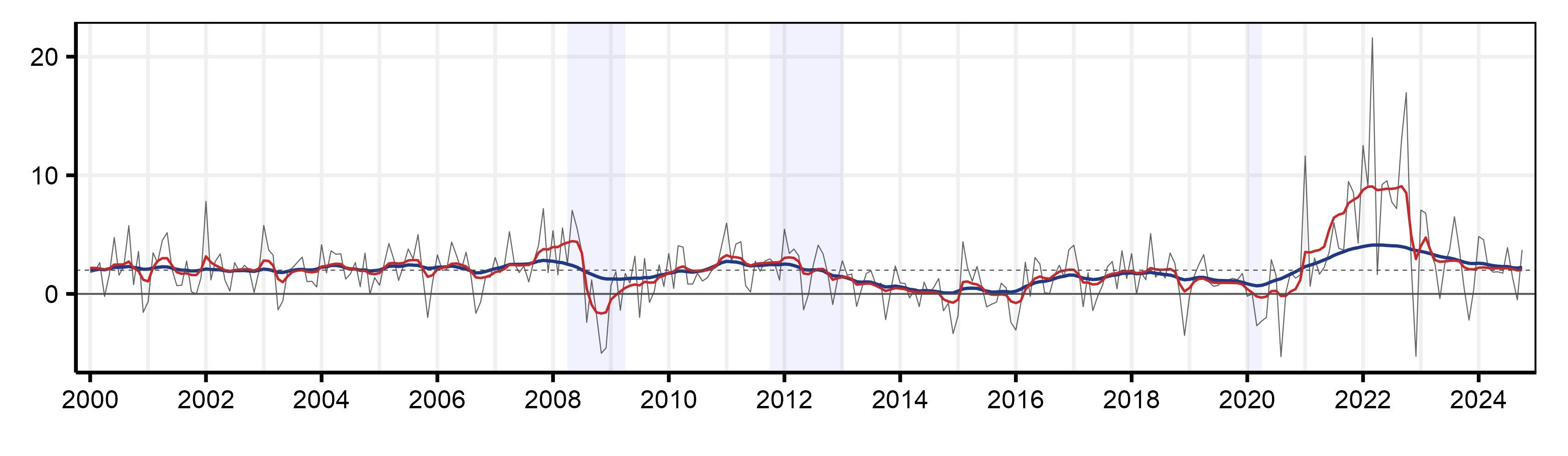}
    \end{minipage}%
 
    \vspace*{-1.1em}
     \begin{minipage}[t]{\textwidth}
      \centering
      (d) EA HICP Core Inflation
    \end{minipage}    

    \begin{minipage}[t]{\textwidth}
      \centering
       \includegraphics[width=\textwidth, trim = 1mm -5mm 2mm 0mm, clip]{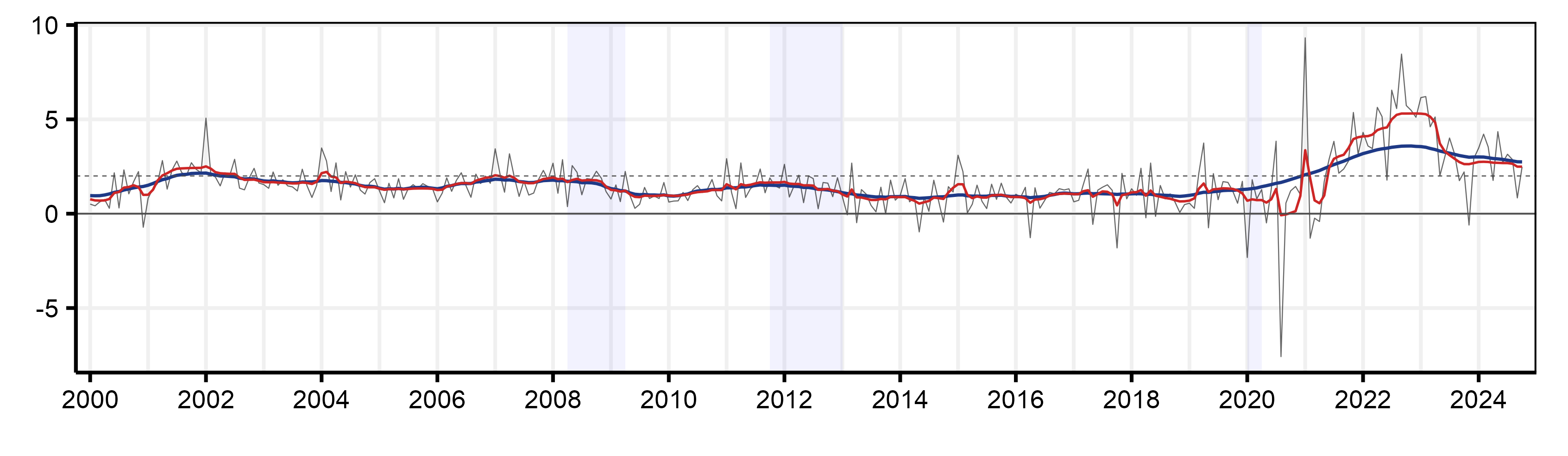}
    \end{minipage}%

    \vspace*{-1.8em}
    \begin{minipage}[t]{\textwidth}
      \centering
      \includegraphics[width=0.8\textwidth, trim = -100mm 0mm 0mm 0mm, clip]{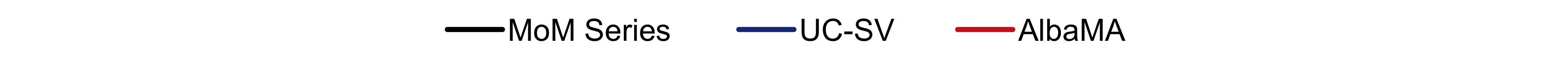}
    \end{minipage}%

    \begin{threeparttable}
    \centering
    \begin{minipage}{\textwidth}
      \begin{tablenotes}[para,flushleft]
    \setlength{\lineskip}{0.1ex}
    \notsotiny 
  {\textit{Notes}: The figure compares AlbaMA to \cite{stock2007has}'s unobserved components model with stochastic volatility (UC-SV). The panels refer to inflation and core inflation series in the US and the Euro Area.}
    \end{tablenotes}
  \end{minipage}
  \end{threeparttable}
\end{figure}

\begin{table}[ht]
  \footnotesize
  \centering
  \begin{threeparttable}
  \caption{\normalsize {AlbaMA for the US} \label{tab:results_us}
    \vspace{-0.3cm}}
    \setlength{\tabcolsep}{0.7em} 
    \setlength\extrarowheight{2.9pt}

    \begin{tabular}{l| rrrrrrr}
    \toprule \toprule
    \addlinespace[2pt]
    \multicolumn{1}{l}{} & \multicolumn{1}{c}{AlbaMA} & \multicolumn{1}{c}{SG MA} & \multicolumn{1}{c}{MA(3)} & \multicolumn{1}{c}{MA(6)} & \multicolumn{1}{c}{MA(6) vs (3)} & \multicolumn{1}{c}{MA(12)} & \multicolumn{1}{c}{MA(12) vs(6)} \\ 
    \midrule
    \rowcolor{gray!15} 
  \multicolumn{6}{l}{CPI Core Inflation} &   &\cellcolor{gray!15} \\ \addlinespace[2pt] 
  Full Sample & \textbf{0.93} & 0.46 & 0.89 & 0.86 & 0.88 & 0.86 & 0.92 \\ 
  Full Sample ex Cov & \textbf{0.93} & 0.45 & 0.89 & 0.87 & 0.89 & 0.86 & 0.92 \\ 
  Post 1990 & \textbf{0.91} & -0.29 &  0.78 &  0.82 &  0.79 &  0.81 &  0.88 \\ 
  Post 2020 & \textbf{0.83} & 0.08 & 0.69 & 0.52 & 0.41 & 0.41 & 0.73 \\ 
  Post 2021 & \textbf{0.77} & -0.91 &  0.62 &  0.28 &  0.07 & -0.32 &  0.51 \\ 
  Great Recession & \textbf{0.63} & -0.25 &  0.61 &  0.44 &  0.32 &  0.17 &  0.61 \\ 
  \midrule
  \rowcolor{gray!15} 
  \multicolumn{6}{l}{CPI Inflation} &   &\cellcolor{gray!15} \\ \addlinespace[2pt] 
  Full Sample & \textbf{0.87} & 0.53 & 0.81 & 0.72 & 0.76 & 0.77 & 0.83 \\ 
  Full Sample ex Cov & \textbf{0.88} & 0.52 & 0.82 & 0.73 & 0.77 & 0.78 & 0.84 \\ 
  Post 1990 & 0.51 & 0.42 & \textbf{0.57} & 0.08 & 0.17 & 0.12 & 0.12 \\ 
  Post 2020 & \textbf{0.91} & 0.45 & 0.75 & 0.59 & 0.62 & 0.31 & 0.68 \\ 
  Post 2021 & \textbf{0.95} & 0.22 & 0.77 & 0.63 & 0.71 & 0.12 & 0.66 \\ 
  Great Recession &  0.25 & \textbf{0.66} &  0.65 & -0.15 &  0.20 & -0.79 & -0.90 \\ 
  \midrule
\rowcolor{gray!15} 
\multicolumn{6}{l}{Industrial Production}  &  &\cellcolor{gray!15} \\ \addlinespace[2pt] 
  Full Sample & \textbf{0.72} & 0.66 & 0.55 & 0.27 & 0.32 & 0.23 & 0.40 \\ 
  Full Sample ex Cov & \textbf{0.81} & 0.72 & 0.68 & 0.47 & 0.63 & 0.28 & 0.53 \\ 
  Post 1990 & \textbf{0.85} & 0.73 & 0.70 & 0.57 & 0.71 & 0.34 & 0.57 \\ 
  Post 2020 & \textbf{0.55} &  0.44 &  0.27 & -0.55 & -1.02 & -1.02 & -1.78 \\ 
  Post 2021 & \textbf{0.70} &  0.34 & -0.47 &  0.11 & -0.60 & -2.00 &  0.40 \\ 
  Great Recession & \textbf{0.85} & 0.81 & 0.79 & 0.63 & 0.82 & 0.17 & 0.62 \\ 
  \midrule
\rowcolor{gray!15} 
\multicolumn{6}{l}{Unemployment Rate} &   &\cellcolor{gray!15} \\ \addlinespace[2pt] 
 Full Sample &  0.30 & \textbf{0.55} &  0.32 & -0.16 & -0.20 & -0.07 & -0.19 \\ 
  Full Sample ex Cov & \textbf{0.70} & 0.68 & 0.63 & 0.55 & 0.67 & 0.16 & 0.51 \\ 
  Post 1990 & \textbf{0.81} & 0.58 & 0.62 & 0.66 & 0.68 & 0.30 & 0.41 \\ 
  Post 2020 &  0.24 & \textbf{0.51} &  0.27 & -0.48 & -0.58 & -0.96 & -1.92 \\ 
  Post 2021 &  0.53 &  0.76 &  0.72 &  0.05 & \textbf{0.78} & -1.70 &  0.71 \\ 
  Great Recession & \textbf{0.81} & 0.76 & 0.81 & 0.72 & 0.80 & 0.45 & 0.75 \\ 
  \midrule
\rowcolor{gray!15} 
\multicolumn{6}{l}{PMI} &   &\cellcolor{gray!15} \\ \addlinespace[2pt] 
  Full Sample & \textbf{0.69} &  0.64 &  0.46 & -0.03 &  0.07 & -0.38 & -0.12 \\ 
  Full Sample ex Cov & \textbf{0.69} &  0.64 &  0.48 & -0.03 &  0.09 & -0.39 & -0.13 \\ 
  Post 1990 &  0.63 & \textbf{0.66} &  0.47 & -0.02 &  0.20 & -0.37 & -0.18 \\ 
  Post 2020 & \textbf{0.64} &  0.50 &  0.22 &  0.14 & -0.28 & -0.12 &  0.33 \\ 
  Post 2021 &  0.60 & \textbf{0.62} &  0.35 &  0.28 &  0.34 & -0.76 &  0.44 \\ 
  Great Recession &  0.60 & \textbf{0.74} &  0.60 &  0.02 &  0.42 & -0.40 & -0.46 \\ 
   \bottomrule \bottomrule
\end{tabular}
\begin{tablenotes}[para,flushleft]
  \scriptsize 
    \textit{Notes}: The table gives R$^2$ between one-sided and two-sided estimates. SG MA refers to the Savitzgy-Golay filter, while MA(3), MA(6) and MA(12) denote three-, six-, and twelve-months moving averages. MA(6) vs (3) indicates the R$^2$ between the two-sided MA(6) versus the one-sided MA(3), likewise MA(12) vs (6) shows the R$^2$ between the two-sided MA(12) versus the one-sided MA(6). Full sample runs from 1963m1 to 2024m10. Full sample excluding Covid excludes observations in the year 2020. Post 1990 runs from 1990m1 to 2024m10, likewise for Post 2020 and Post 2021. The evaluation sample for the Great Recession is set from 2008m1 to 2011m12. 
  \end{tablenotes}
\end{threeparttable}
\end{table}

\begin{table}[h]
  \footnotesize
  \centering
  \begin{threeparttable}
  \caption{\normalsize {AlbaMA for the EA} \label{tab:results_ea}
    \vspace{-0.3cm}}
    \setlength{\tabcolsep}{0.7em} 
    \setlength\extrarowheight{2.9pt}

    \begin{tabular}{l| rrrrrrr}
    \toprule \toprule
    \addlinespace[2pt]
    \multicolumn{1}{l}{} & \multicolumn{1}{c}{AlbaMA} & \multicolumn{1}{c}{SG MA} & \multicolumn{1}{c}{MA(3)} & \multicolumn{1}{c}{MA(6)} & \multicolumn{1}{c}{MA(6) vs (3)} & \multicolumn{1}{c}{MA(12)} & \multicolumn{1}{c}{MA(12) vs (6)} \\ 
    \midrule
    \rowcolor{gray!15} 
  \multicolumn{6}{l}{HICP Core Inflation} &   &\cellcolor{gray!15} \\ \addlinespace[2pt] 
  Full Sample & 0.85 & 0.43 & 0.77 & 0.80 & 0.81 & 0.72 & \textbf{0.86} \\ 
  Full Sample ex Cov & \textbf{0.89} & 0.42 & 0.81 & 0.83 & 0.85 & 0.73 & 0.88 \\ 
  Post 1990 &  0.71 & -0.46 &  0.55 &  0.60 &  0.63 &  0.61 & \textbf{0.72} \\ 
  Post 2020 & 0.80 & 0.39 & 0.73 & 0.74 & 0.73 & 0.55 & \textbf{0.80} \\ 
  Post 2021 & \textbf{0.76} & -0.06 &  0.70 &  0.60 &  0.65 &  0.02 &  0.62 \\ 
  Great Recession & \textbf{0.80} & -1.48 &  0.48 &  0.68 &  0.70 &  0.33 &  0.74 \\ 
  \midrule
  \rowcolor{gray!15} 
  \multicolumn{6}{l}{HICP Inflation} &   &\cellcolor{gray!15} \\ \addlinespace[2pt] 
  Full Sample & \textbf{0.78} & 0.60 & 0.72 & 0.66 & 0.68 & 0.56 & 0.75 \\ 
  Full Sample ex Cov & \textbf{0.78} & 0.59 & 0.72 & 0.67 & 0.69 & 0.57 & 0.77 \\ 
  Post 1990 & \textbf{0.65} & 0.40 & 0.60 & 0.29 & 0.34 & 0.36 & 0.45 \\ 
  Post 2020 & \textbf{0.75} & 0.58 & 0.73 & 0.72 & 0.74 & 0.42 & 0.75 \\ 
  Post 2021 & 0.64 & 0.40 & 0.65 & 0.64 & 0.64 & 0.07 & \textbf{0.65} \\ 
  Great Recession &  0.55 &  0.59 & \textbf{0.68} &  0.21 &  0.55 & -0.17 &  0.07 \\ 
  \midrule
\rowcolor{gray!15} 
\multicolumn{6}{l}{Industrial Production} &   &\cellcolor{gray!15} \\ \addlinespace[2pt] 
  Full Sample & \textbf{0.74} &  0.53 &  0.24 & -0.17 & -0.28 & -0.24 & -0.19 \\ 
  Full Sample ex Cov & \textbf{0.70} &  0.66 &  0.38 &  0.32 &  0.50 & -0.07 &  0.35 \\ 
  Post 1990 & \textbf{0.78} & 0.68 & 0.53 & 0.39 & 0.57 & 0.09 & 0.38 \\ 
  Post 2020 & \textbf{0.73} &  0.41 &  0.14 & -0.64 & -1.03 & -1.40 & -2.31 \\ 
  Post 2021 & -0.22 & \textbf{0.36} & -0.67 & -0.62 & -0.47 & -2.62 & -0.05 \\ 
  Great Recession & \textbf{0.86} &  0.78 &  0.76 &  0.40 &  0.67 & -0.07 &  0.34 \\ 
  \midrule
\rowcolor{gray!15} 
\multicolumn{6}{l}{Unemployment Rate} &   &\cellcolor{gray!15} \\ \addlinespace[2pt] 
  Full Sample & \textbf{0.94} & 0.78 & 0.83 & 0.66 & 0.74 & 0.50 & 0.75 \\ 
  Full Sample ex Cov & \textbf{0.96} & 0.81 & 0.89 & 0.78 & 0.86 & 0.53 & 0.82 \\ 
  Post 1990 & \textbf{0.96} & 0.81 & 0.89 & 0.80 & 0.87 & 0.64 & 0.84 \\ 
  Post 2020 & \textbf{0.87} &  0.69 &  0.70 &  0.17 &  0.27 & -0.23 &  0.24 \\ 
  Post 2021 & \textbf{0.91} &  0.68 &  0.79 &  0.42 &  0.69 & -1.52 &  0.45 \\ 
  Great Recession & \textbf{0.93} &  0.70 &  0.87 &  0.52 &  0.78 & -0.16 &  0.49 \\ 
   \bottomrule \bottomrule
\end{tabular}
\begin{tablenotes}[para,flushleft]
  \scriptsize 
    \textit{Notes}: The table gives R$^2$ between one-sided and two-sided estimates. SG MA refers to the Savitzgy-Golay filter, while MA(3), MA(6) and MA(12) denote three-, six-, and twelve-months moving averages. MA(6) vs (3) indicates the R$^2$ between the two-sided MA(6) versus the one-sided MA(3), likewise MA(12) vs (6) shows the R$^2$ between the two-sided MA(12) versus the one-sided MA(6). Full sample runs from 1963m1 to 2024m10. Full sample excluding Covid excludes observations in the year 2020. Post 1990 runs from 1990m1 to 2024m10, likewise for Post 2020 and Post 2021. The evaluation sample for the Great Recession is set from 2008m1 to 2011m12. 
  \end{tablenotes}
\end{threeparttable}
\end{table}

\end{document}